\documentclass[prd,twocolumn,showpacs,nofootinbib]{revtex4}
\pdfoutput=1

\usepackage{amsfonts}
\usepackage{amsmath}
\usepackage{amssymb}
\usepackage{aas_macros}
\usepackage{upgreek}
\usepackage{bm}
\usepackage{dcolumn}
\usepackage{epsfig}
\usepackage{graphicx}
\usepackage{graphics}
\usepackage{placeins}
\usepackage{subfigure}
\usepackage{color}

\def\be{\begin{equation}}
\def\ee{\end{equation}}
\def\bi{\begin{itemize}}
\def\ei{\end{itemize}}
\def\ben{\begin{enumerate}}
\def\een{\end{enumerate}}

\def\edth{\check\partial}

\begin{document}

\title{Mapping gravitational-wave backgrounds in modified theories of gravity using pulsar timing arrays}

\author{Jonathan R.~Gair}
\affiliation{Institute of Astronomy, University of Cambridge, Madingley Road, Cambridge, CB3 0HA, UK}
\author{Joseph D.~Romano}
\affiliation{Department of Physics and Astronomy 
and Center for Gravitational-Wave Astronomy,
University of Texas at Brownsville, Brownsville, TX 78520, USA}
\author{Stephen R.~Taylor}
\affiliation{Jet Propulsion Laboratory, 
California Institute of Technology, 
4800 Oak Grove Drive, Pasadena, CA 91106, USA}

\date{\today}

%%%%%%%%%%%%%%%%%%%%%%%%%%%%%%%%%%%%%%%%%%%%%%%

\begin{abstract}
We extend our previous work on applying CMB techniques to the 
mapping of gravitational-wave backgrounds
to backgrounds which have non-GR polarisations. 
Our analysis and results are presented in the context
of pulsar-timing array observations, but the overarching methods are
general, and can be easily applied to LIGO or eLISA observations
using appropriately modified response functions.
Analytic expressions for the pulsar-timing response to gravitational waves with 
non-GR polarisation are given for each mode of a spin-weighted
spherical-harmonic decomposition of the background, which
permit the signal to be mapped across the sky to any desired
resolution. We also derive the pulsar-timing overlap reduction 
functions for the various non-GR polarisations, finding analytic forms
for anisotropic backgrounds with scalar-transverse (``breathing'') and
vector-longitudinal polarisations, and a semi-analytic form for scalar-longitudinal backgrounds. Our results indicate that pulsar-timing
observations will be completely insensitive to scalar-transverse mode
anisotropies in the polarisation amplitude beyond dipole, and
anisotropies in the power beyond quadrupole. Analogously to our
previous findings that pulsar-timing observations lack sensitivity
to tensor-curl modes for a transverse-traceless tensor background, we 
also find insensitivity to vector-curl modes for a 
vector-longitudinal background. 
\end{abstract}

%%%%%%%%%%%%%%%%%%%%%%%%%%%%%%%%%%%%%%%%%%%%%%%

\maketitle

\section{Introduction}
\label{s:intro}
A massive international effort is currently underway to observe gravitational waves across a wide range of frequencies. The second-generation of ground-based
gravitational-wave interferometers are about to start collecting data, with Advanced
LIGO~\cite{aLIGO} observation runs expected to begin before the end of 2015. 
The two Advanced LIGO detectors will form part of a global network of
kilometre-scale laser interferometers, with other instruments due to come online during the rest of this decade. These detectors will employ advanced technologies
to detect gravitational waves from stellar-mass compact binary
systems emitting gravitational radiation in the kHz 
band~\cite{kagra2012,indigoupdate2013,GEO-600,AdvVirgo}. 
The European Space Agency recently selected a science theme based around a $\sim \!10^9~{\rm m}$ arm-length space-based gravitational-wave interferometer 
(eLISA) for the L$3$ mission slot, due to launch in 2034. 
Such a detector will observe gravitational waves in the millihertz band, which are generated by binaries involving the massive black holes that reside in the centres of galaxies, with mass about one million times the mass of the Sun. These observations will permit tests of
fundamental physics to exquisite precision, whilst
also affording detailed demographic studies of massive black-hole
populations~\cite{eLISA:2012}.

Complementary to these experiments are ongoing efforts to
characterise nanohertz gravitational waves through their perturbation to the
arrival-times of radio signals from precisely timed ensembles of
millisecond pulsars spread throughout our 
galaxy~\cite{vanHaasteren:2011,Demorest:2013,Shannon:2013,iptareview2013}. 
As a gravitational wave transits
between the Earth and a pulsar, it induces a change in their proper
separation, leading to a redshift in the arrival rate of the
pulsar signals \cite{sazhin-1978,detweiler-1979,estabrook-1975,burke-1975}. 
It is the exceptional stability of the integrated
pulse profiles of millisecond pulsars, and the resulting accuracy of
the models for the pulse times of arrival (TOAs), that allow gravitational waves to be detected in this way.

The differences between the modelled TOAs and the actual observed TOAs
are known as the \textit{timing residuals}. These residuals contain
the influence of all unmodelled phenomena, such as additional receiver
noise, interstellar medium effects, errors resulting from drifts in
clock standards
or ephemeris inaccuracies, and, most tantalisingly, gravitational
radiation. The signature of gravitational waves in these residuals may be
deterministic or stochastic. The gravitational-wave sources expected to dominate the signal in the nanohertz frequency band are the early adiabatic inspirals of
supermassive black-hole binary (SMBHB) 
systems~\cite{rajaromani1995,jaffe-backer-2003,wyithe-loeb-2003}. Such systems are expected to form following the (suspected ubiquitous) mergers of massive galaxies
during the hierarchical formation of structure. If there is a system which is particularly loud
in gravitational-wave emission then this signal may be individually resolved and
detected with pipelines dedicated to searches for the deterministic
signals of single 
sources~\cite{ellisoptimal2012,ellis-bayesian-2013,taylor-cw-2014}. 
If, however, there are many sources which
pile up in the frequency-domain beyond the ability of our techniques
to separately resolve them, then the combined signal will form  
a stochastic background of gravitational waves. 
Although there are other mechanisms which may
contribute to a stochastic nHz gravitational-wave background (decay of cosmic-string
networks~\cite{vilenkin-1981a,vilenkin-1981b,olmez-2010,sanidas-2012} or
primordial remnants~\cite{grishchuk-1976,grishchuk-2005}), this 
incoherent superposition of signals from many SMBHB systems is expected 
to dominate the signal. 

Standard pipelines in use today employ cross-correlation techniques to
search for stochastic backgrounds. The presence of a common background
of gravitational waves affecting the TOAs of all pulsars in an array (a so-called
pulsar-timing array, PTA~\cite{foster-backer-1990}) makes a cross-correlation
search effective in leveraging the signal against uncorrelated
noise processes. The concept of an \textit{overlap reduction function} 
is common to stochastic background searches for all types of gravitational-wave 
detectors, and describes the sky-averaged overlap of the antenna pattern
functions of the two detectors whose data are being 
correlated~\cite{Flanagan:1993}. 
In PTA analysis, the overlap reduction function for a Gaussian, stationary,
unpolarised, isotropic stochastic background composed of
transverse-traceless (TT) gravitational-wave modes is a smoking-gun signature of the
signal, known as the Hellings and Downs curve~\cite{HellingsDowns:1983}. 
It is a function of one variable: the angular separation between a pair 
of pulsars.

For anisotropic distributions of gravitational-wave power on the sky, 
the overlap reduction function is no longer merely a function of the pulsars'
angular-separation. It will also depend on the positions of the
pulsars on the sky relative to the distribution of gravitational-wave power, and thus
will be a rich source of information in the precision-science-era of
PTAs~\cite{Mingarelli:2013,TaylorGair:2013}. 
Furthermore, the overlap reduction function can be shown to vary when describing backgrounds 
where the graviton
is permitted to have a small but non-zero mass \cite{lee2014}. 
The same is true when describing the overlap reduction functions induced by gravitational-wave polarisation 
%states which are only present in the most general (i.e. beyond-GR) 
%metric description of gravity. 
states present in {\em modified} (metric) theories of gravity.
In addition to the usual GR transverse-traceless tensor polarisation states, the
beyond-GR polarisations consist of a scalar-transverse (``breathing'') state, a
scalar-longitudinal state, and two vector-longitudinal states, each
inducing correlation signatures which are markedly distinct from the
Hellings and Downs curve~\cite{lee2008,chamberlin2012}.

In this paper we focus on the response of pulsar timing observations
to gravitational wave backgrounds with non-GR polarisation states. By decomposing a background of given polarisation in terms of spin-weighted spherical harmonics, we are able
to derive analytic expressions for the detector response functions for
each mode of each non-GR polarisation state as a function of the harmonic multipole. We discuss
the implications of these results for mapping non-GR backgrounds to
any desired angular resolution. We are also able to present analytic
expressions for the overlap reduction functions of anisotropic
scalar-transverse and vector-longitudinal backgrounds, whilst
significant analytic headway is made for the corresponding function for
scalar-longitudinal backgrounds.

In Sec.~\ref{s:response-section} we introduce the concept of the
measured signal in a gravitational-wave detector being a convolution of the metric
perturbations with the response tensor of the detector. We discuss the
six distinct polarisation states of gravitational waves which are permitted within a
general metric theory of gravity by virtue of obeying Einstein's
Equivalence Principle. 
The basis tensors for these polarisations are explicitly given. 
We also discuss the decomposition of the metric perturbations in terms
of appropriate spin-weighted spherical harmonics.
In \cite{gair-2014}, the Fourier amplitudes of a plane-wave expansion
of the metric perturbations for an arbitrary transverse-traceless 
gravitational-wave background were decomposed in terms of a basis of 
spin-weight~$\pm 2$ spherical harmonics. 
In the case of scalar-transverse and scalar-longitudinal polarisations
discussed in this paper, we decompose the Fourier amplitudes
in terms of ordinary (spin-weight $0$) spherical harmonics. 
For the vector-longitudinal polarisations, we decompose the 
Fourier amplitudes in terms of spin-weight~$\pm 1$ spherical harmonics. 
In Sec.~\ref{s:response-section}, we also give expressions for 
the pulsar timing response functions, for either the polarisation
or spin-weighted spherical harmonic expansion coefficients.
The polarisation basis response functions for a pair of  pulsars
are given explicitly in the computational frame, where one pulsar lies
along the $z$-axis and the other lies in the $xz$-plane. 
These are needed for the overlap reduction functions calculations
given in the following section.

The overlap reduction functions for the different polarisation states 
are studied in Sec.~\ref{s:orf-sec}. This function describes the 
response of a pair of pulsars to a gravitational-wave background in a
cross-correlation analysis, and is computed by integrating the overlap
of the response of each pulsar to a particular gravitational-wave 
polarisation over the entire sky. 
For a gravitational-wave background with arbitrary angular structure,
this sky integral must be weighted by the gravitational-wave power at each
sky location. 
We find an analytic expression for the overlap reduction function for a background with scalar-transverse (breathing) polarisation, and show that a PTA will lack sensitivity to angular
structure beyond quadrupole in a cross-correlation analysis for this
type of background. 
We also make significant analytic headway for the overlap reduction function
of a scalar-longitudinal background, and find analytic forms for
the limiting value in the case of co-directional and anti-directional
pulsars. The overlap reduction function for
a vector-longitudinal background with arbitrary angular structure is
found analytically, with superficially perceived divergences in the
overlap reduction function for co-directional pulsars resolved by correctly incorporating the
pulsar term in our calculations.

In Sec.~\ref{s:mapping-sec} we extend our previous work on mapping gravitational-wave backgrounds using CMB techniques
\citep{gair-2014} to non-GR polarisations. 
We derive analytic expressions for the response of a pulsar to each mode (corresponding to a particular spin-weighted spherical harmonic) of the background, including the contribution
from the pulsar term.
In the process of doing these calculations, we find that the
reason for the PTA insensitivity to angular structure beyond
quadrupole in the gravitational-wave power of a scalar-transverse background is due entirely to the
corresponding lack of sensitivity of a single pulsar response to structure in
the polarisation amplitudes beyond dipole. We verify this analytic
result with numerical map making and recovery. The pulsar response to
individual modes of a scalar-longitudinal and vector-longitudinal background are
given analytically, where in the latter case we find that PTAs 
completely lack sensitivity to vector curl modes, analogous to our 
previous finding that PTAs lack sensitivity to tensor curl
modes of a transverse-traceless background \citep{gair-2014}. We
discuss these findings further in Sec.~\ref{s:conclusion}, along with
suggestions for future study and implications for the forthcoming analysis of real
PTA data.

Finally, we include several appendices 
(Apps.~\ref{s:spinweightedY}--\ref{s:appRecoverBreathingORF}), 
containing relevant information (e.g., definitions, identities, recurrence relations) 
for spin-weighted and tensor spherical harmonics, Legendre polynomials, 
Bessel functions, etc., as well as providing technical details for 
the overlap reduction function and response 
function calculations described in Secs.~\ref{s:orf-sec} and \ref{s:mapping-sec}.

\section{Response functions}
\label{s:response-section}

\subsection{Detector response}
\label{s:detresponse}

The response of a detector to a passing gravitational wave is given by
the convolution of the metric perturbations $h_{ab}(t,\vec x)$ with
the impulse response $R^{ab}(t,\vec x)$ of the detector:
\be
r(t) 
= \int_{-\infty}^\infty {\rm d}\tau
\int {\rm d}^3y\>
R^{ab}(\tau,\vec y)
h_{ab}(t-\tau,\vec x-\vec y)\,.
\ee
If we write the metric perturbations as a superposition of
plane waves
\be
h_{ab}(t,\vec x)
=\int_{-\infty}^\infty {\rm d}f
\int_{S^2}{\rm d}^2\Omega_{\hat k}\>
h_{ab}(f,\hat k)e^{i2\pi f(t-\hat k\cdot\vec x/c)}\,,
\ee
then 
\be
r(t)=\int_{-\infty}^\infty {\rm d}f
\int_{S^2} {\rm d}^2\Omega_{\hat k}\>
R^{ab}(f,\hat k)h_{ab}(f,\hat k)
e^{i2\pi f t}\,,
\ee
where
\begin{multline}
R^{ab}(f,\hat k)
=e^{-i2\pi f\hat k\cdot\vec x/c}
\\
\times\int_{-\infty}^\infty {\rm d}\tau
\int {\rm d}^3 y\> 
R^{ab}(\tau,\vec y)\,
e^{-i2\pi f(\tau-\hat k\cdot\vec y/c)}\,.
\end{multline}
Further specification of the response function
depends on the choice of gravitational-wave
detector as well as on the basis tensors used to 
expand $h_{ab}(f,\hat k)$, as we explain below.

%%%%%%%%%%%%%%%%
\subsection{Polarisation basis}
\label{s:polbasis}

In standard GR, the Fourier components 
$h_{ab}(f,\hat k)$ are typically expanded in terms of 
the $+$ and $\times$ polarisation basis tensors:
\be
h_{ab}(f,\hat k)
=h_+(f,\hat k) e^+_{ab}(\hat k) + h_\times (f,\hat k) e^\times_{ab}(\hat k)\,,
\ee
where
\be
\begin{aligned}
e_{ab}^+(\hat k)
&=\hat\theta_a\hat\theta_b-\hat\phi_a\hat\phi_b,
\\
e_{ab}^\times(\hat k)
&=\hat\theta_a\hat\phi_b+\hat\phi_a\hat\theta_b,
\end{aligned}
\ee
and $\hat\theta$, $\hat\phi$ are the standard
unit vectors tangent to the sphere:
\be
\begin{aligned}
\hat k
&=\sin\theta\cos\phi\,\hat x+
\sin\theta\sin\phi\,\hat y+
\cos\theta\,\hat z\,,
\\
\hat\theta
&=\cos\theta\cos\phi\,\hat x+
\cos\theta\sin\phi\,\hat y-
\sin\theta\,\hat z\,, 
\\
\hat\phi
&=-\sin\phi\,\hat x+
\cos\phi\,\hat y\,.
\end{aligned}
\ee
In this paper, we also consider modified metric theories 
of gravity, which admit four other types of polarisation:
a scalar-transverse (or breathing) mode ($B$), 
a scalar-longitudinal mode ($L$), and 
two vector-longitudinal modes ($X$, $Y$). 
The polarisation basis tensors for these modes are:
\begin{align}
e_{ab}^{B}(\hat k)
&=\hat\theta_a\hat\theta_b+\hat\phi_a\hat\phi_b\,,
\label{e:eSTdef}
\\
e_{ab}^{L}(\hat k)
&=\sqrt{2}\,\hat k_a\hat k_b\,,
\label{e:eSLdef}
\\
e_{ab}^{X}(\hat k)
&=\hat\theta_a\hat k_b+\hat k_a\hat\theta_b\,,
\label{e:eVLxdef}
\\
e_{ab}^{Y}(\hat k)
&=\hat\phi_a\hat k_b + \hat k_a\hat\phi_b\,.
\label{e:eVLydef}
\end{align}
In terms of the polarisation tensors, the Fourier 
components $h_{ab}(f,\hat k)$ can be expanded 
generally as
\be
h_{ab}(f,\hat k) 
= \sum_A h_A(f,\hat k) e^A_{ab}(\hat k)
\ee
where $A$ is some subset of $\{+,\times,B,L,X,Y\}$.
The associated response function for a plane wave
with frequency $f$, propagation direction $\hat k$,
and polarisation $A$ is given by
\be
R^A(f,\hat k) = R^{ab}(f,\hat k)e^A_{ab}(\hat k)\,,
\ee
and is related to the detector response $r(t)$ via:
\be
r(t)=\int_{-\infty}^\infty {\rm d}f
\int_{S^2} {\rm d}^2\Omega_{\hat k}
\sum_A
R^A(f,\hat k)h_A(f,\hat k)e^{i2\pi f t}\,.
\label{e:responseRA}
\ee
We will work with the polarisation basis response
functions when calculating the various overlap
reduction functions in Sec.~\ref{s:orf-sec}.

%%%%%%%%%%%%%%%%%%%%%%%%%%%%%%%%%%%%
\subsection{Spherical harmonic basis}
\label{s:sphbasis}

Alternatively, we can expand the Fourier components
$h_{ab}(f,\hat k)$ in terms
of the appropriate spin-weighted spherical harmonics, as
was done in~\cite{gair-2014}. A spin-weighted function is a 
function of both position on the sphere, labelled $\hat{k}$, 
and of a choice of an orthonormal basis, labelled $\hat{l}, \hat{m}$, 
at points on the sphere. Under a rotation of the orthonormal basis, 
spin-weight functions transform in a particular way
\begin{align}
f(\hat{k},\cos\psi \hat{l} - \sin\psi\hat{m},\sin\psi\hat{l}+\cos\psi\hat{m}) = 
{\rm e}^{is\psi}f(\hat k,\hat l,\hat m)
\end{align}
where $s$ is the spin-weight of the function. 
Any spin-weight $s$ function can be expanded as a combination 
of spin-weighted spherical harmonics of the same weight, ${}_sY_{lm}(\hat{k})$. 
A spin-weight $s$ spherical-harmonic can be related to $s$ derivatives of an
ordinary spherical harmonic, as described in App.~\ref{s:spinweightedY}.

For the standard GR tensor modes, if we define $\hat{m}_{\pm}^a = \hat{l}^a \pm i \hat{m}^a$, 
we see that the combinations $\hat{m}_{\pm}^a \hat{m}_{\pm}^b h_{ab}(f,\hat{k})$are spin-weight $\pm2$ functions on the sphere. 
This allows the GR tensor modes to be expanded as combinations of 
spin-weight $\pm2$ 
spherical harmonics, or equivalently in terms of the rank-2 gradient and curl 
spherical harmonics, $Y^G_{(lm)ab}(\hat k)$, $Y^C_{(lm)ab}(\hat k)$, defined by Eq.~(\ref{e:YGClmab}) in App.~\ref{s:grad-curl-tensor}:
\begin{multline}
h_{ab}(f,\hat k)
=\sum_{l=2}^\infty \sum_{m=-l}^l
\left[a^G_{(lm)}(f)Y^G_{(lm)ab}(\hat k)
\right.
\\
\left.
+a^C_{(lm)}(f)Y^C_{(lm)ab}(\hat k)\right]\,.
\end{multline}

For the breathing and scalar-longitudinal modes, the functions $\hat{m}_{\pm}^a \hat{m}_{\pm}^b h_{ab}(f,\hat{k})$ 
are spin-weight $0$ and so we can expand 
$h_{ab}(f,\hat k)$ in terms of ordinary (scalar) spherical 
harmonics:
\begin{align}
h_{ab}(f,\hat k) 
&=\frac{1}{\sqrt{2}}
\sum_{l=0}^\infty \sum_{m=-l}^l
a^B_{(lm)}(f)Y_{lm}(\hat k)e^B_{ab}(\hat k)\,,
\\
h_{ab}(f,\hat k) 
&=\frac{1}{\sqrt{2}}
\sum_{l=0}^\infty \sum_{m=-l}^l
a^L_{(lm)}(f)Y_{lm}(\hat k)e^L_{ab}(\hat k)\,,
\end{align}
since the polarisation tensors
$e^B_{ab}(\hat k)$ and $e^L_{ab}(\hat k)$ are invariant
under a rotation of $\hat\theta$, $\hat\phi$.
For the vector-longitudinal modes, $\hat{m}_{\pm}^a \hat{m}_{\pm}^b h_{ab}(\hat{k})$ have 
spin-weight $\pm1$ and so we can expand $h_{ab}(f,\hat k)$ in terms of 
spin-weight $\pm1$ spherical harmonics or, equivalently, in terms of tensor fields $Y^{V_G}_{(lm)ab}(\hat k)$,
$Y^{V_C}_{(lm)ab}(\hat k)$ constructed from the 
rank-1 vector spherical harmonics 
$Y^G_{(lm)a}(\hat k)$, $Y^C_{(lm)a}(\hat k)$
defined by Eqs.~(\ref{e:YGClma}) and (\ref{e:YVGVClmab}) in App.~\ref{s:grad-curl-vector}:
\begin{multline}
h_{ab}(f,\hat k)
= \sum_{l=1}^\infty \sum_{m=-l}^l
\left[a^{V_G}_{(lm)}(f)Y^{V_G}_{(lm)ab}(\hat k)
\right.
\\
\left.
+a^{V_C}_{(lm)}(f)Y^{V_C}_{(lm)ab}(\hat k)\right]\,.
\label{e:hab-vector-GC}
\end{multline}

The above expressions for $h_{ab}(f,\hat k)$ can be written 
in compact form 
\be
h_{ab}(f,\hat k) = \sum_{(lm)}\sum_P
a_{(lm)}^P(f) Y^P_{(lm)ab}(\hat k)
\ee
if we take $P$ to be a subset of $\{G,C,B,L,V_G,V_C\}$, and 
define 
\be
Y^{B,L}_{(lm)ab}(\hat k)\equiv 
\frac{1}{\sqrt{2}}Y_{lm}(\hat k)e^{B,L}_{ab}(\hat k)
\label{e:YBLlmab}
\ee
to unify the notation for the spherical harmonic basic
tensors.
(The factor of $1/\sqrt{2}$ is needed for the tensor
spherical harmonics $Y^{B,L}_{(lm)ab}(\hat k)$ to satisfy
orthonormality relations similar to 
Eqs.~(\ref{e:YVGVCOrthog}) and (\ref{e:YGCOrthog}).)
The associated response function for a given spherical harmonic
mode is%, with component $a^P_{(lm)}(f)$, is:
\be
R^P_{(lm)}(f) 
= \int_{S^2}{\rm d}^2\Omega_{\hat k}\>
R^{ab}(f,\hat k)Y^{P}_{(lm)ab}(\hat k)\,,
\ee
and are related to the detector response $r(t)$ via:
\be
r(t)=\int_{-\infty}^\infty {\rm d}f
\sum_{(lm)}
\sum_P
R^P_{(lm)}(f)a^P_{(lm)}(f)e^{i2\pi f t}\,.
\label{e:responseRP}
\ee
We will work with these response functions
for the mapping discussion in Sec.~\ref{s:mapping-sec}.

%%%%%%%%%%%%%%%%%%%%%%%%%%%%%%%%%%%%%%%%%%%%%%%%%%%%%%%%%%%%%%%%%%

\subsection{Pulsar timing response}
\label{s:pulsarresponse}

A gravitational wave transiting an Earth-pulsar line of 
sight creates a perturbation in the 
intervening metric.
This causes a change in their proper separation, which is
manifested as a redshift in the pulse
frequency~\cite{sazhin-1978,detweiler-1979,estabrook-1975,burke-1975}:
\begin{eqnarray}
z(t,\hat k) \equiv 
\frac{\Delta v(t)}{\nu_0} &=& \frac{1}{2} 
\frac{ u^a u^b}{1+\hat k \cdot \hat u} \Delta h_{ab}(t,\hat{k})\,,
%&=& \frac{d\Delta T(t)}{dt}\,,
\end{eqnarray}
where $\hat k$ is the direction of propagation of the gravitational wave,
$\hat u$ is the direction to the pulsar, and $\Delta h_{ab}(t,\hat k)$
is the difference between the metric perturbation at Earth, $(t,\vec
x)$, and at the pulsar some distance $L$ from the Earth, $(t_p,\vec x_p)= (t-L/c, \vec x+L\hat u)$:
\begin{widetext}
\begin{align}
\label{eq:ETPT}
\Delta h_{ab}(t,\hat k) 
&\equiv \int_{-\infty}^\infty {\rm d}f\>
\sum_A h_A(f,\hat k)e^A_{ab}(\hat k) 
\left[e^{i2\pi f(t-\hat k\cdot\vec x/c)}-e^{i2\pi f(t_p-\hat k\cdot\vec x_p/c)}\right]
\\
&=\int_{-\infty}^\infty {\rm d}f\>
\sum_A h_A(f,\hat k)e^A_{ab}(\hat k) e^{i2\pi f(t-\hat k\cdot\vec x/c)}
\left[1-e^{-i2\pi fL(1+\hat k\cdot\hat u)/c}\right]\,.
\end{align}
For a gravitational wave background, which is a superposition of waves 
from all directions on the sky, 
the pulsar redshift integrated over $\hat k$ is given by
\be
\label{eq:res_wpulsar}
z(t) = \int_{-\infty}^\infty {\rm d}f\int_{S^2} {\rm d}^2\Omega_{\hat k}\sum_A
\frac{1}{2}\frac{u^a u^b}{1+\hat k\cdot \hat u}e_{ab}^A(\hat k)
\left[ 1-e^{-i2\pi fL(1+\hat k\cdot \hat u)/c}\right]
h_A(f,\hat k)
e^{i2\pi f (t-\hat k\cdot\vec x/c)}\,.
\ee
Comparing the above expression with Eq.~(\ref{e:responseRA}), we see that
the detector response function $R^A(f,\hat k)$ for a Doppler frequency
measurement $r(t)\equiv z(t)$ is given by
\begin{equation}
R^A(f,\hat k) = 
\frac{1}{2}\frac{u^a u^b}{1+\hat k\cdot \hat u}e_{ab}^A(\hat k)
e^{-i2\pi f\hat k\cdot\vec x}
\left[1-e^{-i2\pi fL(1+\hat k\cdot\hat u)/c}\right]\,.
\label{e:RA_pulsars_exact}
\end{equation}
For a timing residual measurement $r(t)\equiv \int_0^tdt'\> z(t')$,
the above response function $R^A(f,\hat k)$ would need to be
multiplied by a factor of $1/(i2\pi f)$.
The response functions for individual spherical harmonic modes are similarly given by 
\begin{equation}
R^P_{(lm)}(f) =
\int_{S^2}{\rm d}^2\Omega_{\hat k}\> 
\frac{1}{2}\frac{u^a u^b}{1+\hat k\cdot \hat u}Y_{(lm)ab}^P(\hat k)
e^{-i2\pi f\hat k\cdot\vec x}
\left[1-e^{-i2\pi fL(1+\hat k\cdot\hat u)/c}\right]\,.
\label{e:RP_pulsars_exact}
\end{equation}
%

%%%%%%%%%%%%%%%%%%%%%%%%%%%%%%%%%%%%%%%%%%%%%%%%%%%%%%%%%%%%%%%%%%
\subsection{Response functions for a pair of pulsars in the computational frame}
\label{s:response-computationalframe}

In the following section, we will calculate the correlated 
response of a pair of pulsars to a gravitational wave background.
This calculation is most easily done in the so-called
{\em computational frame} 
\citep{allen-ottewill,Mingarelli:2013,gair-2014}, in which the 
two pulsars are in the directions
\be
\begin{aligned}
\hat u_1 &= (0, 0, 1)\,,
\\
\hat u_2 &= (\sin\zeta, 0, \cos\zeta)\,.
\end{aligned}
\ee
In addition, we can choose the origin of the computational 
frame to be at the solar-system barycentre (SSB), 
for which a detector (i.e., a radio telescope on Earth)
has $\vec x\approx\vec 0$.
In this frame the polarisation basis response functions given in
Eq.~(\ref{e:RA_pulsars_exact}) simplify to
\begin{align}
R^{+}_1(f,\hat{k})&= \frac{1}{2} (1-\cos\theta) 
\left(1-{\rm e}^{-2\pi i fL_1(1+\cos\theta)/c}\right)\,, \\
R^{+}_2(f,\hat{k})&= \frac{1}{2} \left[
(1-\sin\zeta\sin\theta\cos\phi-\cos\theta\cos\zeta) 
-\frac{2\sin^2\zeta\sin^2\phi}
{1+\sin\zeta\sin\theta\cos\phi+\cos\theta\cos\zeta}\right]\nonumber\\
&\hspace{4cm} \times 
\left(1-{\rm e}^{-2\pi i fL_2(1+\sin\zeta\sin\theta\cos\phi+\cos\theta\cos\zeta)/c}\right)\,,\\
R^{\times}_1(f,\hat{k})&= 0\,, \\
R^{\times}_2(f,\hat{k})&= -\frac{1}{2} \left(
\frac{\sin^2\zeta\cos\theta\sin(2\phi)-\sin(2\zeta)\sin\theta\sin\phi} 
{1+\sin\zeta\sin\theta\cos\phi+\cos\theta\cos\zeta}\right)
\left(1-{\rm e}^{-2\pi i fL_2(1+\sin\zeta\sin\theta\cos\phi+\cos\theta\cos\zeta)/c}\right)\,,\\
R^{B}_1(f,\hat{k})&= \frac{1}{2} (1-\cos\theta) \left(1-{\rm e}^{-2\pi i fL_1(1+\cos\theta)/c}\right)\,, \\
R^{B}_2(f,\hat{k})&= \frac{1}{2} (1-\sin\zeta\sin\theta\cos\phi-\cos\theta\cos\zeta) \left(1-{\rm e}^{-2\pi i fL_2(1+\sin\zeta\sin\theta\cos\phi+\cos\theta\cos\zeta)/c}\right)\,,\\
R^{L}_1(f,\hat{k})&=\frac{1}{\sqrt{2}}
\frac{\cos^2\theta}{1+\cos\theta}\left(1-{\rm e}^{-2\pi i fL_1(1+\cos\theta)/c}\right)\,, \label{e:CFresSL}\\ 
R^{L}_2(f,\hat{k})&=\frac{1}{\sqrt{2}}
\frac{(\sin\zeta\sin\theta\cos\phi+\cos\theta\cos\zeta)^2}{1+\sin\zeta\sin\theta\cos\phi+\cos\theta\cos\zeta}\left(1-{\rm e}^{-2\pi i fL_2(1+\sin\zeta\sin\theta\cos\phi+\cos\theta\cos\zeta)/c}\right)\,,\\
R^{X}_1(f,\hat{k})&=\frac{-\cos\theta\sin\theta}{1+\cos\theta} \left(1-{\rm e}^{-2\pi i fL_1(1+\cos\theta)/c}\right)\,, \label{e:CFresVLx}\\
R^{X}_2(f,\hat{k})&= \frac{ (\sin\zeta\sin\theta\cos\phi+\cos\theta\cos\zeta)(\sin\zeta\cos\theta\cos\phi-\sin\theta\cos\zeta)}{1+\sin\zeta\sin\theta\cos\phi+\cos\theta\cos\zeta} \nonumber \\
&\hspace{4cm} \times 
\left(1-{\rm e}^{-2\pi i fL_2(1+\sin\zeta\sin\theta\cos\phi+\cos\theta\cos\zeta)/c}\right),\\
R^{Y}_1(f,\hat{k})&=0\,, \label{e:CFresVLy}\\
R^{Y}_2(f,\hat{k})&= \frac{-\sin\phi\sin\zeta(\sin\zeta\sin\theta\cos\phi+\cos\theta\cos\zeta)}{1+\sin\zeta\sin\theta\cos\phi+\cos\theta\cos\zeta} \left(1-{\rm e}^{-2\pi i fL_2(1+\sin\zeta\sin\theta\cos\phi+\cos\theta\cos\zeta)/c}\right)\,.
\end{align}
\end{widetext}
The second (exponential) term inside the bracketed term at the end of each of these expressions is the contribution from the pulsar term. We are in 
general interested in the
regime $y_I\equiv 2\pi fL_I/c \gg 1$ $(I=1,2)$, 
and we will present results below to leading
order in this limit. In the GR case, this limit is equivalent to
setting the pulsar term equal to $0$ in the above expressions, i.e., replacing the whole bracketed term by $1$. This is also
the correct thing to do for the breathing modes, but more care is needed for
the other non-GR modes as the term multiplying the pulsar term is
singular at $\cos\theta=-1$, so we leave this term in for now.
We will use the above expressions for the response functions in
Sec.~\ref{s:orf-sec}, when deriving the overlap reduction functions 
for the different polarisation states.

\section{Overlap reduction functions}
\label{s:orf-sec}

The statistical properties of a Gaussian-stationary background are
encoded in the quadratic expectation values of the Fourier components
of the waveform, e.g., 
$\langle h_A(f,\hat{k})h^*_{A'}(f',\hat{k}')\rangle$, where 
$A=\{+,\times,B,L,X,Y\}$, in a decomposition with respect to
the polarisation basis tensors.
For an uncorrelated, anisotropic
background these quadratic expectation values take the form
\be
\langle h_A(f,\hat{k}) h^*_{A'}(f',\hat{k}')\rangle = 
H_A(f) P_A(\hat{k})
\delta_{AA'} \delta^2(\hat{k},\hat{k'}) \delta(f-f'),
\label{eq:<hh>}
\ee
where $H_A(f)$ and $P_A(\hat{k})$ encode the spectral and angular
properties of the $A^\mathrm{th}$ gravitational-wave 
polarisation, respectively.
[We are assuming here that the spectral and angular dependence of 
the background factorize as $P_A(\hat k)H_A(f)$.]
If the background is unpolarised then there is the restriction
$P_+=P_\times$ and $P_{X}=P_{Y}$, and similarly for 
$H_+$, $H_\times$, and $H_X$, $H_Y$.  

The functions $P_A(\hat{k})$ define the 
anisotropic gravitational-wave power distribution on the sky for
polarisation $A$, and can be
expanded as sums of scalar spherical harmonics
\be \label{eq:Pexpand}
P_A(\hat{k}) = \sum_{l=0}^\infty \sum_{m=-l}^l P^A_{lm} Y_{lm}(\hat{k})\,.
\ee
The expectation value of the correlation between two detectors,
labelled 1 and 2, can be written in the form
\be \langle r_1(t) r_2(t')\rangle = \sum_A \int_{-\infty}^\infty {\rm d}f\>
{\rm e}^{2\pi i f (t-t')} H_A(f) \Gamma^A(f),
\ee
where the overlap reduction function, $\Gamma^A(f)$, is 
given by 
\be
\Gamma^A(f) = \sum_{l=0}^\infty\sum_{m=-l}^l P^A_{lm}
\Gamma^A_{lm} (f)\,,
\ee 
with 
\be \Gamma^A_{lm} (f) = 
\int_{S^2} {\rm d}^2\Omega_{\hat{k}}\>
Y_{lm}(\hat{k}) R^A_1( f,\hat{k})R^{A*}_2 (f,\hat{k})\,.
\label{e:GammaAlm}
\ee
Note that a repeated polarisation index $A$, as in the last two 
equations, is not summed over, unless explicitly indicated with 
a summation sign.
Note also that to simplify the notation, we have not included 
a $12$ subscript on the overlap reduction functions, as we did 
in \cite{gair-2014}, to indicate the two pulsars.

In the following subsections we 
calculate the overlap
reduction functions, $\Gamma^A_{lm}(f)$, for each mode of the power distribution and for each polarisation state, by evaluating the right-hand side of
Eq.~(\ref{e:GammaAlm}) and using the expressions for the response functions 
$R^A(f,\hat k)$ given at the end of Sec.~\ref{s:response-section}.
It turns out that we are able to derive analytic expressions
for the overlap reduction functions for the
$+$, $\times$, breathing, and two vector-longitudinal
polarisation modes.
For scalar-longitudinal backgrounds, we are able to do the 
$\phi$-integration of (\ref{e:GammaAlm}) analytically, but 
need to resort to numerical integration to do the integral over
$\theta$.
Details of the calculations are given in several appendices. 
Plots of $\Gamma^A_{lm}(f)$ as a function of the angle between 
the two pulsars are given 
in Figs.~\ref{f:gammaP}, \ref{f:gammaB}, \ref{f:gammaL}, and 
\ref{f:gammaX}.
We only show plots for $m\ge 0$, since 
$\Gamma^A_{lm}=(-1)^m \Gamma^A_{l,-m}$ as
a consequence of $Y_{lm}(\hat k) = (-1)^m Y_{l,-m}(\hat k)$.

%%%%%%%%%%%%%%%%%%%%%%%%%%%%%%%%%%%%%%%%%%%%%%%%%%
\subsection{Transverse tensor backgrounds}
\label{s:tensor-ORF-sec}

\begin{figure*}[htbp]
\begin{center}
\subfigure[]{\includegraphics[width=.49\textwidth]{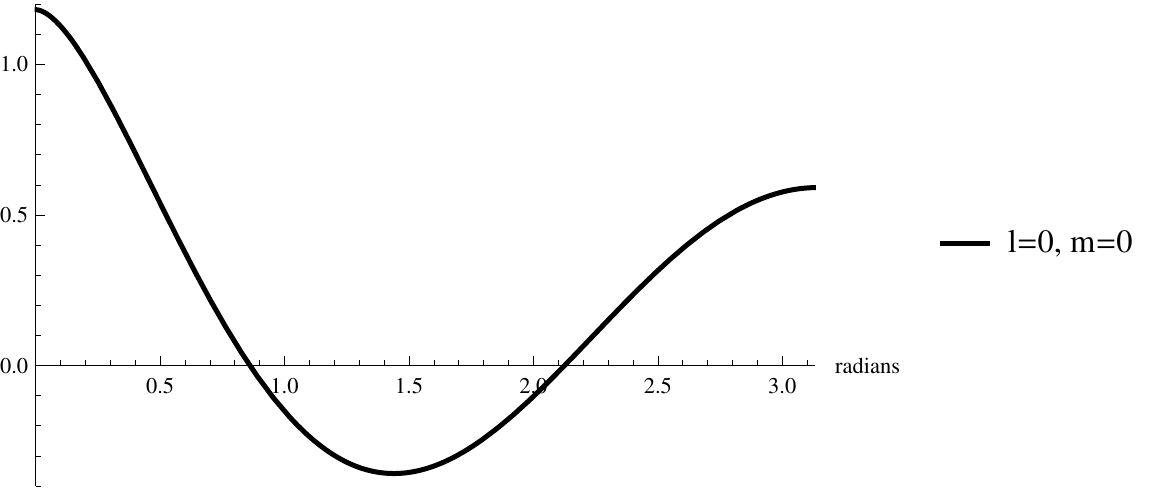}}
\subfigure[]{\includegraphics[width=.49\textwidth]{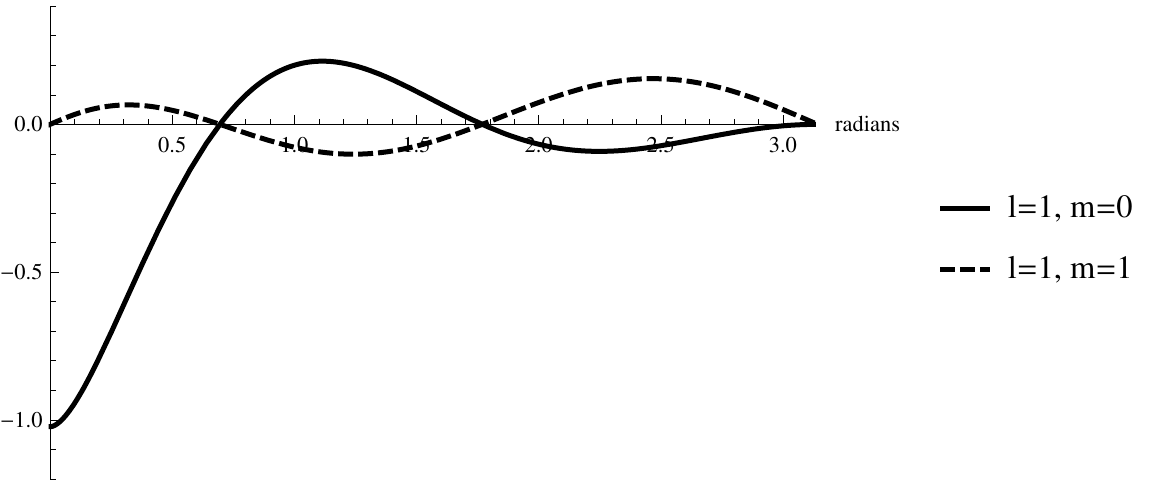}}
\subfigure[]{\includegraphics[width=.49\textwidth]{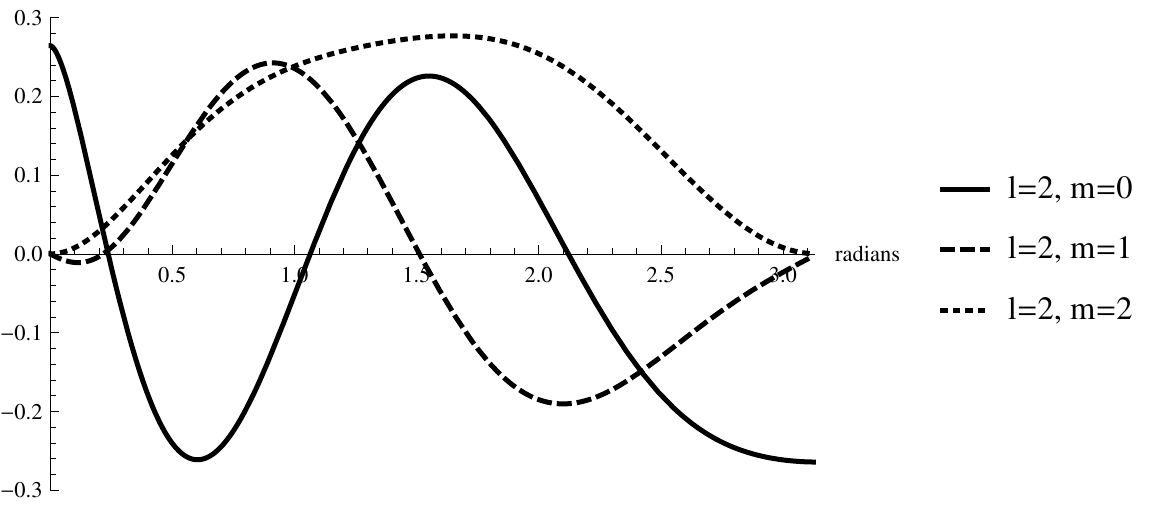}}
\subfigure[]{\includegraphics[width=.49\textwidth]{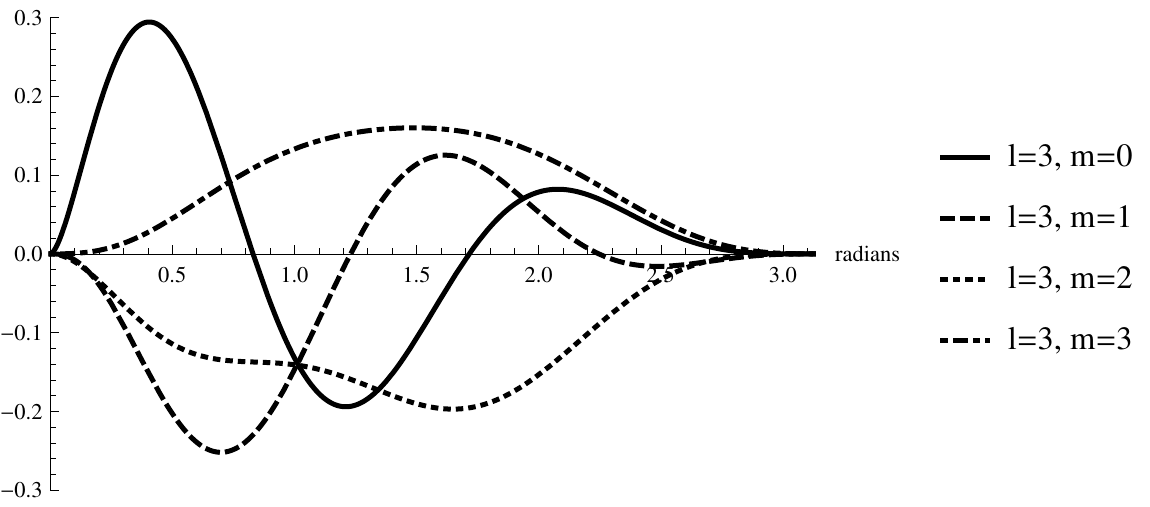}}
%\subfigure[]{\includegraphics[width=.49\textwidth]{gammaP_4m}}
%\subfigure[]{\includegraphics[width=.49\textwidth]{gammaP_5m}}
\caption{Plots of $\Gamma^+_{lm}$ for $l=0,1,2,3$ as a function of
the angle between the two pulsars for an uncorrelated, anisotropic
background.}
\label{f:gammaP}
\end{center}
\end{figure*}

Analytic expressions for the overlap reduction functions
$\Gamma^A_{lm}(f)$  
for uncorrelated, anisotropic $(+,\times)$ 
tensor backgrounds in GR were derived in \cite{gair-2014}.
For such backgrounds, we can work in the limit 
$2\pi f L/c\gg 1$ and set the pulsar terms to zero (for which
the frequency dependence goes away), obtaining 
finite expressions for the overlap reduction function, 
even for potentially troublesome cases such as $\cos\zeta=\pm 1$.
Appendix~\ref{s:appTensor} summarizes the key analytic
expressions derived in that paper.
Plots of $\Gamma^+_{lm}$ for $l=0,1,2,3$ and $m\ge 0$ as a function of 
the angle between the two pulsars are shown in Fig.~\ref{f:gammaP}.
[$\Gamma^\times_{lm}=0$ as a consequence of $R_1^\times(f,\hat k)=0$
in the computational frame.]

\subsection{Scalar-transverse backgrounds}
\label{s:breathing-ORF-sec}

\begin{figure*}[htbp]
\begin{center}
\subfigure[]{\includegraphics[width=.49\textwidth]{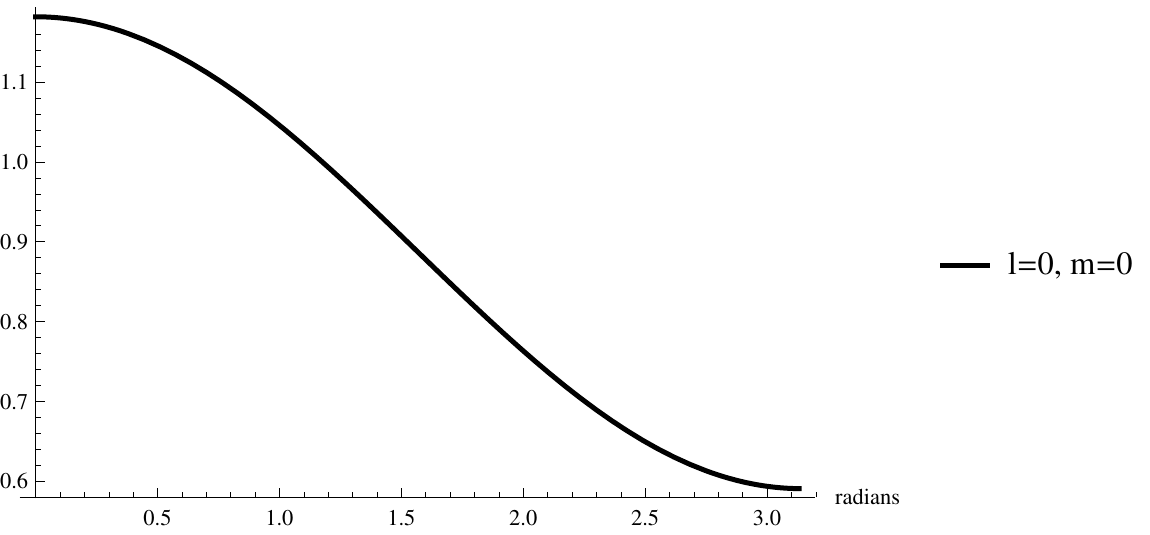}}
\subfigure[]{\includegraphics[width=.49\textwidth]{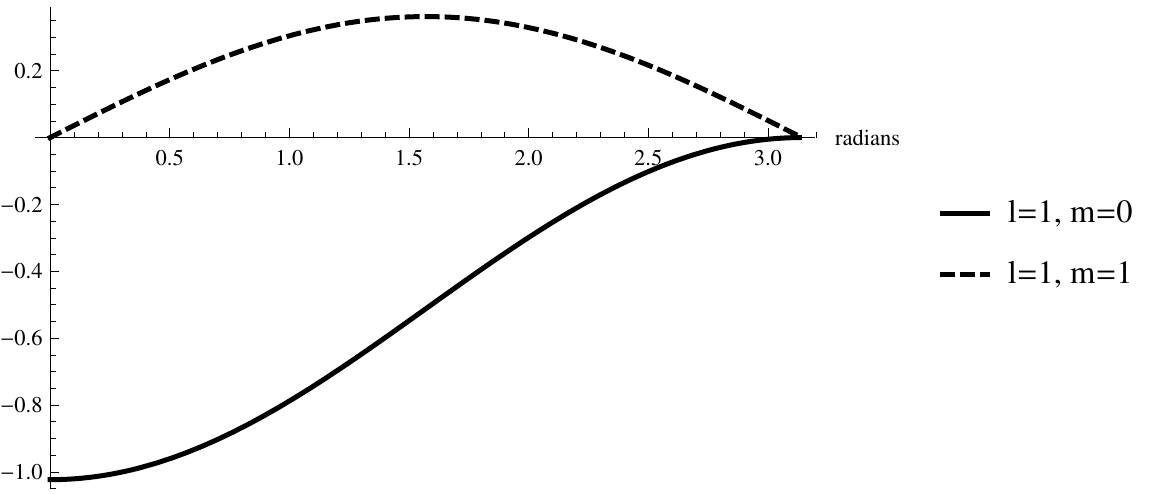}}
\subfigure[]{\includegraphics[width=.49\textwidth]{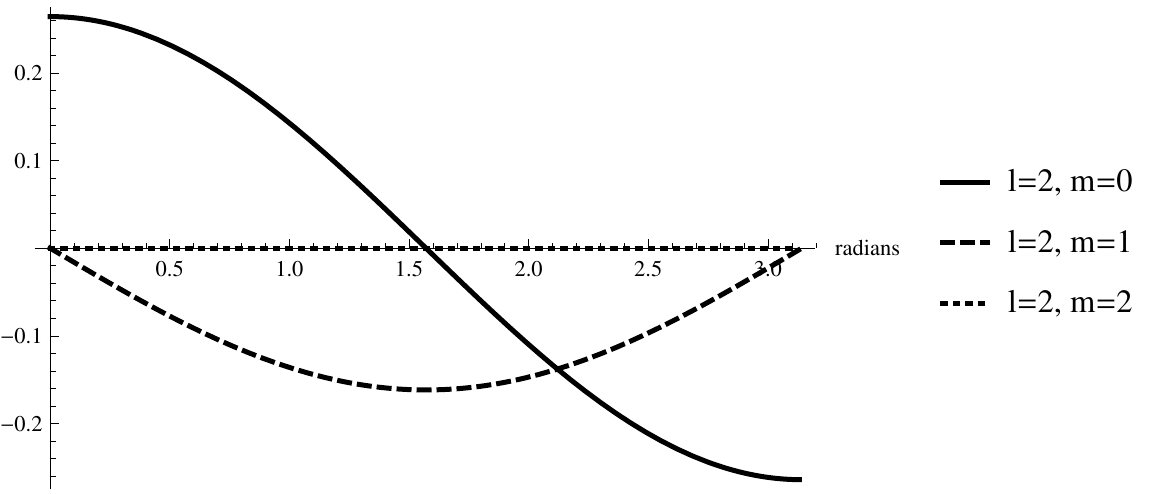}}
\subfigure[]{\includegraphics[width=.49\textwidth]{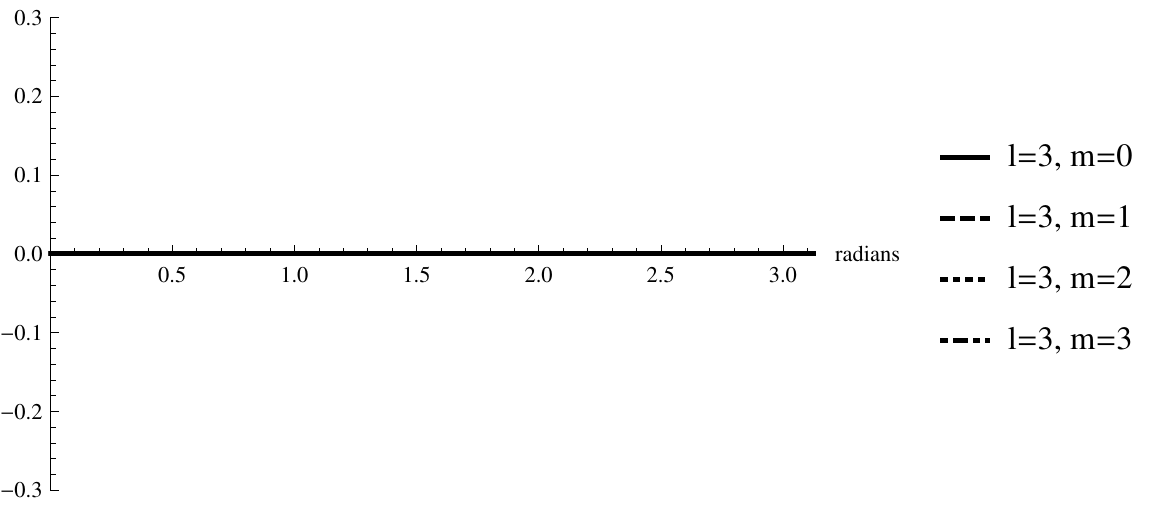}}
\caption{Plots of $\Gamma^B_{lm}$ for $l=0,1,2,3$ as a function of the 
angle between the two pulsars for an uncorrelated, anisotropic background.
As mentioned in the text, the overlap functions are identically
zero for $l\ge 3$ or $|m|\ge 2$.}
\label{f:gammaB}
\end{center}
\end{figure*}

For scalar-transverse (breathing mode) backgrounds, we can again 
make the assumption $2\pi fL/c \gg 1$ and set the pulsar term to zero. 
It then follows that
\begin{widetext}
\be
\begin{aligned} 
\Gamma^B_{lm} 
&= \frac{1}{4} \int_{-1}^1{\rm d}x \int_{0}^{2\pi}{\rm d}\phi\>
(1-x) \left(1-x \cos\zeta - \sqrt{1-x^2} \cos\phi \sin\zeta \right)
N_l^m P_l^m(x) {\rm e}^{i m \phi}
\\ 
&= \frac{\pi N_l^m}{4} \int_{-1}^1 {\rm d}x\>
\left[2 \delta_{m0} (1-x)(1-x\cos\zeta) P_l(x) - (\delta_{m1} P_l^1(x) +
\delta_{m,-1}P_l^{-1}(x)) \sqrt{1-x^2} (1-x)\sin\zeta \right]
\\
&= \pi N_l^m \delta_{m0} \left[
\left(1+\frac{1}{3} \cos\zeta\right) \delta_{l0} - \frac{1}{3}
(1+\cos\zeta) \delta_{l1} + \frac{2}{15} \cos\zeta \delta_{l2} \right] 
+ \pi N_l^{|m|}
(-1)^{\frac{m-|m|}{2}} \delta_{|m|,1} \sin\zeta \left( \frac{1}{3}
\delta_{l1} - \frac{1}{5} \delta_{l2}\right),
\label{eq:gammaST}
\end{aligned}
\ee
\end{widetext}
where we have used the definition of the scalar spherical harmonics given in
Eq.~(\ref{e:Nlm}) of App.~\ref{s:spinweightedY} and properties of the
associated Legendre polynomials summarised in
App.~\ref{s:legendre_polynomials}. We see that we are only
sensitive to modes of the background with $l \leq 2$ and $|m| \leq 1$.
Plots of $\Gamma^B_{lm}$ for $l=0,1,2,3$ and $m\ge0$ are shown
in Fig.~\ref{f:gammaB}.

\subsection{Scalar-longitudinal backgrounds}
\label{s:longitudinal-ORF-sec}

\begin{figure*}[htbp]
\begin{center}
\subfigure[]{\includegraphics[width=.49\textwidth]{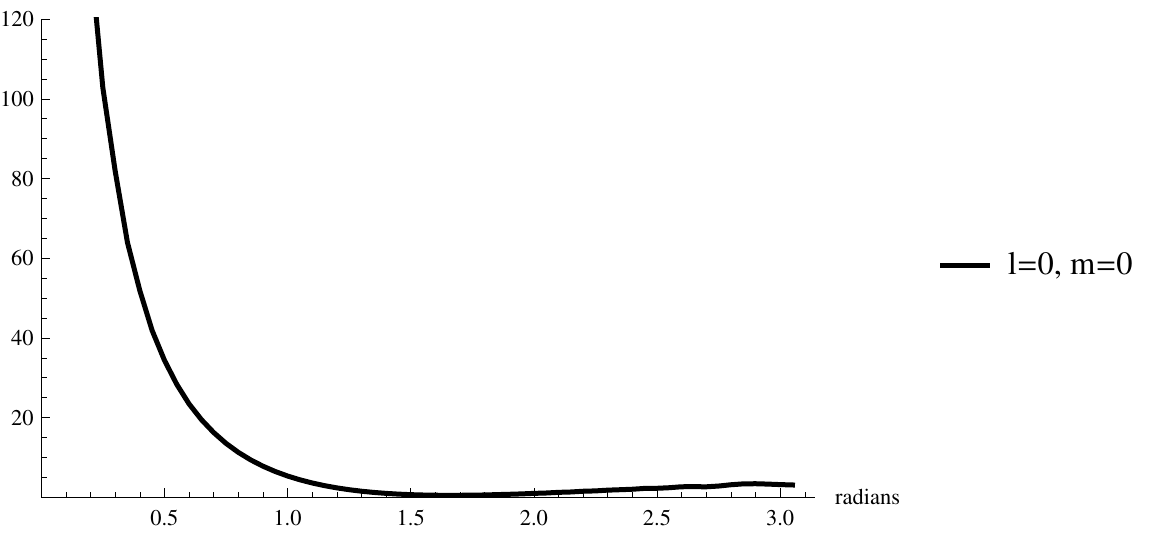}}
\subfigure[]{\includegraphics[width=.49\textwidth]{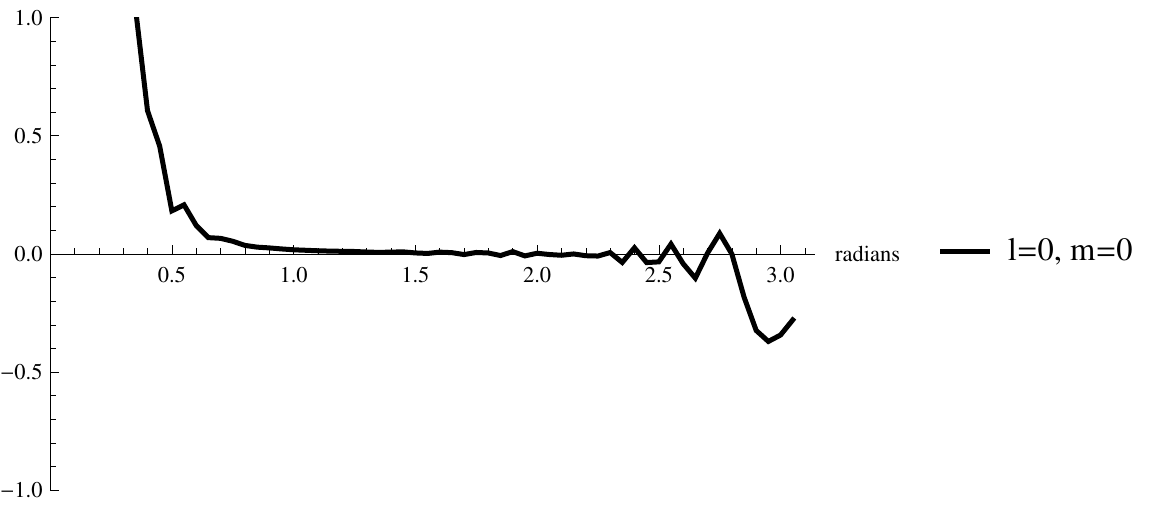}}
\subfigure[]{\includegraphics[width=.49\textwidth]{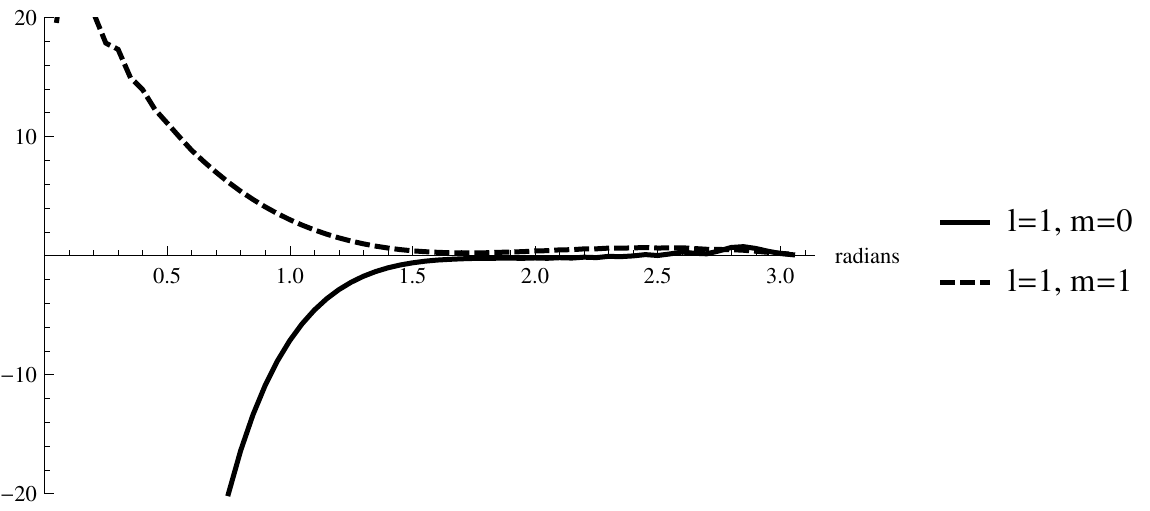}}
\subfigure[]{\includegraphics[width=.49\textwidth]{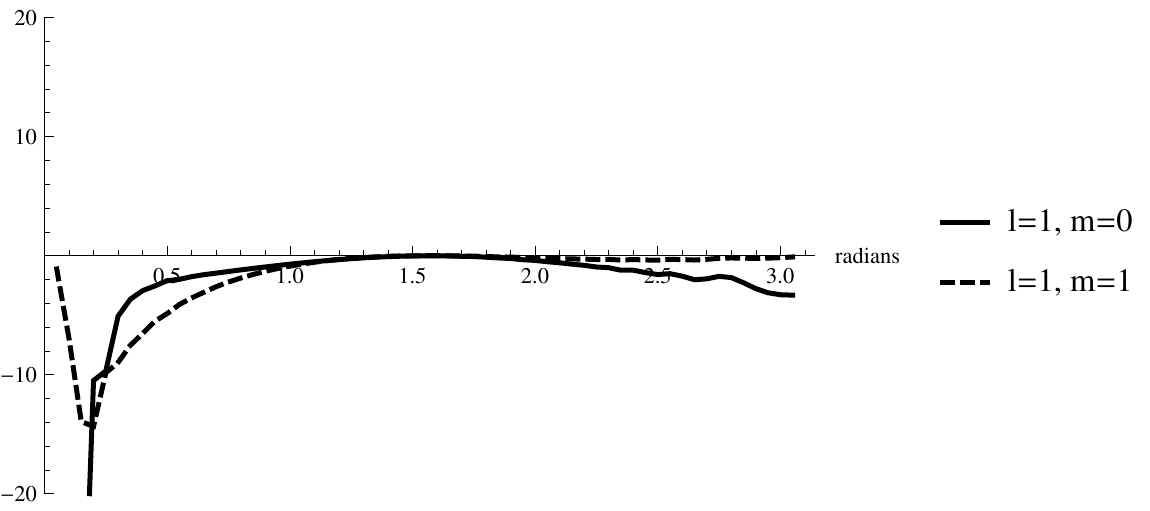}}
\subfigure[]{\includegraphics[width=.49\textwidth]{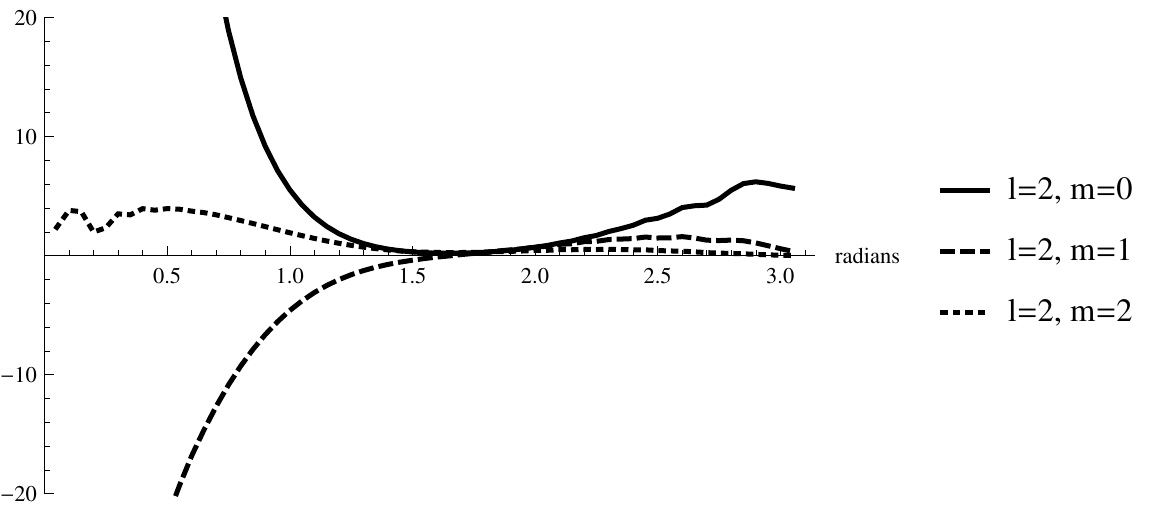}}
\subfigure[]{\includegraphics[width=.49\textwidth]{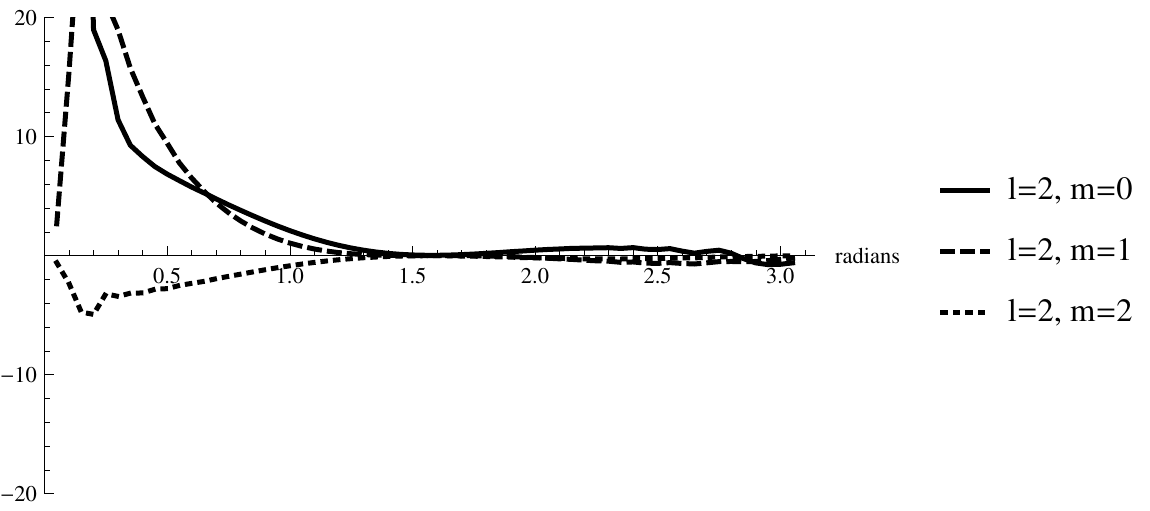}}
\subfigure[]{\includegraphics[width=.49\textwidth]{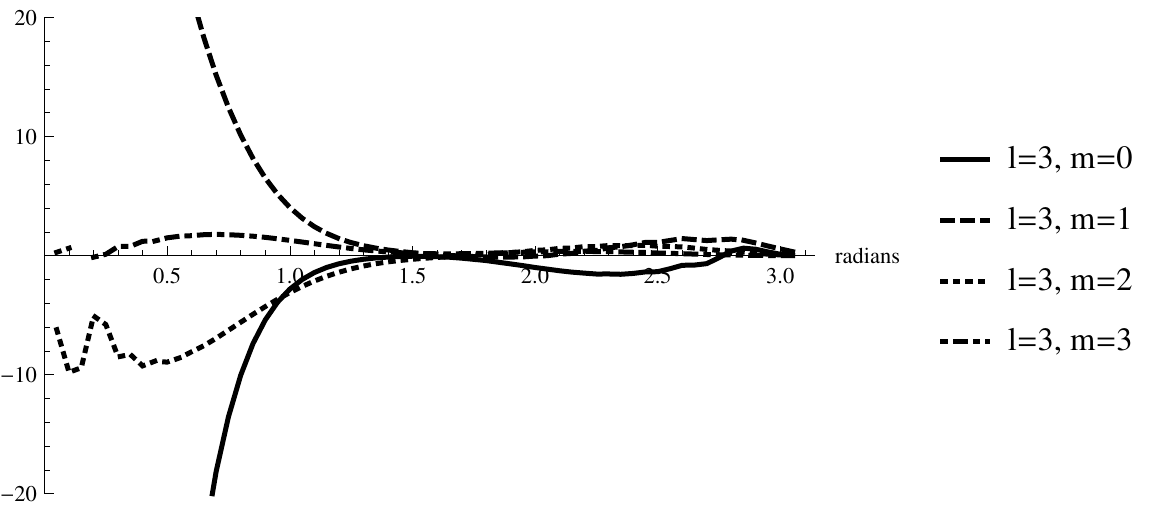}}
\subfigure[\ $l=3$, imag part]{\includegraphics[width=.49\textwidth]{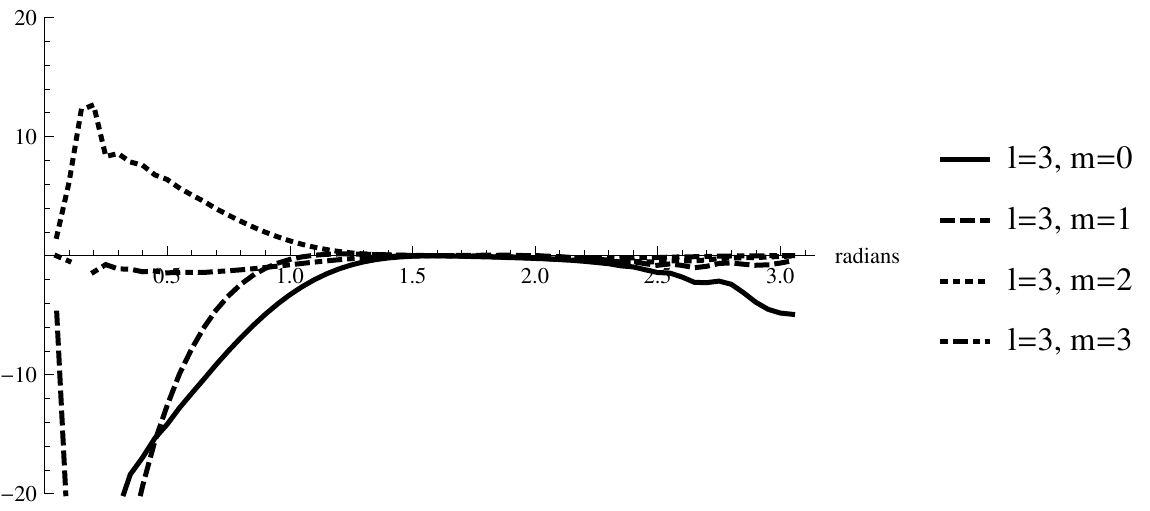}}
\caption{Plots of the real part (left column) and imaginary part 
(right column) of $\Gamma^L_{lm}(f)$ for $l=0,1,2,3$ as a function 
of the angle between the two pulsars for an uncorrelated, anisotropic background.
These were calculated using the  semi-analytic approximation described
in the main text.
For these plots we have chosen $y_1=100$ and $y_2=200$, where $y_I=2\pi f L_I/c$
and $L_I$ is the distance to pulsar $I$.}
\label{f:gammaL}
\end{center}
\end{figure*}

The response for a scalar-longitudinal background, 
Eq.~(\ref{e:CFresSL}), is singular
at $\cos\theta=-1$ if the pulsar term is not included. We must
therefore include the pulsar term when evaluating the overlap
reduction function for backgrounds of this form. Using the notation
$y_1 = 2 \pi f L_1/c$, $y_2 = 2 \pi f L_2/c$, where $L_I$ is the
distance to pulsar $I$, %we will keep terms up to constant order $(y_1)^0$, $(y_2)^0$. 
the overlap reduction function for a given $(lm)$, is given explicitly by
\begin{widetext}
\be
\Gamma^L_{lm}(f) = \frac{1}{2}N_l^m  \int_{-1}^1 {\rm d}x\>
\left[ \frac{x^2}{1+x} 
\left(1 - {\rm e}^{-iy_1(1+x)}\right) 
I_{m}(y_2,x) \right] P_l^m(x),
\label{eq:gamL_int}
\ee
where
\be
I_{m}(y,x)
=\int_0^{2\pi} {\rm d}\phi\>
\frac{(\sqrt{1-x^2}\sin\zeta\cos\phi+x\cos\zeta)^2}
{1+x \cos\zeta + \sqrt{1-x^2} \sin\zeta\cos\phi} 
\left(1 - {\rm e}^{iy(1+x \cos\zeta + \sqrt{1-x^2} \sin\zeta\cos\phi)}\right) {\rm e}^{im\phi}.
\ee
\end{widetext}
The integral for $I_{m}(y,x)$ is challenging to evaluate in general;
however see App.~\ref{s:scalar-longitudinal} for an approximate 
expression, valid for large $y$.
As shown in Apps.~\ref{s:appCoDirectional} and 
\ref{s:appAntiDirectional}, it can be more simply 
evaluated for co-directional pulsars (i.e., $\cos\zeta=1$) and for anti-directional pulsars (i.e., $\cos\zeta=-1$).
Using the approximate expression for $I_m(y,x)$ evaluated in
App.~\ref{s:scalar-longitudinal}, we then do the integration over
$x$ given in Eq.~(\ref{eq:gamL_int}) numerically.
The results of this {\em semi-analytic} calculation for 
$\Gamma^L_{lm}(f)$
for $l=0,1,2,3$ and $|m|>0$ are shown in Fig.~\ref{f:gammaL}.
For these plots we have chosen $y_1=100$ and $y_2=200$.

\begin{figure*}[htbp]
\begin{center}
{\includegraphics[width=.7\textwidth]{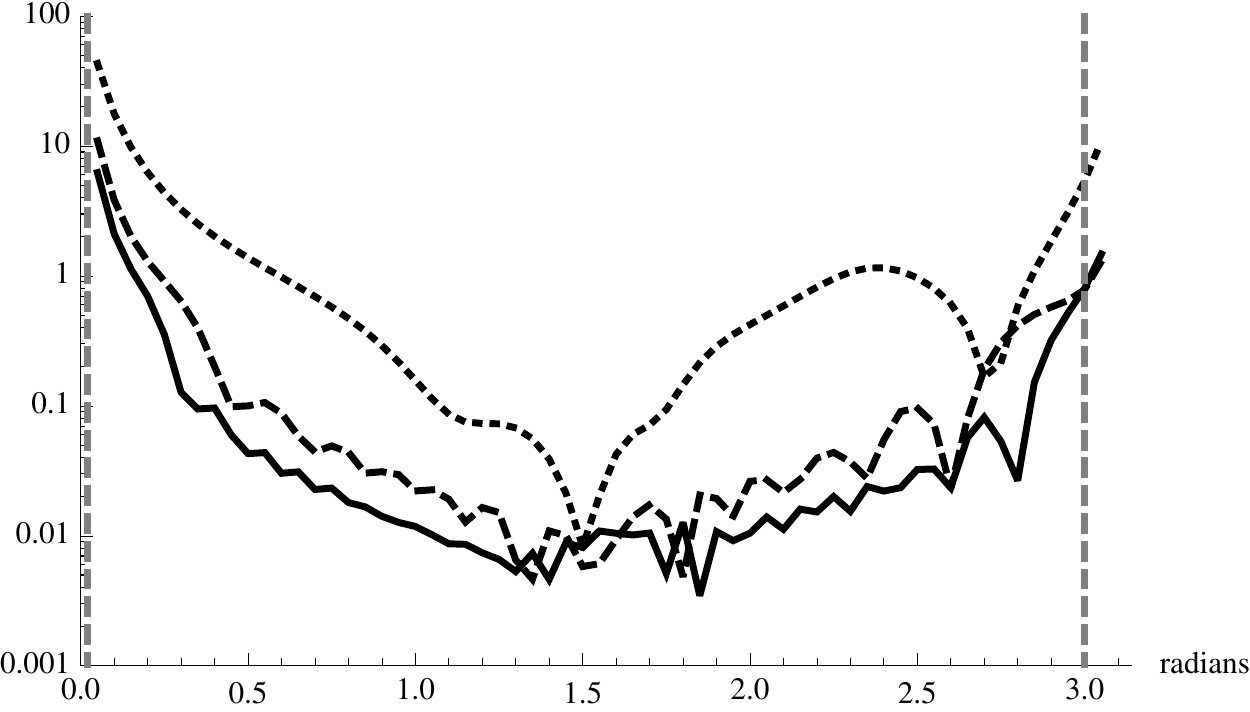}}
\caption{Fractional percentage difference between the values of the 
$l=0$, $m=0$ overlap reduction function $\Gamma^L_{00}(f)$ calculated 
(i) semi-analytically (i.e., using the analytic expression for $I_{lm}(y,x)$ derived
in App.~\ref{s:scalar-longitudinal}, and doing the $x$-integration numerically), and 
(ii) doing the full $(\theta,\phi)$ sky integration numerically using
a HEALPix~\cite{HEALPix} pixelisation of the sky.
The dotted curve is for $y_1=10$, $y_2=20$;
the dashed curve is for $y_1=50$, $y_2=100$; and
the solid curve is for $y_1=100$, $y_2=200$,
where $y_I=2\pi fL_I/c$ and $L_I$ is the distance to pulsar $I$.
Note that the percentage difference decreases as $y_1$ and $y_2$ increase.
The vertical dashed grey lines at the left and right-hand edges of the 
plot correspond to the minimum and maximum angular separation
($0.95$ degrees and 174~degrees, respectively) over all pairs of pulsars in 
the European Pulsar Timing Array (EPTA).}
\label{f:2DvsSemi1d_comparison}
\end{center}
\end{figure*}

The semi-analytic calculation agrees quite well with the
full $(\theta,\phi)$ sky integration, as shown in Fig.~\ref{f:2DvsSemi1d_comparison}.
(The 2-dimensional sky integration was actually done using 
a HEALPix~\cite{HEALPix} pixelisation of the sky.)
This plot shows the fractional percentage difference between 
the values of the $l=0$, $m=0$ overlap reduction function 
$\Gamma_{00}^L(f)$ calculated using these two methods.
As can be seen from the figure, the agreement is best for values
of $\zeta$ that stay away from $\zeta =0$ and $\zeta=\pi$. However, at those special points we can use the analytic expressions given in Apps.~\ref{s:appCoDirectional} and \ref{s:appAntiDirectional}, and these are tabulated for $l=0,1,2,3$ in Table \ref{tab:co-anti-sl}. This allows us to obtain a good approximation to the overlap reduction function for all $\zeta$. We note that Fig.~\ref{f:2DvsSemi1d_comparison} shows that the percentage difference between the numerical and semi-analytic curves becomes smaller for larger values of $y_1$ and $y_2$, which is consistent with the semi-analytic expression being valid for large $y$. 
%The values of the overlap reduction function for the special cases of $\zeta =0$ and $\zeta=\pi$ are given by evaluating

%

\begin{table} 
\centering
\begin{tabular}{c c c c c} 
\hline
$l$ & \multicolumn{2}{c}{$\zeta = 0$} & \multicolumn{2}{c}{$\zeta = \pi$} \\
& Real & Imaginary & Real & Imaginary \\
\hline
\hline
$0$ & $261$ & $117$ & $3.31$ & $0.254$ \\
$1$ & $-445$ & $-201$ & $-6.78$ & $-0.388$ \\
$2$ & $561$ & $254$ & $6.19$ & $0.567$ \\
$3$ & $-639$ & $-290$ & $-6.44$ & $-0.590$ \\
\hline
\hline
\end{tabular}
\caption{\label{tab:co-anti-sl} Values of the co-directional ($\zeta=0$) and anti-directional ($\zeta=\pi$) overlap reduction function for a scalar-longitudinal GW background given for $l=0,1,2,3$. The pulsars have $y_1=100$ and $y_2=200$. The values in the table correspond to $m=0$ modes since all other values of $m$ give zero overlap reduction function values.}
\end{table}

\subsection{Vector-longitudinal backgrounds}
\label{s:vector-ORF-sec}

\begin{figure*}[htbp]
\begin{center}
\subfigure[]{\includegraphics[width=.49\textwidth]{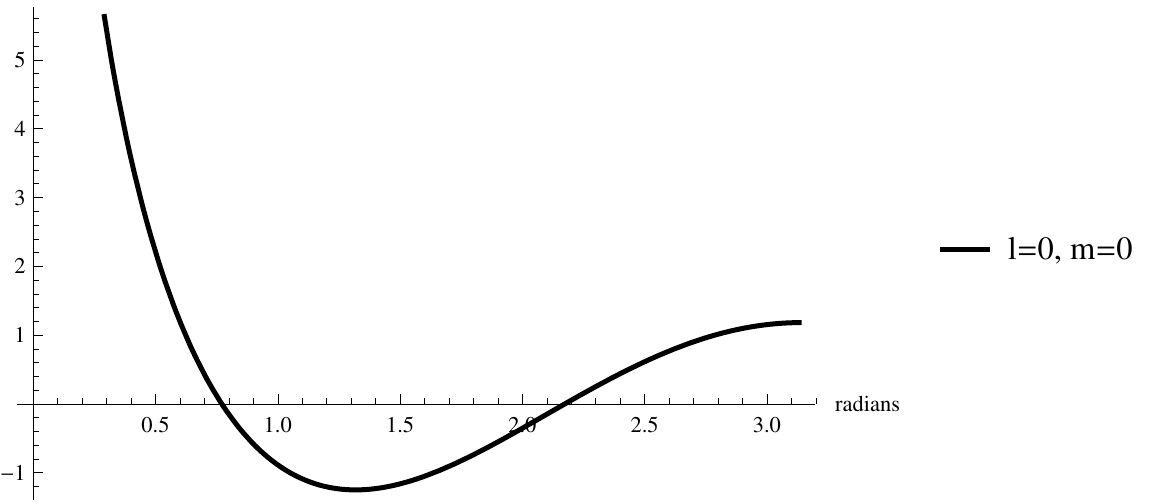}}
\subfigure[]{\includegraphics[width=.49\textwidth]{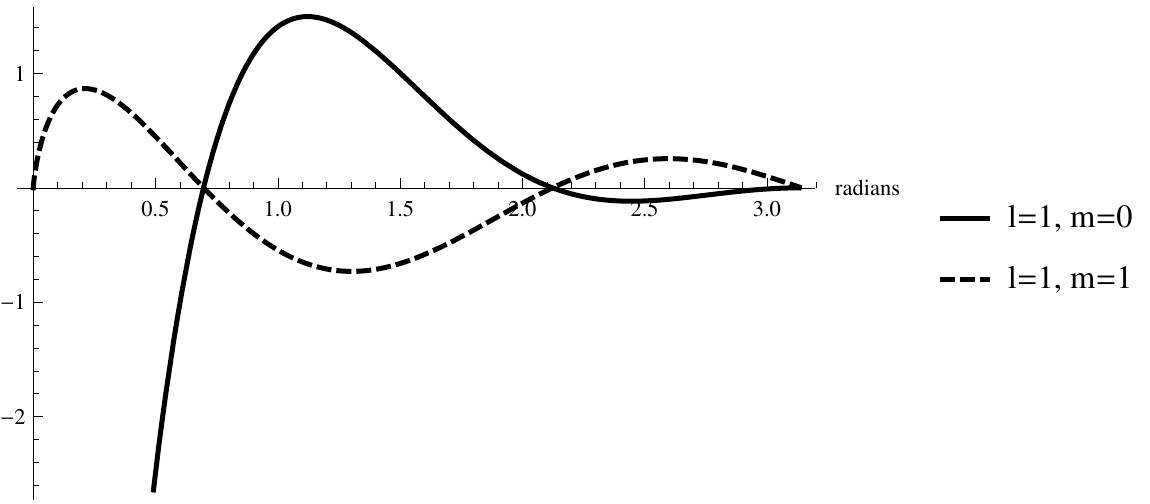}}
\subfigure[]{\includegraphics[width=.49\textwidth]{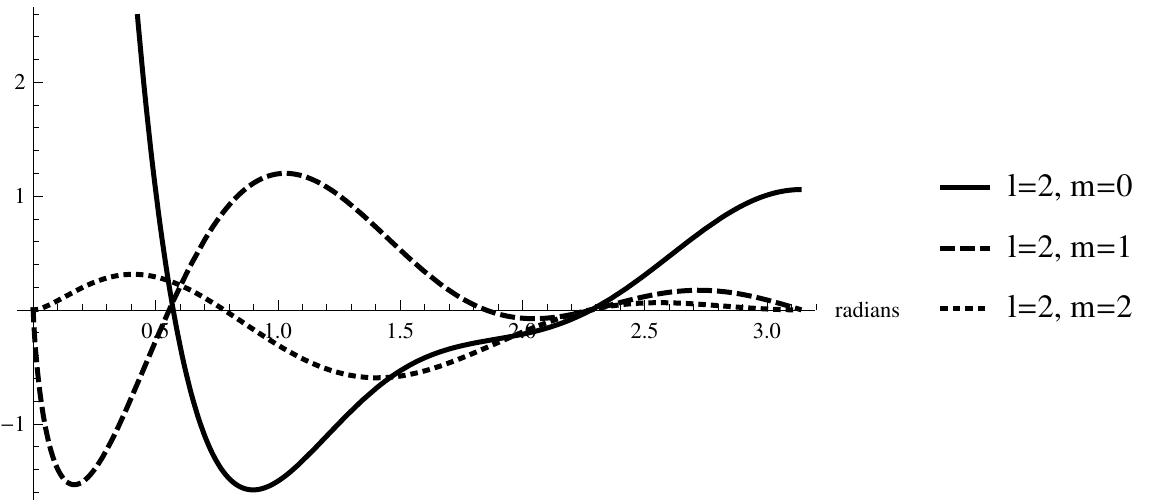}}
\subfigure[]{\includegraphics[width=.49\textwidth]{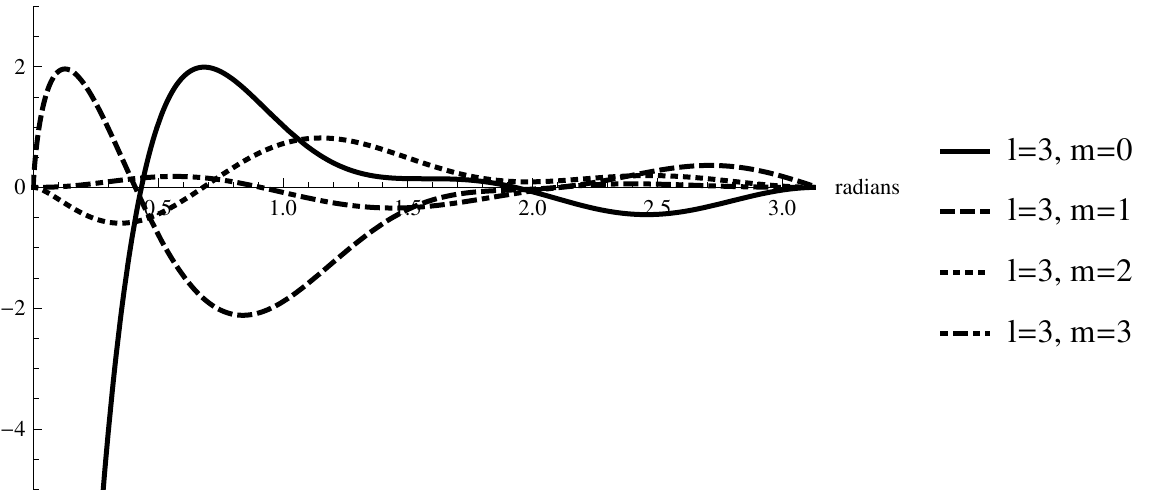}}
\caption{Plots of $\Gamma^X_{lm}$ for $l=0,1,2,3$ as a function of
  the angle between the two pulsars for an uncorrelated, anisotropic background.}
\label{f:gammaX}
\end{center}
\end{figure*}

If we ignore the pulsar term, then the response for a vector-longitudinal 
background,
Eq.~(\ref{e:CFresVLx}), looks singular at $\cos\theta=-1$. However, due to
the factor of $\sin\theta$ in the numerator this is a $1/\sqrt{1+\cos\theta}$ 
type singularity which is integrable. We can therefore also ignore the
pulsar term for these backgrounds and obtain a finite result. 
The analytic calculation is very similar to that in App.~E of \cite{gair-2014} for the standard $(+,\times)$ tensor backgrounds of GR.
Details of the calculation are given in App.~\ref{s:vector-ORF}.
Plots of $\Gamma^X_{lm}$ for $l=0,1,2,3$ and $m\ge0$ are shown
in Fig.~\ref{f:gammaX}.
[$\Gamma^Y_{lm}=0$ as a consequence of $R^Y_1(f,\hat k)=0$
in the computational frame.]

We note that in the limit $\cos\zeta \rightarrow 1$, the $m=0$ overlap 
reduction functions diverge.  This is because in that limit the
singularities at $(1+ \hat{k} \cdot \hat{u}_1) =0$ and $(1+ \hat{k} \cdot
\hat{u}_2) =0$ coincide and behave like $1/(1+\cos\theta)$ rather than
$1/\sqrt{1+\cos\theta}$. Again, this singularity 
is eliminated if the pulsar terms are included in the integrand and 
the pulsars are assumed to be at finite distance. 
Details of that calculation are given in App.~\ref{s:vector-ORF-coszeta1}.

\section{Mapping the background}
\label{s:mapping-sec}

In \cite{gair-2014} we applied the methodology used to characterise
CMB polarisation to describe gravitational-wave backgrounds in general
relativity. This involved expanding a transverse tensor GR 
background in terms of (rank-2) gradients and curls of spherical harmonics, 
which are closely related to spin-weight $\pm2$ spherical harmonics. 
As described in Sec.~\ref{s:sphbasis}, we can use a
similar decomposition to represent arbitrary backgrounds with
alternative polarisation states.
As explained earlier, for scalar-transverse
and scalar-longitudinal backgrounds, we expand in terms of the 
ordinary (scalar) spherical harmonics,
while for vector-longitudinal backgrounds we must expand in terms
of spin-weight $\pm 1$ spherical harmonics. 

In the following subsections, we derive analytic expressions
for the pulsar response functions $R^P_{(lm)}(f)$ defined in 
Eq.~(\ref{e:RP_pulsars_exact}), for each mode of a background with each of the different polarisation
states, labeled by $P=\{G,C,B,L,V_G,V_C\}$.
We calculate the response in the ``cosmic" reference frame, 
where the angular dependence of the gravitational-wave
background is to be described.
The origin of this frame is at the SSB and a pulsar is 
located in direction $\hat u$, with angular coordinates 
$(\zeta,\chi)$, i.e., 
\be
\hat u^a = (\sin\zeta\cos\chi,\sin\zeta\sin\chi,\cos\zeta)\,,
\ee
and is at a distance $L$ from the SSB.
In this frame, we can again make the approximation 
$\vec x\approx\vec 0$ for the detector locations (i.e., radio 
receivers on Earth).
As was done in \cite{gair-2014}, it is simplest to evaluate the response 
in the cosmic frame by making a change of variables 
of the integrand of Eq.~(\ref{e:RP_pulsars_exact}), 
so that $\hat u$ points along the $z$-axis.  
This corresponds to a rotation defined by the Euler angles
$(\alpha,\beta,\gamma)=(\chi,\zeta,0)$.
Using the transformation properties of the tensor spherical
harmonics $Y^P_{(lm)ab}(\hat k)$ under a rotation, it follows
that
\be
R^P_{(lm)}(f) = 
Y_{lm}(\hat u) 
{\cal R}^P_l(2\pi fL/c)\,,
\label{e:RP-mapping}
\ee
where ${\cal R}^P_l(2\pi fL/c)$ is proportional to the 
$m=0$ component of the response function calculated in the 
rotated frame (with the pulsar directed along the $z$-axis):
\be
{\cal R}^P_l(2\pi fL/c) \equiv
\sqrt{\frac{4\pi}{2l+1}}
R^P_{(l0)}(f)\big|_{\hat u=\hat z}.
\label{e:calR}
\ee
Note that we need only consider the $m=0$ component, since the 
pulsar response must be axi-symmetric in the rotated frame,
while the tensor spherical harmonics we consider are all proportional 
to $e^{im\phi}$ in this frame.
Thus, we see from Eq.~(\ref{e:RP-mapping}) that the dependence 
on the direction to the pulsar is given
simply by $Y_{lm}(\hat u)$, while the distance to the pulsar is 
responsible for the frequency-dependence of the response function.
Finally, using Eq.~(\ref{e:RP_pulsars_exact}) with 
$\vec x\approx\vec 0$ and doing the integration over $\phi$, we find
\begin{widetext}
\be
{\cal R}^P_l(2\pi fL/c) 
= 2\pi\sqrt{\frac{4\pi}{2l+1}}
\int_{-1}^1{\rm d}x\>
\frac{1}{2}\frac{1}{1+x}
Y^P_{(l0)zz}(\theta,0)
\left(1-e^{-i2\pi fL(1+x)/c}\right)\,,
\label{e:calR-explicit}
\ee
\end{widetext}
where $x=\cos\theta$.
It is this function that we need to evaluate in the following subsections.

We finish this subsection by noting an important result implicit in
Eq.~(\ref{e:RP-mapping}) connected to the distinguishability of
different background polarisation states. For every polarisation type,
the response of a pulsar factorises into a piece that is dependent on
pulsar position, which is $Y_{lm}(\hat{u})$ for all polarisation
types, and a piece that depends only on the distance to the pulsar.
%In the limit that the pulsar distance tends to infinity or all
%pulsars are at the same distance that second piece is the same for
%all pulsars.
Even if we had infinitely many pulsars distributed across the sky, at
any given frequency, the best we could do would be to construct a
pulsar response map across the sky and decompose it into (scalar)
spherical harmonics. The coefficient of each term would be a sum of
the ${\cal R}^P_l(2\pi fL/c)$'s for all polarisation states, $P$,
which at face value means that it would not be possible to disentangle
the different polarisation states. However, as we will see below, a
scalar-transverse and transverse tensor background can always be
distinguished as current PTAs operate in a regime in which the response functions are effectively independent of the pulsar distance, i.e., the pulsar term can be ignored. In that limit, we are only sensitive to modes with $l < 2$ of scalar-tensor backgrounds, while transverse tensor backgrounds can only contain modes with $l \geq 2$. The
longitudinal modes cannot be distinguished from the transverse modes,
however, unless we have several pulsars, at different distances, in
each direction on the sky. For the longitudinal modes the
finite-distance corrections introduced by the pulsar term are
important for typical pulsar distances of current PTAs, which gives an
additional handle to identify those modes. Alternatively, if we made
some assumption about how the background amplitude was correlated at
different frequencies, e.g., that it followed a power law, we would
also break this degeneracy as the response of the array to
longitudinal modes has a frequency dependence through the same
term. Thus, it is in principle possible to disentangle every component
of the background for each polarisation state at each frequency, given
sufficiently many pulsars at a sufficient variety of distances along
each line of sight. In practice, a pulsar timing array containing
$N_p$ pulsars can only measure $2 N_p$ real components of the
background at any given frequency~\cite{gair-2014,Cornish-vanHaasteren:2014} and so the
resolution of any reconstructed map of the background will be limited
by the size of the pulsar timing array. Roughly speaking, to probe an angular scale of the order $1/l_{\rm max}$ we would require $N_p = (l_{\rm max} + 1)^2 - 4$ pulsars, if we assumed the background was consistent with GR and therefore contained only transverse tensor polarisation modes. If we allow for arbitrary polarisations we would expect to need $N_p = 3 (l_{\rm max} + 1)^2$ pulsars, since we now have structure down to $l=0$, and we effectively have three different possible polarisation states --- transverse (either scalar or tensor, but they are distinguished by the $l$ of the mode), scalar longitudinal or vector longitudinal. A full investigation of what can be measured in practice is beyond the scope of this current work and we leave it for future study.
\begin{widetext}

%%%%%%%%%%%%%%%%%%%%%%%%%%%%%%%%%%%%%%%%%%%%%%%%%%%%%%%%%%%%
\subsection{Standard transverse tensor backgrounds}
\label{s:tensor-mapping}

In \cite{gair-2014}, the standard transverse tensor modes of GR 
were expanded in terms of gradient and curl tensor spherical harmonics, 
and the corresponding response functions were calculated to be
\be
R^{G}_{(lm)}(f)\approx 2\pi (-1)^l\> {}^{(2)}\!N_l Y_{lm}(\hat u)\,,
\quad 
R^{C}_{(lm)}(f)\approx0\,,
\label{e:RGC-G14}
\ee
where ${}^{(2)}\!N_l$ is a normalisation constant defined in Eq.~(\ref{e:2N}) of App~\ref{s:grad-curl-tensor}, and the $\approx$ signs means that the pulsar term was {\em ignored} for this calculation.
Extending the analysis given in \cite{gair-2014} to {\em include} the pulsar term, we find
\be
R^{G}_{(lm)}(f)= Y_{lm}(\hat u){\cal R}^G_l(2\pi fL/c)\,,
\quad 
R^{C}_{(lm)}(f)=0\,,
\ee
where
\be
{\cal R}^G_l(y) 
= 2\pi\,\frac{{}^{(2)}N_l}{4}
\int_{-1}^1 {\rm d}x\>\left[(1-x)(1-x^2)
\left(1-{\rm e}^{-iy(1+x)}\right) \frac{{\rm d}^2P_l}{{\rm d}x^2} \right]\,.
\label{e:tauG}
\ee
Integrating Eq.~(\ref{e:tauG}) by parts twice,
\be
{\cal R}^{G}_l (y) 
=\pi\,{}^{(2)}\!N_l(-i)^l e^{-iy}
\left[(2-2iy+y^2) j_l(y)
-i(6+4iy+y^2)\frac{{\rm d}j_l}{{\rm d}y}
-(6iy - y^2)\frac{{\rm d}^2 j_l}{{\rm d}y^2}
- iy^2\frac{{\rm d}^3 j_l}{{\rm d}y^3}
\right]\,,
\ee
where $j_l(y)$ denotes a spherical Bessel function, as defined 
in App.~\ref{s:bessel}, and ${\rm d}j_l/{\rm d}y$, ${\rm d}^2j_l/{\rm d}y^2$, and ${\rm d}^3j_l/{\rm d}y^3$ can be simplified using Eqs.~(\ref{e:jl'})--(\ref{e:jl'''}). Taking the usual limit that the pulsar is many gravitational-wave 
wavelengths from the Earth ($y\gg 1$), we find
${\cal R}^G_l (y) \approx 2\pi (-1)^l\> {}^{(2)}\!N_l$,
which is consistent with Eq.~(\ref{e:RGC-G14}), where the response functions
were calculated without the pulsar term.

%%%%%%%%%%%%%%%%%%%%%%%%%%%%%%%%%%%%%%%%%%%
\subsection{Scalar-transverse backgrounds} 
\label{s:breathing-mapping}

Repeating the calculation in \cite{gair-2014} for an arbitrary scalar-transverse 
(breathing mode) background, we find
\be 
R^B_{(lm)}(f) = Y_{lm}(\hat u){\cal R}^B_l(2\pi fL_/c)\,,
\label{e:RB-mapping1}
\ee
with
\be 
\begin{aligned} 
{\cal R}^B_l(y)
&= 2\pi \frac{1}{\sqrt{2}}
\int_{-1}^1{\rm d}x\>
\frac{1}{2}(1-x)P_l(x)\left(1-{\rm e}^{-i(1+x)y}\right)
\\
&=2\pi\frac{1}{\sqrt{2}}
\left\{\delta_{l0} - \frac{1}{3} \delta_{l1} - (-i)^l {\rm e}^{-iy} 
\left[ \left(1-i\frac{l}{y}\right)j_l(y) + i j_{l+1}(y)\right]
\right\}\,,
\label{e:RB-mapping2}
\end{aligned}
\ee
\end{widetext}
where we used Eqs.~(\ref{e:sphbess}), (\ref{e:jl'}) from App.~\ref{s:bessel} to get the terms involving the spherical Bessel functions. Since the spherical Bessel functions behave like $1/y$ for large $y$,
the terms in square brackets tend to zero as $y\rightarrow\infty$,
leading to the approximate expression for the response function 
\be
R^B_{(lm)}(f) \approx 2\pi Y_{lm}(\hat u)\,
\frac{1}{\sqrt{2}}
\left[\delta_{l0} - \frac{1}{3} \delta_{l1}\right],
\label{e:RB-mapping-approx}
\ee
which is valid in the limit where we ignore the pulsar term.

\begin{figure*}
\begin{center}
\includegraphics[width=0.99\textwidth]{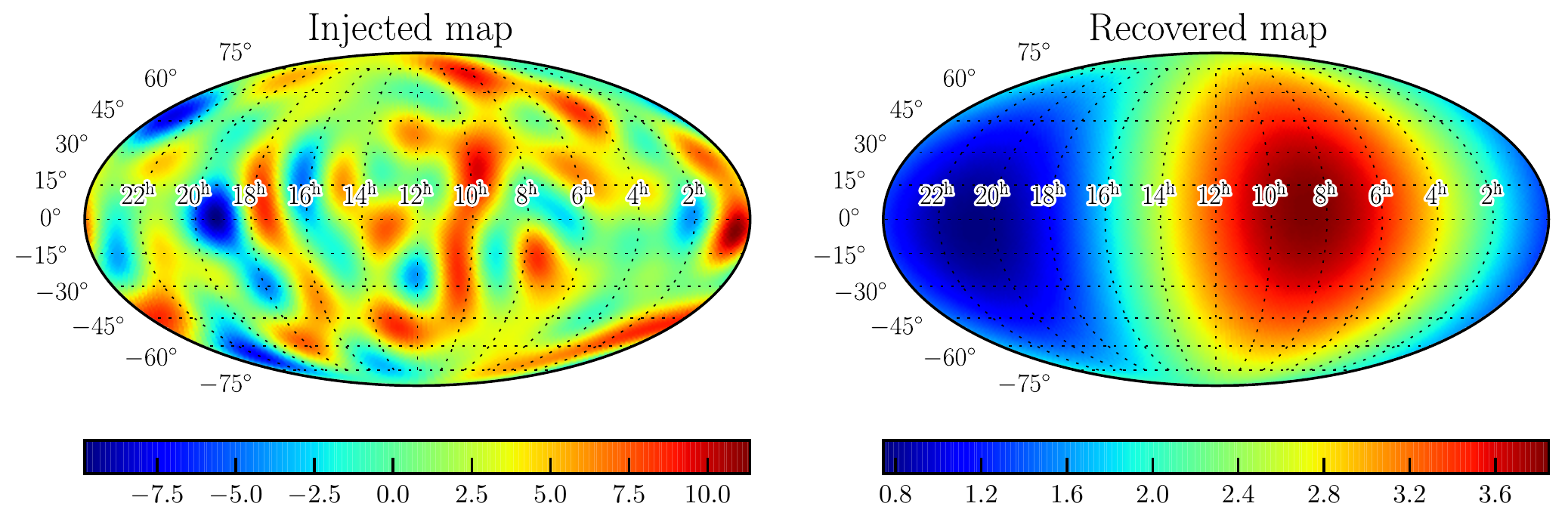}
\caption{Maps of the real amplitude component of a scalar transverse (breathing mode) background. \textit{(Left)} a randomly generated scalar-transverse gravitational-wave sky,
with structure up to and including $l=10$. \textit{(Right)} the
corresponding recovered sky, computed by first forming the observed
signal vector for an array of $N_p=50$ pulsars via 
$\mathbf{r}=\mathbf{R}\mathbf{h}$, where each element of the array
response matrix, $\mathbf{R}$, corresponds to the response of a
particular pulsar to gravitational waves propagating in a given sky direction. 
We perform a noiseless map recovery by computing
$\mathbf{R}^+\mathbf{r}$ (where $\mathbf{R}^+$
is the pseudo-inverse of $\mathbf{R}$) which gives the map in the right panel. We
note the lack of small-scale angular structure in the recovered map
compared to the injected map.}
\label{f:breathing-mode-maps}
\end{center}
\end{figure*}

\begin{figure*}[htbp]
\begin{center}
\subfigure[]{\includegraphics[width=.49\textwidth]{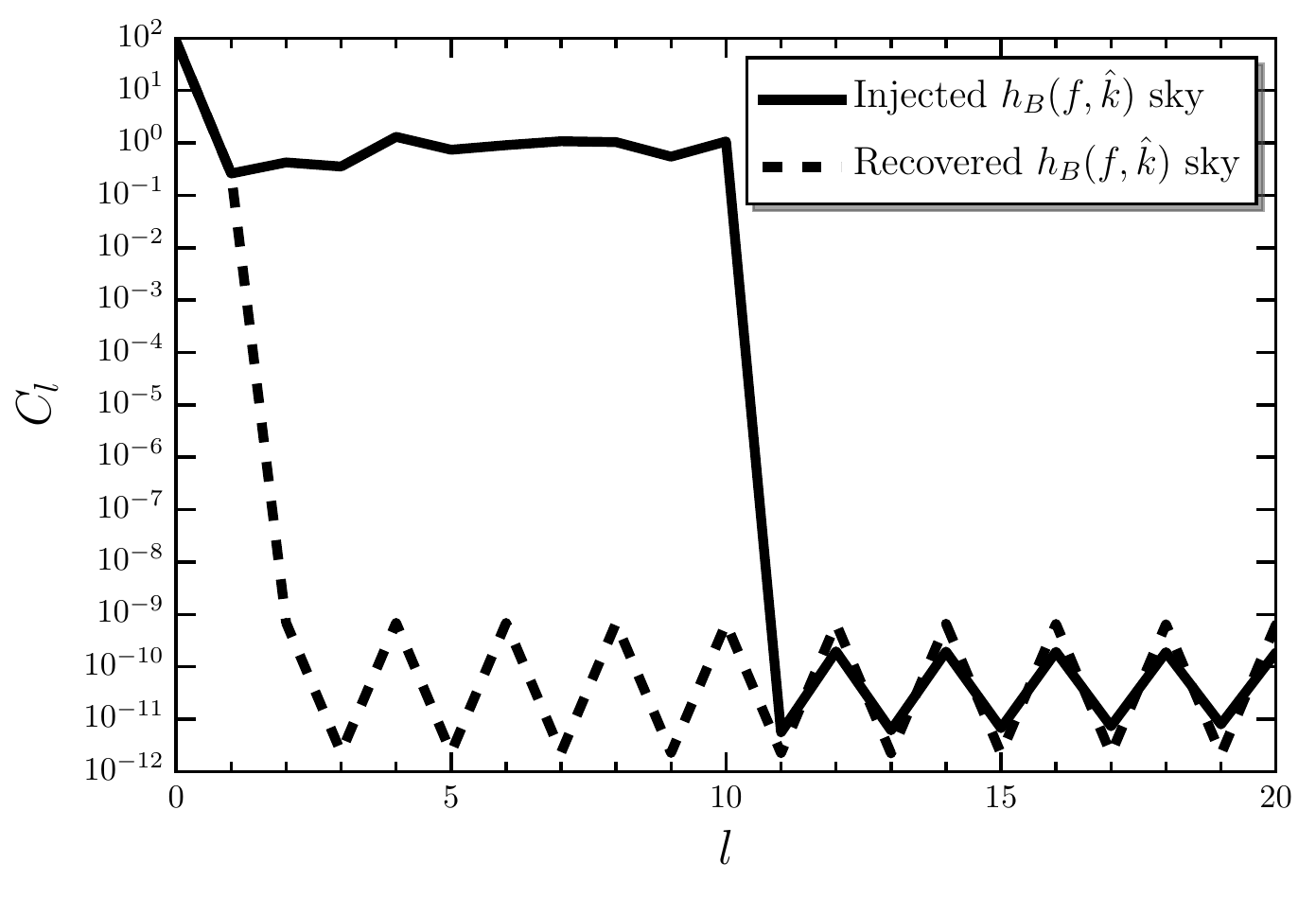}}
\subfigure[]{\includegraphics[width=.49\textwidth]{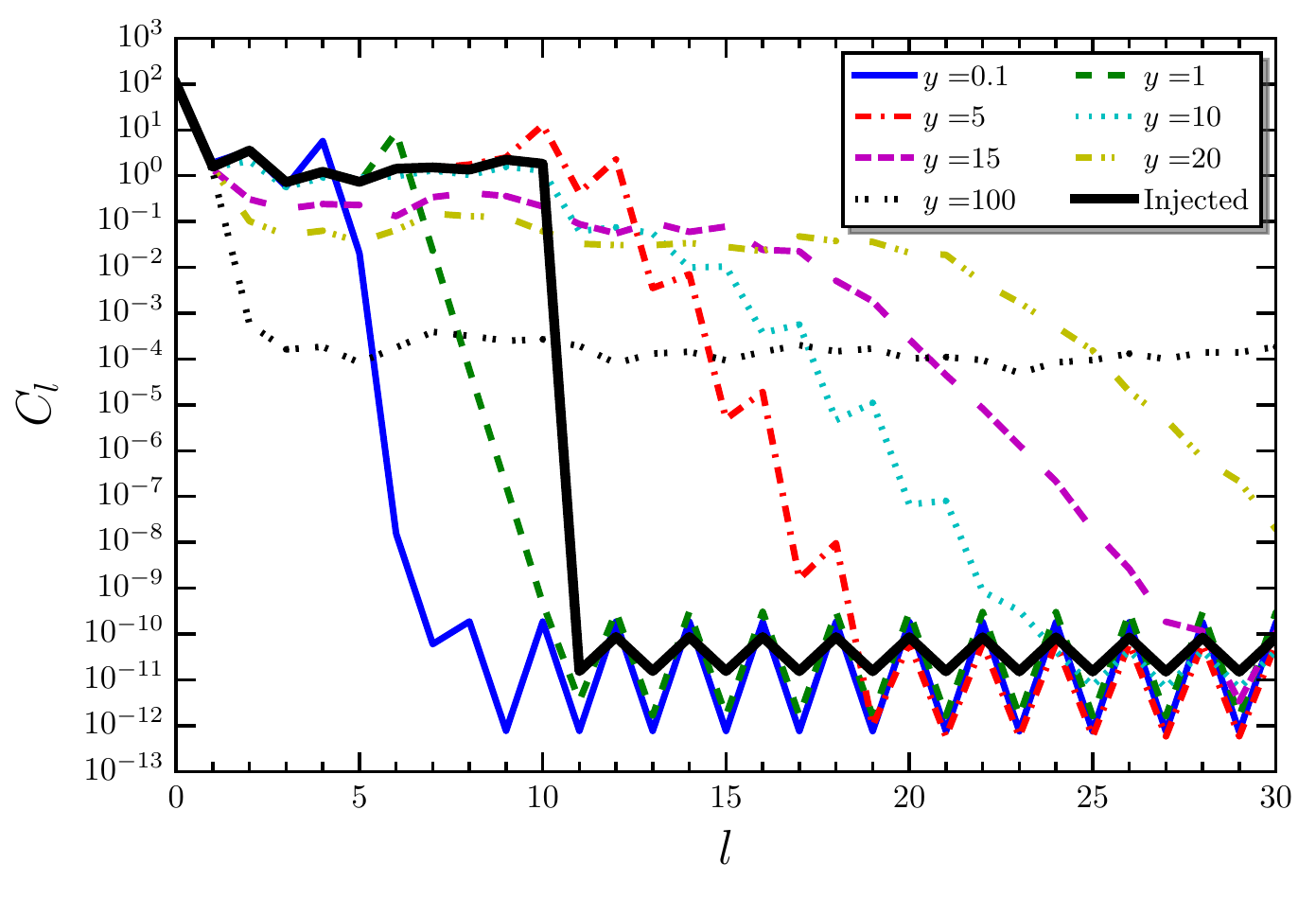}}
\caption{{\it (Left)} A comparison of the angular power spectra of the injected
  scalar-transverse sky map shown in the left-hand panel of 
  Fig.~\ref{f:breathing-mode-maps}, and the PTA-recovered map shown in the
  right-hand panel of the same figure. We see that PTAs will completely
  lack sensitivity to angular structure in a scalar-transverse 
gravitational-wave sky
  beyond the dipole level. This result is confirmed analytically in
  Eq.\ (\ref{e:RB-mapping-approx}). {\it (Right)} We use the full
  Earth and pulsar term response from
  Eq.\ (\ref{e:RA_pulsars_exact}) to investigate map recovery with
  finite $y$. The pulsar-term
  will be highly oscillatory across the sky, so we expect some numerical
  fluctuations in our results. As $y$ is increased the PTA shows greater sensitivity to higher multipole moments in the GW background. At $y=10$ the
  PTA is able to recover all modes in the injected map,
  although the non-zero sensitivity of the pulsar-term response at higher multipoles amplifies noise from the pixelation of the sky. For $y\gtrsim 20$ the Earth term behavior is recovered,
  and we observe a lack of sensitivity to modes beyond dipole. See text
  for further details.}
\label{f:breathing-angular-power}
\end{center}
\end{figure*}

Equation (\ref{e:RB-mapping-approx}) contains a key result of this
paper. In the limit that $y\rightarrow\infty$, where the influence of
the pulsar term tends to zero, we find that PTAs will completely lack
sensitivity to any angular structure beyond $l=1$ in a 
gravitational-wave background with scalar-transverse polarisation. 
We can verify this analytic result through
numerical map making and recovery. Using 
\be
h_B(f,\hat k) = \frac{1}{\sqrt{2}}
\sum_{l=0}^\infty\sum_{m=-l}^l a_{lm}^B(f) Y_{lm}(\hat k)\,,
\label{e:hB}
\ee
which relates the expansion coefficients $h_B(f,\hat k)$ and
$a^B_{(lm)}(f)$ in the polarisation and spherical harmonic bases
(see Secs.~\ref{s:polbasis}, \ref{s:sphbasis}), we generate
a random scalar-transverse (breathing mode) background with angular
structure up to and including $l=10$. This injected map is shown in
the left panel of Fig.~\ref{f:breathing-mode-maps}. To compute the
PTA response to such a background, we generate a
random array of $N_p=50$ pulsars scattered isotropically across the
sky. We work in the polarisation basis rather than the spherical-harmonic basis here, since the PTA response to different angular scales in the GW background is trivial in the latter, and we seek a numerical confirmation of 
Eq.~(\ref{e:RB-mapping2}). The PTA response is computed (with a sky resolution set by a given number of pixels $N_{\rm pix}$) using the
Earth term component of Eq.~(\ref{e:RA_pulsars_exact}), by taking the
dot product of the array response matrix, $\mathbf{R}$, with the vector of amplitude
values at each sky-location, $\mathbf{h}$. The matrix $\mathbf{R}$ has 
dimensions $(N_p\times N_{\rm pix})$, with each element corresponding to the
response of a particular pulsar to gravitational waves propagating in a certain
direction (denoted by a map pixel), as given by the integrand of Eq.~(\ref{e:RA_pulsars_exact}). The resulting vector is the signal
observed by the full array, $\mathbf{r}=\mathbf{R}\mathbf{h}$. 
We can invert this in a noiseless map
recovery by taking the dot product of the Moore-Penrose pseudoinverse of
$\mathbf{R}$ with this observed signal vector. The recovered
scalar-transverse sky is shown in the right-hand panel of Fig.~\ref{f:breathing-mode-maps}, 
where we note a lack of small-scale
angular structure. We compute the angular power spectrum of the
recovered and injected maps via \textsc{HEALPix} \citep{HEALPix},
which is capable of rapid map decompositions. The results are shown in
the left-hand panel of Fig.~\ref{f:breathing-angular-power}, where we see that despite the
injected map having structure up to $l=10$, the recovered map only
contains structure up to and including the dipole. This numerical
result is a confirmation of the corresponding analytic computation in
Eq. (\ref{e:RB-mapping-approx}).

We can also check Eqs.\ (\ref{e:RB-mapping1}) and (\ref{e:RB-mapping2}), 
which imply that the PTA response to a
scalar-transverse background will extend beyond the dipole for pulsars at finite distances. We
do so again with numerical map making and recovery, by using the full
Earth and pulsar term scalar-transverse response function given in
Eq.\ (\ref{e:RA_pulsars_exact}). The pulsar term
will be highly oscillatory across the sky, so we expect some numerical
fluctuations in our results. For this study we inject white Gaussian noise in each pulsar measurement, with an amplitude such that the GW background remains in the strong signal limit.
In the right-hand panel of Fig.~\ref{f:breathing-angular-power} we 
see that the PTA has increasing sensitivity to higher multiple moments in the background as $y$ is increased. At $y\sim 5-10$ the PTA is able to recover the full angular structure of the background, but also suffers from noise leakage at higher multipoles, since the non-zero response of the pulsar term at these higher multipoles amplifies noise arising from the pixelation of the sky. The pulsar term response peaks at $l\sim y$, such that for PTAs with $y=15, 20$ we see a drop-off in sensitivity at $l\sim 15, 20$, even though the response is merely amplifying pixel noise at these multipoles. For $y\gtrsim 20$ the Earth term behaviour is recovered,
and we observe a lack of sensitivity to modes beyond dipole. To put
these results into context, we recall that $y=2\pi fL/c$ and peak PTA sensitivity
to a gravitational-wave background occurs at $f\sim 1/T$ where $T$ is the total
observation time. For $T=20$ years, this gives $f\sim 1.6~{\rm nHz}$. 
Thus in order for a PTA to have sensitivity to structure in a
scalar-transverse sky beyond dipole, we need $y\lesssim 10$, which
corresponds to all pulsars in our array being at a distance of $\lesssim
0.01$ kpc from Earth. Given that most timed millisecond pulsars have distances
$\gtrsim 0.2$ kpc, it is unlikely that this extended reach to sensitivity beyond dipole
modes will be possible with current arrays.

Using the mapping response functions $R^B_{(lm)}(f)$ calculated above,
we can also compute the overlap reduction function for an uncorrelated,
anisotropic background, recovering the result given 
in Sec.~\ref{s:breathing-ORF-sec}.
Details of that calculation are given in App.~\ref{s:appRecoverBreathingORF}.

%%%%%%%%%%%%%%%%%%%%%%%%%%%%%%%%%%%%%%%%%%%%%%%%%%
\subsection{Scalar-longitudinal backgrounds}
\label{s:longitudinal-mapping}

For an arbitrary scalar-longitudinal background we find
\begin{align}
R^L_{(lm)}(f) &=Y_{lm}(\hat{u}) {\cal R}^L_l(2\pi fL/c)\,, 
\end{align}
where
\begin{widetext}
\be
\begin{aligned}
{\cal R}^L_l(y) 
&\equiv 2\pi \int_{-1}^{1}{\rm d}x\>
\frac{1}{2}
\frac{x^2}{1+x} P_l(x) \left(1- {\rm e}^{-i y (1+x)} \right) 
\\
&=2\pi \int_{-1}^{1}{\rm d}x\>
\frac{1}{2}
\left[-1 + x +\frac{1}{1+x}\right]
P_l(x) \left(1- {\rm e}^{-i y (1+x)} \right) 
\\
&= 2\pi\left\{
-\delta_{l0} + \frac{1}{3} \delta_{l1}
+(-i)^l {\rm e}^{-iy} \left[ \left(1-i\frac{l}{y}\right)j_l(y) 
+ i j_{l+1}(y)\right]+ \frac{1}{2}H_l(y)\right\}\,,
\end{aligned}
\ee
where $H_l(y)$ is defined by Eq.~(\ref{e:Hl}) in App.~\ref{s:appCoDirectional}.
Since the spherical Bessel functions behave like $1/y$ for large $y$, the terms 
in the square brackets above tend to zero as $y\rightarrow\infty$,
yielding
\be
R^L_{(lm)}(f) \approx
2\pi Y_{lm}(\hat{u}) 
\left[-\delta_{l0} + \frac{1}{3} \delta_{l1} + \frac{1}{2} H_l(y)\right]\,.
\ee
This is valid for $y\gg 1$, but $y$ finite.

%%%%%%%%%%%%%%%%%%%%%%%%%%%%%%%%%%%%%%%%%%%%%%
\subsection{Vector-longitudinal backgrounds}
\label{s:vector-mapping}

As discussed in Sec.~\ref{s:polbasis}, we can expand each Fourier
component of a vector-longitudinal background in terms of 
gradient and curl tensor spherical harmonics 
$Y^{V_G}_{(lm)ab}(\hat k)$, 
$Y^{V_C}_{(lm)ab}(\hat k)$, which are simply related to the
spin-weight $\pm 1$ spherical harmonics defined in App.~\ref{s:grad-curl-vector}.
It is convenient to relate this expansion 
\be
h_{ab}(f,\hat k)
= \sum_{l=1}^\infty \sum_{m=-l}^l
\left[a^{V_G}_{(lm)}(f)Y^{V_G}_{(lm)ab}(\hat k)
+a^{V_C}_{(lm)}(f)Y^{V_C}_{(lm)ab}(\hat k)\right]
\ee
to a similar expansion in terms of the polarisation basis:
\be
h_{ab}(f,\hat k) = 
h_{X}(f,\hat{k}) e^{X}_{ab}(\hat{k}) + h_{Y}(f,\hat{k}) e^{Y}_{ab}(\hat{k})\,.
\ee
The relationship is
\be
h_{X}(f,\hat{k})\pm ih_{Y} (f,\hat{k}) 
= \mp\frac{1}{\sqrt{2}}\sum_{lm} 
\left( a^{V_G}_{(lm)}(f) \pm i a^{V_C}_{(lm)}(f)\right) {}_{\pm 1}Y_{lm}(\hat{k})\,,
\ee
or, equivalently, 
\be
\begin{aligned}
h_{X}(f,\hat{k})
&=\frac{1}{2\sqrt{2}}\sum_{lm} \left[ a^{V_G}_{(lm)}(f) \left({}_{-1}Y_{lm}(\hat{k}) - {}_{1}Y_{lm}(\hat{k}) \right) 
-i a^{V_C}_{(lm)}(f) \left({}_{-1}Y_{lm}(\hat{k}) + {}_{1}Y_{lm}(\hat{k}) \right) \right],
\\ 
h_{Y}(f,\hat{k})
&=\frac{1}{2\sqrt{2}}\sum_{lm} \left[ a^{V_C}_{(lm)}(f) \left({}_{-1}Y_{lm}(\hat{k}) - {}_{1}Y_{lm}(\hat{k}) \right) 
+i a^{V_G}_{(lm)}(f) \left({}_{-1}Y_{lm}(\hat{k}) + {}_{1}Y_{lm}(\hat{k}) \right) \right]\,,
\end{aligned}
\ee
where ${}_{\pm 1}\!Y_{lm}(\hat k)$ are the spin-weight $\pm1$ spherical harmonics
defined in App.~\ref{s:spinweightedY}.

The expressions for the grad and curl response functions 
for an arbitrary vector-longitudinal 
background can be calculated using the same methods as in the preceding subsections.
We find
\be
R^{V_G}_{I(lm)}(f) 
=Y_{lm}(\hat{u}_I) {\cal R}^{V_G}_l(2\pi fL_I/c)\,,
\qquad
R^{V_C}_{I(lm)}(f)=0\,,
\ee
where
\begin{align}
{\cal R}^{V_G}_l (y) &=\pi\, {}^{(1)}\!N_l
\int_{-1}^1 {\rm d}x\>\left[x(1-x) \left(1-{\rm e}^{-iy(1+x)}\right) \frac{{\rm d}P_l}{{\rm d}x} \right]\,.
\label{e:nuG}
\end{align}
Thus, the response to vector curl modes is identically zero for pulsar timing
arrays, as is the case for tensor curl modes, as shown in \cite{gair-2014}.
Evaluating the integral in Eq.~(\ref{e:nuG}) by parts we find
\begin{align}
R^{V_G}_l (y) 
&=\pi\,{}^{(1)}\!N_l 
\left[-2\delta_{l0}+\frac{4}{3} \delta_{l1} +(-1)^l {\rm e}^{-iy} 
\int_{-1}^1 {\rm d}x\>(1+(2+iy)x +iyx^2){\rm e}^{iyx} P_l(x)\right] 
\\
&=\pi\,{}^{(1)}\!N_l 
\left\{\frac{4}{3} \delta_{l1} + 2 (-i)^l {\rm e}^{-iy} 
\left[\left(1-\frac{il}{y}\right) (l+1)j_l(y) - (y-i(2l+3))j_{l+1}(y) - iy j_{l+2}(y) \right] \right\},
\end{align}
\end{widetext}
where we have dropped the $\delta_{l0}$ term since for spin-weight
$\pm1$ harmonics we have $l\geq1$.
Taking the usual limit that the pulsar is many
gravitational-wave wavelengths from the Earth, $y \gg 1$, and using the
asymptotic result
\be
j_l(y) \approx \frac{1}{y} \sin \left(y-\frac{l\pi}{2}\right) 
+O\left(\frac{1}{y^{\frac{3}{2}}}\right)\,,
\quad {\rm for\ } y\gg 1,
\label{e:jl_asymptotic}
\ee
we find
\be
R^{V_G}_{lm}(f)
\approx 
2\pi Y_{lm}(\hat u)
\left[\frac{2}{3} \delta_{l1} + (-1)^{l}\,{}^{(1)}\!N_l\right]\,.
\ee
As expected, this agrees with the result obtained by evaluating the
integral in Eq.~(\ref{e:nuG}) without the pulsar term, i.e., making the
replacement $\{1-\exp[-iy(1+x)]\} \rightarrow 1$.

%%%%%%%%%%%%%%%%%%%%%%%%%%
\subsection{Overlap reduction function for statistically isotropic backgrounds}
\label{s:ORF-mapping}

For a statistically isotropic, unpolarized and parity-invariant
background (see, for example, Eqs.~(52)--(54) of~\cite{gair-2014})
\be
\Gamma(f) = \sum_l C_l \Gamma_{l}(f),
\label{e:Gamma_statiso}
\ee
where
\be
\Gamma_{l}(f) = \sum_{m=-l}^l \sum_P 
R_{1(lm)}^P(f) R_{2(lm)}^{P*}(f).
\ee
Here $\sum_P$ is a sum over the polarization 
states for a particular type of background 
(e.g., $P=\{V_G,V_C\}$ or $P=\{G,C\}$ for vector-longitudinal
or transverse tensor backgrounds).
Using the results of the previous subsections, 
we have in the limit $y_1\gg 1$, $y_2\gg 1$ (where $y_I = 2\pi f L_I/c$ as before):

\noindent
Transverse tensor modes ($l\ge 2$):
\be
\Gamma^T_l(f)\approx 
\pi(2l+1)(N^T_l)^2 P_l(\cos\zeta)\,,
\ee
which was found in \cite{gair-2014}.\\
Scalar-transverse mode ($l\ge 0$):
\be
\Gamma^B_l(f)\approx
\pi (2l+1)\frac{1}{2}\left[\delta_{l0} + \frac{1}{9}\delta_{l1}\right]\,P_l(\cos\zeta)\,.
\ee
Scalar-longitudinal mode ($l\ge 0$):
\begin{widetext}
\be
\Gamma^L_l(f) \approx 
\pi(2l+1)
\left\{\delta_{l0}\left[1-\frac{1}{2}\left(H_0(y_1)+H_0^*(y_2)\right)\right]
+\delta_{l1}\left[\frac{1}{9} +\frac{1}{6}\left(H_1(y_1) + H_1^*(y_2)\right)\right]
+\frac{1}{4}H_l(y_1)H_l^*(y_2)\right\}P_l(\cos\zeta)\,.
\ee
\end{widetext}
Vector-longitudinal modes ($l\ge 1$):
\be
\Gamma^V_l(f) 
\approx \pi(2l+1)({}^{(1)}\!N_l)^2
\left[-\frac{8}{9}\delta_{l1}+1\right]
P_l(\cos\zeta)\,.
\ee
Note that only the scalar-longitudinal overlap reduction
functions $\Gamma^L_l(f)$ are actually 
frequency-dependent in the large $y$ limit, via
their dependence on $H_l(y_I)$.
The other overlap reduction functions depend only on the 
angular separation $\zeta$ between the pair of pulsars.

\begin{figure*}[htbp]
\begin{center}
\subfigure[]{\includegraphics[width=.49\textwidth]{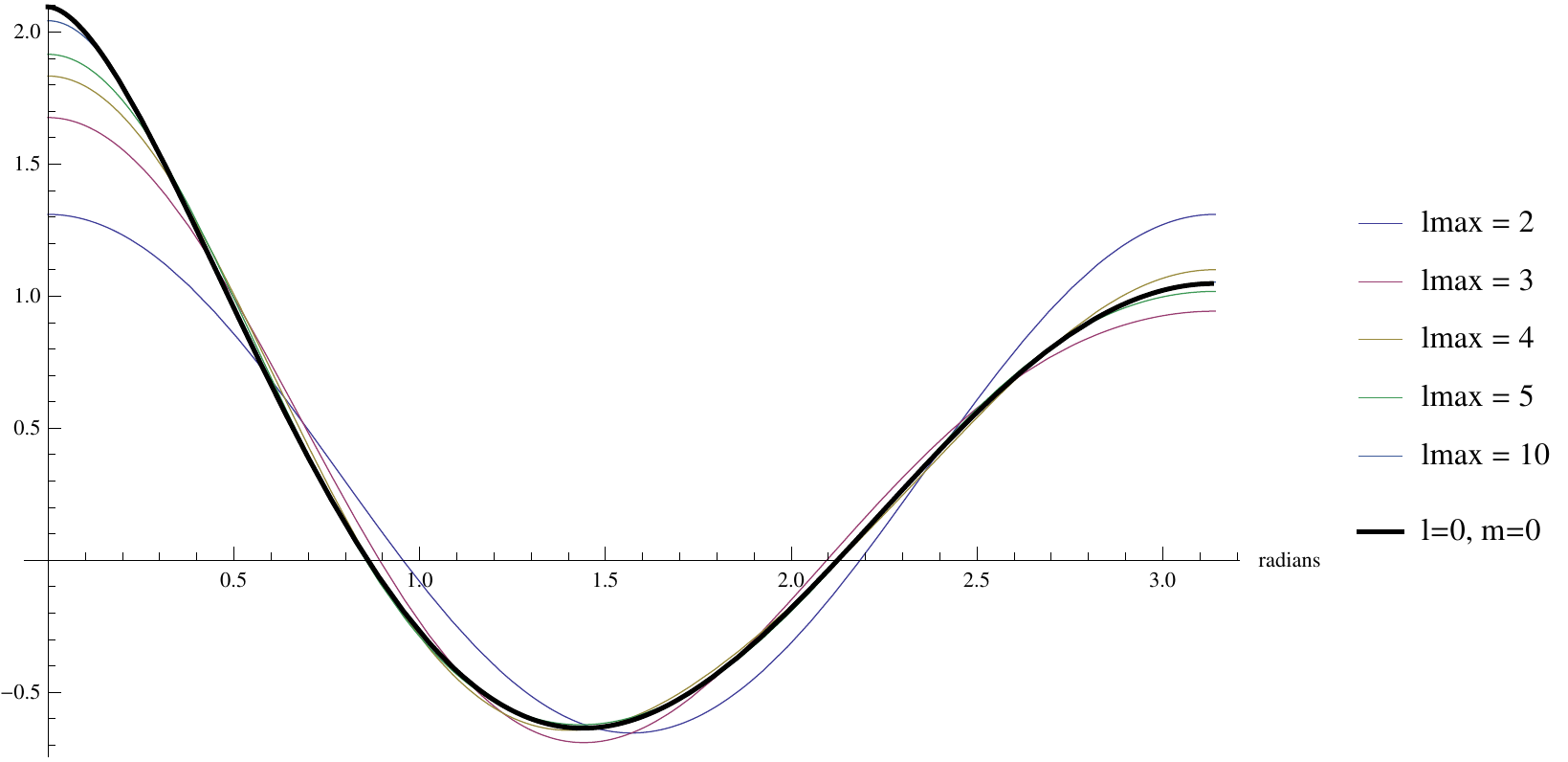}}
\subfigure[]{\includegraphics[width=.49\textwidth]{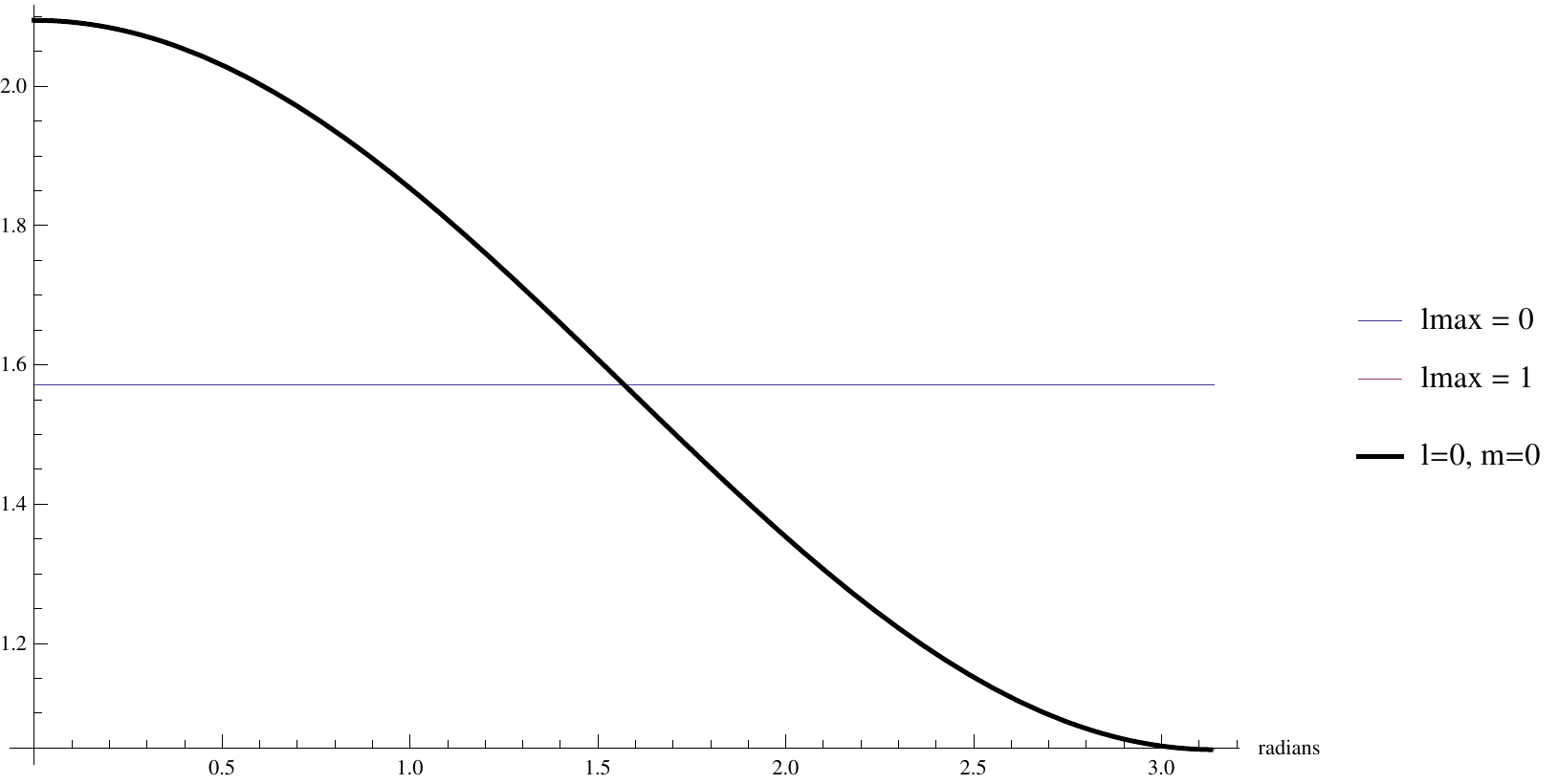}}
\subfigure[]{\includegraphics[width=.49\textwidth]{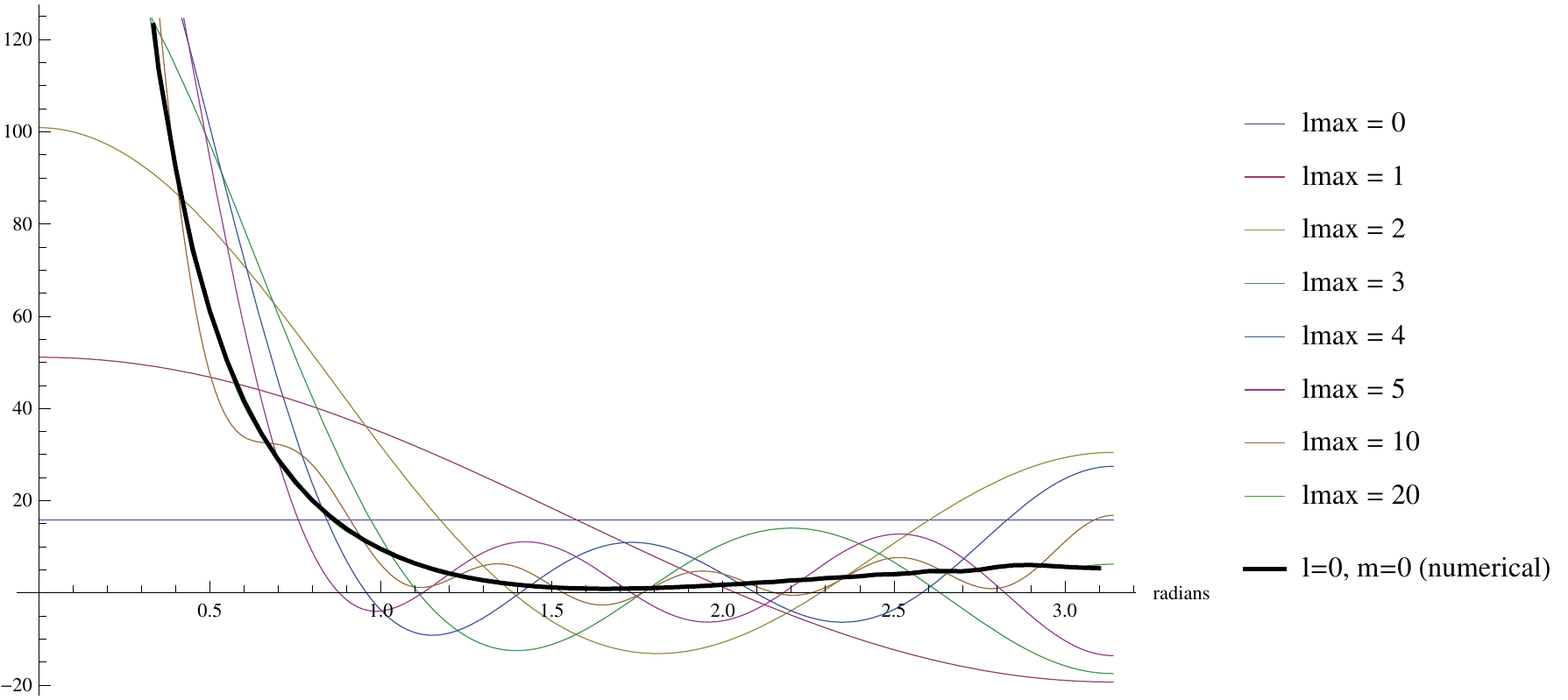}}
\subfigure[]{\includegraphics[width=.49\textwidth]{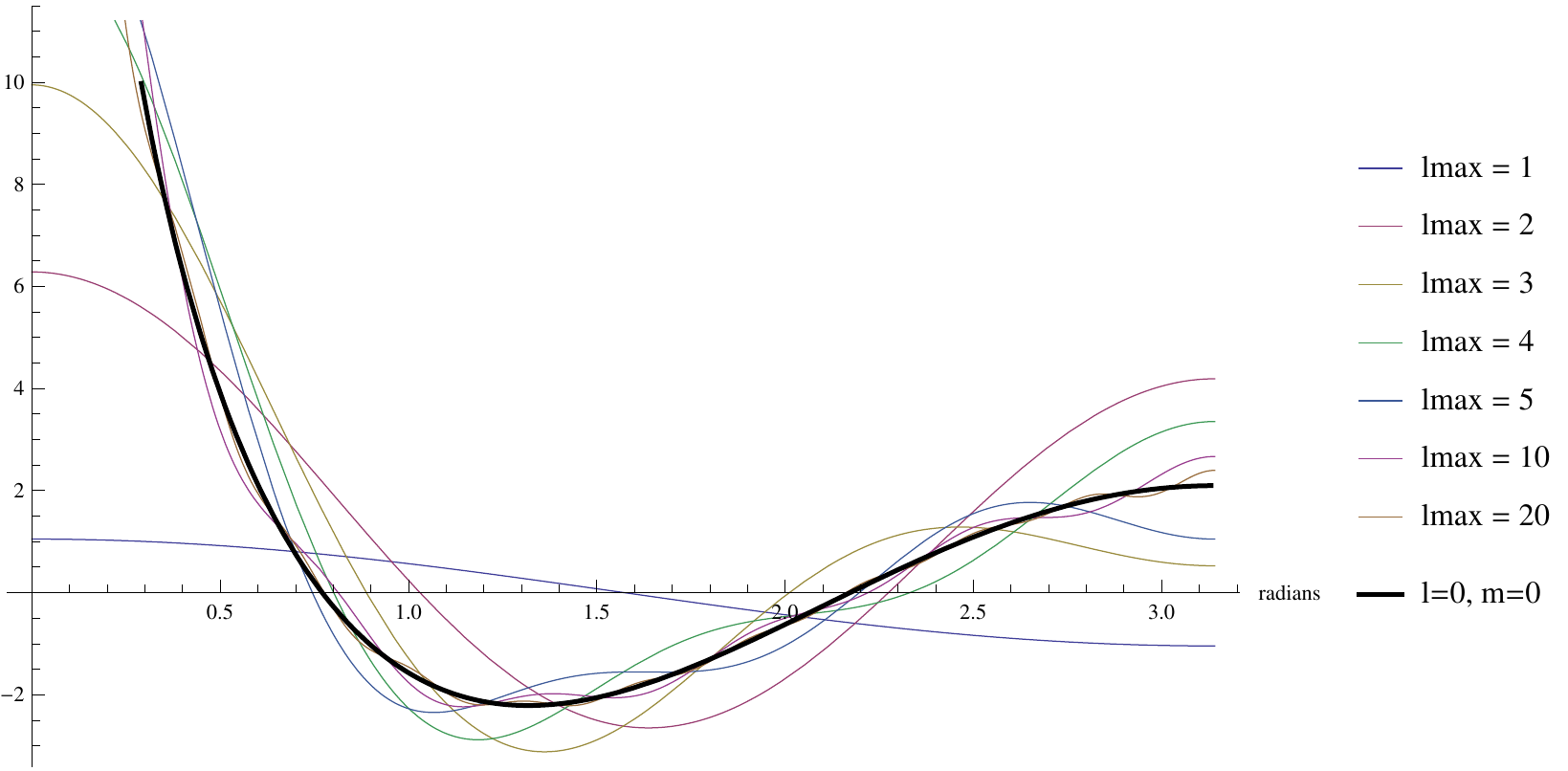}}
\caption{Approximations to the overlap reduction functions for an 
isotropic, unpolarized and uncorrelated stochastic background,
plotted as a function of the angle between a pair of pulsars.
The approximations are obtained by summing products of the response 
functions over $l$ for different values of $l_{\rm max}$.
Panel (a): transverse tensor background;
Panel (b): scalar-transverse (breathing) background;
panel (c): scalar-longitudinal background;
panel (d): vector-longitudinal background.
We are working in the large $y$ limit for all of these cases.
For the scalar-longitudinal background, we have taken $y_1=100$ and $y_2=200$.
The thick black line in each plot is the ``full" expression for
the overlap reduction function, corresponding to the limit 
$l_{\rm max}\rightarrow\infty$.
(These limiting expressions equal $\sqrt{4\pi}/2$ times the 
$l=0$, $m=0$ component of the overlap reduction functions 
calculated in Sec.~\ref{s:orf-sec}.)
For the scalar-longitudinal case, the full expression was 
calculated numerically.}
\label{f:gamma_approx}
\end{center}
\end{figure*}

As shown in \cite{gair-2014}, 
an isotropic, unpolarized and uncorrelated background
has $C_l=1$ for all $l$.
In Fig.~\ref{f:gamma_approx} we plot approximations 
to $\Gamma^B$, $\Gamma^L$, $\Gamma^V$, and $\Gamma^T$ 
corresponding to different values of $l_{\rm max}$ in 
the summation of Eq.\~(\ref{e:Gamma_statiso}), 
taking $C_l=1$ for all $l$ up to $l_{\rm max}$.
(Recall that for the vector overlap reduction function,
the summation starts at $l=1$, while for the tensor overlap
reduction function, it starts at $l=2$.)
These finite $l_{\rm max}$ expressions are compared
to the $l=0$, $m=0$ components of the overlap
reduction functions calculated in Sec.~\ref{s:orf-sec}
and plotted in Figs.~\ref{f:gammaP}, \ref{f:gammaB},
\ref{f:gammaL}, \ref{f:gammaX}.
The normalization is different than in those figures,
since the $l=0$, $m=0$ components need to be multiplied
by $\sqrt{4\pi}/2$ in order to obtain the isotropic overlap
reduction function.
(The factor of $\sqrt{4\pi}$ comes from $Y_{00}(\hat k)=1/\sqrt{4\pi}$;
the factor of $1/2$ is needed to get agreement between 
Eq.~(\ref{eq:<hh>}) and Eq.~(32) of \cite{gair-2014} for 
isotropic, unpolarized backgrounds.)

Figure~\ref{f:gamma_approx} confirms what was found for the transverse tensor modes in~\cite{gair-2014}, namely that a good approximation to the full overlap reduction function can be obtained by including only a relatively small number of modes in the sum. The maximum $l$ required in the sum is approximately $1, 4, 10$ and $20$ for the scalar-transverse, transverse tensor, vector-longitudinal and scalar-longitudinal backgrounds respectively. 

\section{Conclusion}
\label{s:conclusion}

%A general metric theory of gravity obeying Einstein's equivalence
%principle can have as many as six distinct polarisations of
%gravitational waves. These split evenly into three transverse and three
%longitudinal modes, where in this case transverse and longitudinal refer to
%the plane or direction in which a configuration of test particles are
%deformed due to the passage of the gravitational wave. The three transverse
%polarisations correspond to the familiar {\it plus} and {\it cross}
%modes permitted within GR, and a scalar (breathing) mode whose
%influence is a purely radial deformation in the transverse plane. The
%three longitudinal modes correspond to two vector modes ($X$ and $Y$),
%and a scalar mode.

In this paper we have investigated the overlap reduction functions
and response functions of PTAs for non-GR polarisations of
gravitational waves. The overlap reduction function describes the sensitivity of a pair
of pulsars to a gravitational-wave background in a cross-correlation analysis. 
The cross-correlation signature traced out by the overlap reduction function 
from an entire array of precisely-timed millisecond pulsars will aid in isolating any
gravitational-wave signal from other stochastic processes which may have similar
spectral properties. Hence, current searches for stochastic gravitational-wave
backgrounds rely on models of the overlap reduction function as the smoking-gun 
signature of a signal. 
For an isotropic stochastic background in GR, the overlap reduction function is 
known as the {\it Hellings and Downs} curve, and depends only on the angular 
separation between pulsars in the array. The overlap reduction functions for 
arbitrary anisotropic stochastic backgrounds in GR were investigated in
\citet{gair-2014,Mingarelli:2013}, where it was shown that these
functions are now dependent on the positions of each pulsar relative
to the distribution of gravitational-wave power on the sky. 

The gravitational wave
polarisation has a strong influence on the overlap reduction function through the form of the
pulsar response functions. \citet{chamberlin2012} studied
the form of the overlap reduction functions for isotropic backgrounds of gravitational waves 
for scalar-transverse, scalar-longitudinal, and vector-longitudinal polarisation modes. 
In this paper, we have extended that analysis to find analytic
expressions for the overlap reduction functions for anisotropic non-GR backgrounds. 
A key result of this work is that PTAs will completely lack sensitivity to structure
beyond quadrupole in the power of a scalar-transverse background.
This result holds regardless of the number of pulsars, timing-precision, or
observational schedules---it is a property of the geometric sensitivity
of PTAs to gravitational-wave signals of scalar-transverse
polarisation. Additionally, we have found analytic expressions for the
overlap reduction functions for arbitrary anisotropic 
vector-longitudinal backgrounds. 
We also derived a semi-analytic expression for the overlap reduction functions of anisotropic scalar-longitudinal backgrounds, in which case a consideration of the pulsar-term is crucial to 
avoid divergences.

In the second half of this paper, we extended the formalism of our previous work in
\citet{gair-2014}, where the Fourier amplitudes in a plane-wave
expansion of the GR metric perturbation were decomposed
with respect to a basis of gradient and curl spherical harmonics, 
which are related to spin-weight $\pm 2$ spherical harmonics. 
By determining the components of the background in such a decomposition it is possible to construct a map of both the amplitude and the phase of the gravitational wave background across the sky, rather than simply reconstructing the power distribution. 
The decomposition in terms of spin-weight $\pm 2$ spherical harmonics is 
made possible by the transverse-traceless nature of the GR gravitational-wave 
metric perturbations.
Here we have appealed to the structure of the gravitational-wave metric 
perturbations for non-GR polarisations to perform the same procedure---the 
Fourier amplitudes of scalar modes can be expanded in terms of ordinary 
spin-weight $0$ spherical harmonics, while the vector mode amplitudes can
be expanded in terms of a spin-weight $\pm 1$ spherical harmonic basis. 
In so doing,
we found that PTAs lack sensitivity to structure in the polarisation
amplitude of a scalar-transverse background beyond dipole anisotropy,
which can be used to explain the lack of sensitivity to power anisotropies
beyond quadrupole. This result was verified through numerical map
making and recovery, where we found some sensitivity to modes beyond
dipole when $y=2\pi fL/c$ was very small, but this would require all
pulsars to lie within a distance of $0.01$ kpc from Earth. We also found
that PTAs will lack sensitivity to vector curl modes for a 
vector-longitudinal background, which is analogous to the finding in 
\citet{gair-2014} that PTAs are insensitive to the tensor curl modes of 
gravitational-wave backgrounds in GR.

This paper provides several ready-to-use expressions for overlap reduction functions 
for non-GR stochastic backgrounds with arbitrary anisotropy. These expressions
can be trivially plugged into any current or planned PTA stochastic background
search pipeline to obtain limits on the strain amplitude of a non-GR
gravitational-wave sky. We also provide several ready-to-use expressions for the response
functions of a single pulsar to anisotropies in a non-GR gravitational-wave background. 
The implications of this are that we can use an array of pulsars to perform a Bayesian or
frequentist search for the angular dependence of the Fourier modes of a plane-wave
expansion of the gravitational-wave metric perturbations, and in so doing produce maps
of the polarisation content of the sky that include phase
information rather than simply map the distribution of power. %This
%will allow us to fully reconstruct the waveform of the gravitational waves influencing
%each pulsar. 

The results in this paper also indicate what is possible to
measure in principle with a sufficiently extensive pulsar timing
array. The dependence of the response on the pulsar location on the
sky is proportional to $Y_{lm}(\hat{u})$, where $\hat{u}$ is the
direction to the pulsar, for all polarisation types. By decomposing
the pulsar response map, at a particular frequency, into regular (scalar) spherical harmonics, the coefficients of each $Y_{lm}(\hat{u})$ mode of the response map can be
determined, but these coefficients will be a sum of the contributions from each of
the polarisation types. Scalar-transverse and transverse tensor
backgrounds can be distinguished because PTAs typically operate in a regime in which the pulsar term is negligible and so the response is independent of the distance to the pulsar. In that regime, PTAs are only sensitive to modes of the scalar-transverse background with $l < 2$, while transverse tensor backgrounds can only contain modes with $l \geq 2$. However, longitudinal backgrounds can only be
distinguished from transverse backgrounds if there are multiple
pulsars along a given line of sight, or if there is a known
correlation (e.g., a power law) between the background amplitudes at
different frequencies. In either of these scenarios, we can exploit the 
dependence of the pulsar term on $2\pi fL/c$, which is much more
significant for the longitudinal modes of the background. Thus, in the limit of
infinitely many pulsars distributed across the sky at a range of
distances, we would expect to be able to measure the entire content of
the background in each polarisation state and at each frequency. In
practice, of course, a pulsar timing array of $N_p$ pulsars can only
measure $2N_p$ real components of the background~\cite{gair-2014,Cornish-vanHaasteren:2014}, and
so the resolution of any map that we produce will be limited by the
number of pulsars in the array. Roughly speaking, to produce a map of the gravitational wave sky in all polarisation states to an angular resolution of $1/l_{\rm max}$ would require $N_p = 3 (l_{\rm max}+1)^2$ pulsars, but this should be explored more carefully in the future.

For a further discussion of the prospects of this type of mapping analysis,
see~\citet{gair-2014,Cornish-vanHaasteren:2014}. We plan to apply the results of this paper to the analysis of real data, 
to map the amplitude and phase content of non-GR gravitational-wave 
backgrounds influencing the arrival times of millisecond pulsars. 
This will allow us to place constraints on beyond-GR polarisations 
of nanohertz gravitational waves.

%%%%%%%%%%%%%%%%%%%%%%%%%%%%%%%%%%%%%%%%%%%%%%%
\acknowledgments 
JG's work is supported by the Royal Society. 
This research
was in part supported by ST's appointment to the NASA Postdoctoral
Program at the Jet Propulsion Laboratory, administered by Oak Ridge
Associated Universities through a contract with NASA.
JDR acknowledges support from NSF Awards PHY-1205585, PHY-1505861,
HRD-1242090, and the NANOGrav Physics Frontier Center, NSF PFC-1430284.
This research has made use of Python and its standard libraries: 
numpy and matplotlib.  
We have also made use of MEALPix (a Matlab implementation of 
HEALPix~\cite{HEALPix}), developed by the GWAstro Research Group 
and available from {\tt http://gwastro.psu.edu}.
This work was performed using the Darwin Supercomputer of the 
University of Cambridge High Performance Computing Service 
(http://www.hpc.cam.ac.uk/), provided by Dell Inc.~using Strategic
Research Infrastructure Funding from the Higher Education Funding Council 
for England and funding from the Science and Technology Facilities Council.
The authors also acknowledge support of NSF Award PHY-1066293 and the 
hospitality of the Aspen Center for Physics, where this work was completed.

\begin{appendix}

\begin{widetext}

\section{Spin-weighted spherical harmonics}
\label{s:spinweightedY}

This appendix summarizes some useful relations involving spin-weighted
and ordinary spherical harmonics, ${}_sY_{lm}(\hat k)$ and $Y_{lm}(\hat k)$.
For more details, see e.g., 
\citet{Goldberg:1967} and \citet{delCastillo}.
Note that we use a slightly different normalization convention than in
\citet{Goldberg:1967}.
Namely, we put the Condon-Shortley factor $(-1)^m$ in the definition of the
associated Legendre functions $P_l^m(x)$, and thus do not explicitly 
include it in the definition of the spherical harmonics.
Also, for our analysis, we can restrict attention to spin-weighted 
spherical harmonics having {\em integral} spin weight $s$, even though 
spin-weighted spherical harmonics with half-integral spin weight do exist.

Ordinary spherical harmonics:
\be
Y_{lm}(\hat k)=
Y_{lm}(\theta,\phi) = N_l^m
P_l^m(\cos\theta)e^{im\phi}\,,
\quad
{\rm where}\ 
N_l^m = \sqrt{\frac{2l+1}{4\pi}\frac{(l-m)!}{(l+m)!}}.
\label{e:Nlm}
\ee
Relation of spin-weighted spherical harmonics 
to ordinary spherical harmonics:
\be
\begin{aligned}
{}_sY_{lm}(\theta,\phi) 
&=\sqrt{\frac{(l-s)!}{(l+s)!}}\,
\edth^s Y_{lm}(\theta,\phi)
\quad{\rm for}\quad
0\le s\le l\,,
\\
{}_sY_{lm}(\theta,\phi) 
&=\sqrt{\frac{(l+s)!}{(l-s)!}}\,
(-1)^s
\overline{\edth}{}^{-s} Y_{lm}(\theta,\phi)
\quad{\rm for}\quad
-l\le s\le 0\,,
\end{aligned}
\ee
where
\be
\begin{aligned}
\edth \eta
&=-(\sin\theta)^s
\left[\frac{\partial}{\partial\theta}
+i\csc\theta\frac{\partial}{\partial\phi}\right]
(\sin\theta)^{-s}\eta\,,
\\
\overline{\edth} \eta
&=-(\sin\theta)^{-s}
\left[\frac{\partial}{\partial\theta}
-i\csc\theta\frac{\partial}{\partial\phi}\right]
(\sin\theta)^{s}\eta\,,
\end{aligned}
\ee
and $\eta=\eta(\theta,\phi)$ is a spin-$s$ scalar field.\\

\noindent
%Special case for $m=0$:
%
%\be
%{}_sY_{l0}(\theta,\phi) 
%= e^{-is\phi}\,{}_0Y_{ls}(\theta,\phi)
%= N_l^s P_l^s(\cos\theta)
%\ee
%
Series representation:
\be
{}_sY_{lm}(\theta,\phi)
=(-1)^m\left[\frac{(l+m)!(l-m)!}{(l+s)!(l-s)!}\frac{2l+1}{4\pi}
\right]^{1/2}(\sin\theta/2)^{2l}
\sum_{k=0}^{l-s}
\binom{l-s}{k}\binom{l+s}{k+s-m}(-1)^{l-k-s} 
e^{im\phi}(\cot\theta/2)^{2k+s-m}.
\ee
Complex conjugate:
\be
{}_sY_{lm}^*(\theta,\phi) = (-1)^{m+s}\,{}_{-s}Y_{l,-m}(\theta, \phi).
\ee
Relation to Wigner rotation matrices:
\be
D^l{}_{m'm}(\phi,\theta,\psi)=
(-1)^{m'} \sqrt{\frac{4\pi}{2l+1}}\,{}_mY_{l,-m'}(\theta,\phi)e^{-im\psi},
\label{e:WignerD}
\ee
or
\be
\left[D^l{}_{m'm}(\phi,\theta,\psi)\right]^*=
(-1)^{m} \sqrt{\frac{4\pi}{2l+1}}\,{}_{-m}Y_{l,m'}(\theta,\phi)e^{im\psi}.
\label{e:WignerD_CC}
\ee
Parity transformation:
\be
{}_sY_{lm}(\pi-\theta,\phi+\pi) = (-1)^l\,{}_{-s}Y_{lm}(\theta, \phi).
\ee
\noindent
Orthonormality (for fixed $s$):
\be
\int_{S^2} {\rm d}^2\Omega_{\hat k}\>
{}_sY_{lm}(\hat k) \,{}_{s}Y_{l'm'}^*(\hat k)
\equiv\int_0^{2\pi} {\rm d}\phi\int_0^\pi \sin\theta\, {\rm d}\theta\>
{}_sY_{lm}(\theta,\phi) \,{}_sY_{l'm'}^*(\theta,\phi)
= \delta_{ll'}\delta_{mm'}.
\ee
Addition theorem for spin-weighted spherical harmonics:
\be
\sum_{m=-l}^l {}_sY_{lm}(\theta_1,\phi_1)\,{}_{s'}Y_{lm}^*(\theta_2,\phi_2)
=(-1)^{-s'}\sqrt{\frac{2l+1}{4\pi}}\,{}_{-s'}Y_{ls}(\theta_3,\phi_3)
e^{is'\chi_3},
\ee
where
\be
\cos\theta_3 = \cos\theta_1\cos\theta_2 + \sin\theta_1\sin\theta_2\cos(\phi_2-\phi_1),
\ee
and
\be
\begin{aligned}
e^{-i(\phi_3+\chi_3)/2}
&=\frac{\cos\frac{1}{2}(\phi_2-\phi_1)\cos\frac{1}{2}(\theta_2-\theta_1)
-i\sin\frac{1}{2}(\phi_2-\phi_1)\cos\frac{1}{2}(\theta_1+\theta_2)}
{\sqrt{\cos^2\frac{1}{2}(\phi_2-\phi_1)\cos^2\frac{1}{2}(\theta_2-\theta_1)
+\sin^2\frac{1}{2}(\phi_2-\phi_1)\cos^2\frac{1}{2}(\theta_1+\theta_2)}},
\\
e^{i(\phi_3-\chi_3)/2}
&=\frac{\cos\frac{1}{2}(\phi_2-\phi_1)\sin\frac{1}{2}(\theta_2-\theta_1)
+i\sin\frac{1}{2}(\phi_2-\phi_1)\sin\frac{1}{2}(\theta_1+\theta_2)}
{\sqrt{\cos^2\frac{1}{2}(\phi_2-\phi_1)\sin^2\frac{1}{2}(\theta_2-\theta_1)
+\sin^2\frac{1}{2}(\phi_2-\phi_1)\sin^2\frac{1}{2}(\theta_1+\theta_2)}}.
\end{aligned}
\ee
Addition theorem for ordinary spherical harmonics:
\be
\sum_{m=-l}^l Y_{lm}(\hat k_1) Y_{lm}^*(\hat k_2)
=\frac{2l+1}{4\pi}\,P_l(\hat k_1\cdot\hat k_2).
\label{e:addition_theorem_0Y}
\ee
Integral of a product of spin-weighted spherical harmonics:
\be
\int_{S^2} {\rm d}^2\Omega_{\hat k}\>
{}_{s_1}Y_{l_1m_1}(\hat k) \,
{}_{s_2}Y_{l_2m_3}(\hat k) \,
{}_{s_3}Y_{l_3m_3}(\hat k) \,
= \sqrt{\frac{(2l_1+1)(2l_2+1)(2l_3+1)}{4\pi}}
\left( \begin{array}{ccc}l_1&l_2&l_3\\m_1&m_2&m_3\end{array} \right)
\left( \begin{array}{ccc}l_1&l_2&l_3\\-s_1&-s_2&-s_3\end{array} \right),
\ee
where 
$\left( \begin{array}{ccc}l_1&l_2&l_3\\m_1&m_2&m_3 \end{array} \right)$
is a Wigner 3-$j$ symbol, which can be written as
\begin{eqnarray}
\left(\begin{array}{ccc} l&l'&L\\m&m'&M \end{array} \right) 
&=& \sqrt{\frac{(l+l'-L)!(l-l'+L)!(-l+l'+L)! (l+m)!(l-m)!(l'+m')!(l'-m')!(L+M)!(L-M)!}{(l+l'+L+1)!} } \times 
\nonumber \\
&&\hspace{1cm} \times \sum_{z \in \mathbb{Z}} 
\frac{(-1)^{z+l+l'-M}}{z!(l+l'-L-z)!(l-m-z)!(l'+m'-z)!(L-l'+m+z)!(L-l-m'+z)!}.
\end{eqnarray}
See, for example, \citet{Wigner:1959}, \citet{Messiah:1962}, \citet{LLnonrelQM} and references therein.
Note that although this sum is over all integers it contains only a 
finite number of non-zero terms since the factorial of a negative 
number is defined to be infinite.

\section{Gradient and curl rank-1 (vector) spherical harmonics}
\label{s:grad-curl-vector}

The gradient and curl rank-1 (vector) spherical harmonics are defined 
for $l\ge 1$ by
\be
\begin{aligned}
Y^{G}_{(lm)a} &\equiv \frac{1}{2} {}^{(1)}\!N_l\partial_a Y_{lm}
=\frac{1}{2} {}^{(1)}\!N_l\left(\frac{\partial Y_{lm}}{\partial\theta}\,\hat\theta_a +
\frac{1}{\sin\theta}\frac{\partial Y_{lm}}{\partial\phi}\,\hat\phi_a\right),
\\
Y^{C}_{(lm)a} &\equiv \frac{1}{2} {}^{(1)}\!N_l(\partial_b Y_{lm})\epsilon^b{}_a
=\frac{1}{2} {}^{(1)}\!N_l\left(-\frac{1}{\sin\theta}\frac{\partial Y_{lm}}{\partial\phi}\,\hat\theta_a
+\frac{\partial Y_{lm}}{\partial\theta}\,\hat\phi_a\right),
\label{e:YGClma}
\end{aligned}
\ee
where $\hat\theta$ and $\hat\phi$ are the standard unit vectors tangent
to the 2-sphere
\be
\begin{aligned}
\hat\theta
&=\cos\theta\cos\phi\,\hat x+
\cos\theta\sin\phi\,\hat y-
\sin\theta\,\hat z\,,
\\
\hat\phi
&=-\sin\phi\,\hat x+
\cos\phi\,\hat y\,,
\label{e:thetahat_phihat}
\end{aligned}
\ee
${}^{(1)}N_l$ is a normalisation constant
\be
{}^{(1)}N_l = \sqrt{\frac{2(l-1)!}{(l+1)!}}\,,
\label{e:1N}
\ee
and $\epsilon_{ab}$ is the Levi-Civita anti-symmetric tensor
\be
\epsilon_{ab} 
= \sqrt{g} \left( \begin{array}{cc}0&1\\-1&0\end{array}\right)\,,
\qquad
g\equiv {\rm det}(g_{ab})\,.
\ee
Following standard practice, we use the metric tensor on the
2-sphere 
$g_{ab}$ and its inverse $g^{ab}$ to ``lower" and ``raise" tensor
indices---e.g., $\epsilon^{c}{}_b \equiv g^{ca}\epsilon_{ab}$. 
In standard spherical coordinates $(\theta,\phi)$,
\be
g_{ab}=\left(
\begin{array}{cc}
1&0\\
0&\sin^2\theta\\
\end{array}
\right)\,,
\qquad
\sqrt{g}=\sin\theta\,.
\ee
The grad and curl spherical harmonics are related to the 
spin-weight $\pm1$ spherical harmonics 
\be
\begin{aligned}
{}_{\pm1}Y_{lm}(\hat{k}) 
&=\sqrt{\frac{(l-1)!}{(l+1)!}}
\frac{N_l^m}{\sqrt{1-x^2}} 
\left( \pm(1-x^2) \frac{{\rm d}P_l^m}{{\rm d}x} + m P_l^m(x)\right) {\rm e}^{im\phi}\,,
\quad{\rm where}\ x=\cos\theta
\end{aligned}
\ee
via
\be
Y^G_{(lm)a}(\hat k) \pm iY^C_{(lm)a}(\hat k)
=\pm\frac{1}{\sqrt{2}}(\hat\theta_a \pm i \hat\phi_a)\,{}_{\mp 1}Y_{lm}(\hat k)
\ee
or, equivalently, 
\be
\begin{aligned}
Y^{G}_{(lm)a}(\hat{k})
&=\frac{1}{2\sqrt{2}}\left[
\left({}_{-1}Y_{lm}(\hat{k}) - {}_{1}Y_{lm}(\hat{k}) \right) \hat\theta_a
+i  \left({}_{-1}Y_{lm}(\hat{k}) + {}_{1}Y_{lm}(\hat{k}) \right) \hat\phi_a\right],
\\
Y^{C}_{(lm)a}(\hat{k})
&=\frac{1}{2\sqrt{2}}\left[
\left({}_{-1}Y_{lm}(\hat{k}) - {}_{1}Y_{lm}(\hat{k}) \right) \hat\phi_a
-i  \left({}_{-1}Y_{lm}(\hat{k}) + {}_{1}Y_{lm}(\hat{k}) \right) \hat\theta_a\right].
\end{aligned}
\ee

For decompositions of vector-longitudinal backgrounds, as discussed in the main text, 
it will be convenient to construct rank-2 tensor fields
\be
\begin{aligned}
Y^{V_G}_{(lm)ab}
&= Y^G_{(lm)a}\hat k_b+ Y^G_{(lm)b}\hat k_a\,,
\\
Y^{V_C}_{(lm)ab}
&= Y^C_{(lm)a}\hat k_b+ Y^C_{(lm)b}\hat k_a\,,
\label{e:YVGVClmab}
\end{aligned}
\ee
where $\hat k$ is the unit radial vector
orthogonal to the surface of the 2-sphere:
\be
\hat k = \sin\theta\cos\phi\,\hat x+\sin\theta\sin\phi\,\hat y
+\cos\theta\,\hat z\,.
\ee
These fields satisfy the following orthonormality relations
\be
\begin{aligned}
\int_{S^2} {\rm d}^2\Omega_{\hat k}\> 
Y^{V_G}_{(lm)ab} (\hat{k}) Y^{V_G}_{(l'm')}{}^{ab\,*}(\hat{k}) 
&=\delta_{ll'}\delta_{mm'}\,, 
\\
\int_{S^2} {\rm d}^2\Omega_{\hat k}\> 
Y^{V_C}_{(lm)ab} (\hat{k}) Y^{V_C}_{(l'm')}{}^{ab\,*}(\hat{k}) 
&=\delta_{ll'}\delta_{mm'}\,,
\\
\int_{S^2} {\rm d}^2\Omega_{\hat k}\> 
Y^{V_G}_{(lm)ab} (\hat{k}) Y^{V_C}_{(l'm')}{}^{ab\,*}(\hat{k}) 
&=0\,.
\label{e:YVGVCOrthog}
\end{aligned}
\ee

\section{Gradient and curl rank-2 (tensor) spherical harmonics}
\label{s:grad-curl-tensor}

The gradient and curl rank-2 (tensor) spherical harmonics are defined for 
$l\ge 2$ by:
\be
\begin{aligned}
Y^G_{(lm)ab} &= N_l 
\left(Y_{(lm);ab} - \frac{1}{2} g_{ab}  Y_{(lm);c}{}^{c} \right)\,,
\\
Y^C_{(lm)ab} &= \frac{N_l}{2} 
\left(Y_{(lm);ac}\epsilon^c{}_b +  Y_{(lm);bc} \epsilon^c{}_a \right)\,,
\label{e:YGClmab}
\end{aligned}
\ee
where a semicolon denotes covariant derivative on the 2-sphere, and
${}^{(2)}N_l$ is a normalisation constant
\be
{}^{(2)}N_l = \sqrt{\frac{2 (l-2)!}{(l+2)!}}\,.
\label{e:2N}
\ee
Using the standard polarisation tensors on the 2-sphere:
\be
\begin{aligned}
e^+_{ab}(\hat k) 
&= \hat\theta_a\hat\theta_b -\hat\phi_a\hat\phi_b\,,
\\
e^\times_{ab}(\hat k) 
&= \hat\theta_a\hat\phi_b +\hat\phi_a\hat\theta_b\,,
\end{aligned}
\ee
where $\hat\theta$, $\hat\phi$ are given by Eq.~(\ref{e:thetahat_phihat}),
we have~\cite{HuWhite:1997}:
\be
\begin{aligned}
Y^G_{(lm)ab}(\hat{k}) 
&= \frac{{}^{(2)}N_l}{2} \left[ W_{(lm)}(\hat{k}) e_{ab}^+(\hat k) 
+ X_{(lm)}(\hat{k}) e_{ab}^\times(\hat k) \right]\,, 
\\
Y^C_{(lm)ab}(\hat{k}) 
&= \frac{{}^{(2)}N_l}{2} \left[ W_{(lm)}(\hat{k})e_{ab}^\times(\hat k) 
- X_{(lm)}(\hat{k}) e_{ab}^+(\hat k) \right]\,,
\label{e:YGCdef}
\end{aligned}
\ee
where
\be
\begin{aligned}
W_{(lm)}(\hat{k}) 
&= \left( \frac{\partial^2}{\partial \theta^2} 
- \cot\theta\frac{\partial}{\partial\theta} 
+\frac{m^2}{\sin^2\theta} \right)Y_{lm}(\hat{k}) 
= \left( 2 \frac{\partial^2}{\partial \theta^2} 
+ l(l+1) \right) Y_{lm}(\hat{k})\,,
\\
X_{(lm)}(\hat{k}) 
&= \frac{2 i m}{\sin\theta} \left( \frac{\partial}{\partial\theta} 
- \cot\theta\right) Y_{lm}(\hat{k})\,.
\label{e:WX}
\end{aligned}
\ee
%
%Equivalently,
%
%\begin{align}
%W_{(lm)}(\hat{k}) 
%&= +2 \sqrt{\frac{2l+1}{4\pi} \frac{(l-m)!}{(l+m)!}} 
%G^+_{(lm)}(\cos\theta) e^{i m \phi}\,,
%\label{e:Wdef}
%\\
%i X_{(lm)}(\hat{k}) 
%&= -2 \sqrt{\frac{2l+1}{4\pi} \frac{(l-m)!}{(l+m)!}} 
%G^-_{(lm)}(\cos\theta) e^{i m \phi}\,,
%\label{e:Xdef}
%\end{align}
%
%where
%
%\begin{align}
%G^+_{(lm)}(\cos\theta) 
%&= -\left(\frac{l-m^2}{\sin^2\theta} 
%+ \frac{1}{2} l(l-1) \right) P^m_l(\cos\theta) 
%+ (l+m) \frac{\cos\theta}{\sin^2\theta} P^m_{l-1}(\cos\theta)\,,
%\label{e:G+def}
%\\
%G^-_{(lm)}(\cos\theta) 
%&= \frac{m}{\sin^2\theta} 
%\left[(l-1) \cos\theta P_l^m(\cos\theta) 
%- (l+m) P^m_{l-1}(\cos\theta) \right]\,.
%\label{e:G-def}
%\end{align}
%
These functions enter the expression for the 
spin-weight~$\pm 2$ spherical harmonics 
\cite{NewmanPenrose:1966, Goldberg:1967}:
\begin{equation}
{}_{\pm2}Y_{lm}(\hat{k})
=\frac{{}^{(2)}N_l}{\sqrt{2}} \left[
W_{(lm)}(\hat{k})\pm i X_{(lm)}(\hat{k})\right]\,,
\label{e:Ypm2}
\end{equation}
which are related to the grad and curl spherical harmonics via
\begin{align}
Y^G_{(lm)ab}(\hat k) \pm i Y^C_{(lm)ab}(\hat k)
&=\frac{1}{\sqrt{2}}
\left(e_{ab}^+(\hat k) \pm i e_{ab}^\times(\hat k)\right)
\,{}_{\mp 2}Y_{lm}(\hat k)\,.
\label{e:YG+iYC}
\end{align}
Note that the grad and curl spherical harmonics satisfy
the orthonormality relations
\be
\begin{aligned}
\int_{S^2} {\rm d}^2\Omega_{\hat k}\> 
Y^{G}_{(lm)ab} (\hat{k}) Y^{G}_{(l'm')}{}^{ab\,*}(\hat{k}) 
&=\delta_{ll'}\delta_{mm'}\,, 
\\
\int_{S^2} {\rm d}^2\Omega_{\hat k}\> 
Y^{C}_{(lm)ab} (\hat{k}) Y^{C}_{(l'm')}{}^{ab\,*}(\hat{k}) 
&=\delta_{ll'}\delta_{mm'}\,,
\\
\int_{S^2} {\rm d}^2\Omega_{\hat k}\> 
Y^{G}_{(lm)ab} (\hat{k}) Y^{C}_{(l'm')}{}^{ab\,*}(\hat{k}) 
&=0\,.
\label{e:YGCOrthog}
\end{aligned}
\ee

\section{Legendre polynomials and associated Legendre functions}
\label{s:legendre_polynomials}

The following is a list of some useful identities involving 
Legendre polynomials $P_l(x)$ and associated Legendre functions 
$P_l^m(x)$.
For additional properties see e.g., \citet{AbramowitzStegun}.
\\

\noindent
Differential equation:
\be
(1-x^2)\frac{{\rm d}^2}{{\rm d}x^2}P_l^m(x) 
- 2x \frac{{\rm d}}{{\rm d}x}P_l^m(x)
+\left[l(l+1)-\frac{m^2}{(1-x^2)}\right]\, P_l^m(x) = 0\,.
\ee
Useful recurrence relations:
\be
\begin{aligned}
(1-x^2)\frac{{\rm d}}{{\rm d}x}P_l^m(x) 
&= \frac{1}{2l+1}\left[(l+1)(l+m)P_{l-1}^m(x)-l(l-m+1)P_{l+1}^m(x)\right]\,,
\\
\sqrt{1-x^2}\frac{{\rm d}}{{\rm d}x}P_l^m(x) 
&= \frac{1}{2}\left[(l+m)(l-m+1)P_l^{m-1}(x)-P_l^{m+1}(x)\right]\,.
\end{aligned}
\ee
Orthogonality relation (for fixed $m$):
\be
\begin{aligned}
\label{eq:PlmOrth}
\int_{-1}^1 {\rm d}x\> P_l^m(x)P_{l'}^m(x) 
&= \frac{2(l+m)!}{(2l+1)(l-m)!}\,\delta_{ll'}\,,
\\
\int_{-1}^1 {\rm d}x\> P_l(x)P_{l'}(x) 
&= \frac{2}{(2l+1)}\,\delta_{ll'}\,.
\end{aligned}
\ee
Relation to ordinary Legendre polynomials, for $m=0,1,\cdots, l$:
\be
\begin{aligned}
P_l^m(x) &= (-1)^m (1-x^2)^{m/2}\frac{{\rm d}^m}{{\rm d}x^m}P_l(x)\,, 
\\
P_l^{-m}(x) &= (-1)^m \frac{(l-m)!}{(l+m)!}P_l^m(x)\,. 
\end{aligned}
\ee
Rodrigues' formula for $P_l(x)$:
\be
P_l(x) = \frac{1}{2^l l!}\frac{{\rm d}^l}{{\rm d}x^l}\left[(x^2-1)^l\right]\,.
\ee
Series representation of Legendre polynomials:
\begin{equation}
P_l(x) 
=\sum_{k=0}^{l} (-1)^k \frac{ (l+k)! }{ (k!)^2 (l-k)! } 
\left(\frac{1-x}{2}\right)^k 
=\sum_{k=0}^l (-1)^{l+k} \frac{ (l+k)! }{ (k!)^2 (l-k)! } 
\left(\frac{1+x}{2}\right)^k\,.
\label{e:rodrigues}
\end{equation}
Useful recurrence relation:
\be
(2l+1)xP_l(x)= (l+1)P_{l+1}(x) + lP_{l-1}(x)\,,
\label{e:xPl}
\ee
which iterated yields
\be
x^2P_l(x)
= \frac{(l+2)(l+1)}{(2l+3)(2l+1)}P_{l+2}(x)
+ \frac{4l^3 + 6l^2 -1}{(2l+3)(2l+1)(2l-1)}P_l(x)
+ \frac{l(l-1)}{4l^2-1} P_{l-2}(x)\,.
\label{e:x^2Pl}
\ee
\section{Bessel functions}
\label{s:bessel}

The following is a list of some useful identities involving 
Bessel functions and spherical Bessel functions of the 
first kind,  $J_\nu(y)$ and $j_l(y)$.
For additional properties, see e.g., \citet{AbramowitzStegun}.
\\

\noindent
Integral representation of ordinary Bessel functions:
\be
J_n(y) = \frac{1}{2\pi}\frac{1}{i^n}
\int_0^{2\pi} {\rm d}\phi\>
{\rm e}^{i(n\phi+y\cos\phi)}\,.
\ee
Integral representation of spherical Bessel functions:
\be
2(-i)^lj_l(y) = 
\int_{-1}^1 {\rm d}x\>P_l(x) {\rm e}^{-i y x}\,.
\label{e:sphbess}
\ee
Relationship between ordinary and spherical Bessel functions:
\be
j_l(y)= \sqrt{\frac{\pi}{2y}} J_{l+\frac{1}{2}}(y)\,.
\label{e:jldef}
\ee
Plane wave expansion:
\be
{\rm e}^{-i2\pi f\hat k\cdot \vec x/c}
={\rm e}^{-iy\cos\theta}
= \sum_{l=0}^\infty(-i)^lj_l(y)(2l+1)P_l(\cos\theta)\,.
\ee
Asymptotic behaviour:
\begin{align}
J_n(y) &\approx \frac{1}{\Gamma(n+1)}\left(\frac{y}{2}\right)^n,
\qquad \mbox{for } 0<y\ll\sqrt{n+1}\,,
\\
J_n(y) &\approx \sqrt{\frac{2}{\pi y}} 
\left[ \cos\left(y - \frac{n \pi}{2} - \frac{\pi}{4}\right) 
+ O\left(\frac{1}{x}\right) \right]\,,
\qquad \mbox{for } y \gg 1\,,
\label{e:Jnasymp}
\\
j_l(y) &\approx \frac{1}{y} \sin \left(y-\frac{l\pi}{2}\right) 
+O\left(\frac{1}{y^{\frac{3}{2}}}\right)\,,
\qquad \mbox{for } y \gg 1\,.
\end{align}
A useful recurrence relation:
\be
j_{l-1}(y) + j_{l+1}(f) = \frac{2l+1}{y}j_l(y)\,.
\label{e:jl-1jl+1}
\ee
Another useful recurrence relation:
\be
\frac{{\rm d}j_l}{{\rm d}y} = \frac{l}{y} j_l(y)-j_{l+1}(y)\,,
\label{e:jl'}
\ee
which iterated once yields
\be
\frac{{\rm d^2}j_l}{{\rm d}y^2} = 
\frac{l(l-1)}{y^2} j_l(y)
-\frac{2l+1}{y}j_{l+1}(y) 
+j_{l+2}(y)\,,
\label{e:jl''}
\ee
and twice yields
\be
\frac{{\rm d^3}j_l}{{\rm d}y^3} = 
\frac{l(l-1)(l-2)}{y^3} j_l(y)
-\frac{3l^2}{y^2}j_{l+1}(y) 
+\frac{3(l+1)}{y}j_{l+2}(y) 
-j_{l+3}(y)\,.
\label{e:jl'''}
\ee

\section{Analytic calculation of the overlap reduction functions
for transverse tensor backgrounds}
\label{s:appTensor}

For completeness, we include here expressions for 
the overlap reduction functions for anisotropic, uncorrelated
backgrounds having the standard transverse tensor 
polarization modes of GR.
These were derived in App.~E of \cite{gair-2014}.
Here we present only the final results; 
readers should consult \cite{gair-2014} for details.
\\

\noindent
For all $l$, $m$:
\be
\Gamma^\times_{lm}(f) = 0\,,
\ee
which trivially follows from the fact that
$R_1^\times(f,\hat k)=0$ in the computational frame.
\\

\noindent
For $m=0$:
\begin{multline}
\Gamma^+_{l0}(f)
= \frac{1}{2}\sqrt{(2l+1)\pi}
\Bigg\{\left(1+\frac{1}{3}\cos\zeta\right)\,\delta_{l0}-\frac{1}{3}\left(1+\cos\zeta\right)\,\delta_{l1} 
+\frac{2}{15}\cos\zeta\,\delta_{l2}
\\
-(1+\cos\zeta){\mathcal F}^-_{0,0,l,0}(\cos\zeta)
-(1-\cos\zeta){\mathcal F}^+_{1,1,l,0}(\cos\zeta)
\Bigg\}\,.
\end{multline}
For $m=1$:
\begin{multline}
\Gamma^+_{l1}(f)
= \frac{1}{4}\sqrt{(2l+1)\pi}\sqrt{\frac{(l-1)!}{(l+1)!}}
\Bigg\{
2\sin\zeta\left(\frac{1}{3}\,\delta_{l1}-\frac{1}{5}\,\delta_{l2}\right)
\\
-\frac{(1+\cos\zeta)^{3/2}}{(1-\cos\zeta)^{1/2}}{\mathcal F}^-_{1,0,l,1}(\cos\zeta)
-\frac{(1-\cos\zeta)^{3/2}}{(1+\cos\zeta)^{1/2}}{\mathcal F}^+_{2,1,l,1}(\cos\zeta)
\Bigg\}\,.
\end{multline}
For $m=2,3,\cdots$:
\begin{multline}
\Gamma^+_{lm}(f)
=-\frac{1}{4}\sqrt{(2l+1)\pi}\sqrt{\frac{(l-m)!}{(l+m)!}}
\Bigg\{
\frac{(1+\cos\zeta)^{\frac{m}{2}+1}}{(1-\cos\zeta)^{\frac{m}{2}}}{\mathcal F}^-_{m,0,l,m}(\cos\zeta) 
-\frac{(1+\cos\zeta)^{\frac{m}{2}}}{(1-\cos\zeta)^{\frac{m}{2}-1}}{\mathcal F}^-_{m-1,-1,l,m}(\cos\zeta)
\\
+\frac{(1-\cos\zeta)^{\frac{m}{2}+1}}{(1+\cos\zeta)^{\frac{m}{2}}}{\mathcal F}^+_{m+1,1,l,m}(\cos\zeta)
-\frac{(1-\cos\zeta)^{\frac{m}{2}}}{(1+\cos\zeta)^{\frac{m}{2}-1}}{\mathcal F}^+_{m,0,l,m}(\cos\zeta)
\Bigg\}\,.
\end{multline}
For $m<0$:
\be
\Gamma^+_{lm}(f) = (-1)^m\Gamma^+_{l, -m}(f)\,.
\ee
The functions 
${\mathcal F}^{\pm}_{q,r,l,m}(\cos\zeta)$
which appear in the 
above equations are defined by
\be
\begin{aligned}
{\mathcal F}^-_{q,r,l,m}(\cos\zeta)
&\equiv\int_{-1}^{-\cos\zeta} dx\> \frac{(1+x)^q}{(1-x)^r}\frac{d^m}{dx^m}P_l(x)\,,
\\
{\mathcal F}^+_{q,r,l,m}(\cos\zeta)
&\equiv\int_{-\cos\zeta}^1 dx\> \frac{(1-x)^q}{(1+x)^r}\frac{d^m}{dx^m}P_l(x)\, .
\end{aligned}
\ee
These functions also
arise when calculating the overlap reduction functions for the 
vector-longitudinal polarization modes.
The ${\mathcal F}^\pm$
integrals can be evaluated analytically as shown in App.~\ref{s:Fpmfncs}
of this paper (or in App.~E of \cite{gair-2014}).

\section{Evaluating the $I_m(y,x)$ integral for 
the overlap reduction function for scalar-longitudinal bacgkrounds}
\label{s:scalar-longitudinal}

The response for a scalar-longitudinal background, 
Eq.~(\ref{e:CFresSL}), is singular
at $\cos\theta=-1$ if the pulsar term is not included. We must
therefore include the pulsar term when evaluating the overlap
reduction function for backgrounds of this form. We use the notation
$y_1 = 2 \pi f L_1/c$, $y_2 = 2 \pi f L_2/c$, where $L_I$ is the
distance to pulsar $I$, that was introduced in the main body of this paper. In the following, we will  ensure that we keep all terms up to constant order $(y_1)^0$, $(y_2)^0$. The final expression, Eq.~(\ref{e:Itildeapprox}), contains some terms of higher order, but these are incomplete. This will be discussed further below. The components of the overlap reduction function are given by
\be
\Gamma^L_{lm} (f) 
= \frac{1}{2}
N_l^m  \int_{-1}^1 {\rm d}x\>
\left[ \frac{x^2}{1+x} 
\left(1 - {\rm e}^{-iy_1(1+x)}\right) 
I_{m}(y_2,x) \right] P_l^m(x)\,,
\label{eq:gamLlmint}
\ee
where
\be
I_{m}(y,x)
=\int_0^{2\pi} {\rm d}\phi\>
\frac{(\sqrt{1-x^2}\sin\zeta\cos\phi+x\cos\zeta)^2}
{1+x \cos\zeta + \sqrt{1-x^2} \sin\zeta\cos\phi} 
\left(1 - {\rm e}^{iy(1+x \cos\zeta + \sqrt{1-x^2} \sin\zeta\cos\phi)}\right) {\rm e}^{im\phi}\,.
\ee
The integral for $I_{m}(y,x)$ can be simplified by writing
\be
\begin{aligned}
I_{m}(y,x)
&= \int_0^{2\pi}{\rm d}\phi\> 
\left[ x \cos\zeta + \sqrt{1-x^2} \sin\zeta\cos\phi - 1\right] 
\left(1 - {\rm e}^{iy(1+x \cos\zeta + \sqrt{1-x^2} \sin\zeta\cos\phi)}\right) {\rm e}^{im\phi}  
+ \tilde{I}_{m}(y,x) 
\\
&= 2\pi\left( x\cos\zeta - 1\right) 
\left( \delta_{m0} - i^m J_m(y\sin\zeta\sqrt{1-x^2}) 
{\rm e}^{iy(1+x \cos\zeta)} \right) 
\\
&\hspace{0.5in} +  \pi \sin\zeta \sqrt{1-x^2} 
\left( \delta_{|m|,1} - i^{m+1} \left[ J_{m+1}(y\sin\zeta\sqrt{1-x^2}) 
- J_{m-1}(y\sin\zeta\sqrt{1-x^2})\right]{\rm e}^{iy(1+x \cos\zeta)} \right) 
\\
&\hspace{1in} + \tilde{I}_{m}(y,x)\,,
\end{aligned}
\ee
where
\begin{equation}
\tilde{I}_{m}(y,x) 
= \int_0^{2\pi}{\rm d}\phi\>
\frac{ \left(1 - {\rm e}^{iy(1+x \cos\zeta + \sqrt{1-x^2} \sin\zeta\cos\phi)}\right)}
{1+x \cos\zeta + \sqrt{1-x^2} \sin\zeta\cos\phi} {\rm e}^{im\phi}\,,
\end{equation}
and $J_n(y)$ denotes the Bessel function of the first kind. For
large values of $y$, Bessel functions have the asymptotic form
given in Eq.~(\ref{e:Jnasymp}),
and we will use this to drop certain terms when we take the limit 
$y_I\rightarrow \infty$ later. 

To evaluate the integral
$\tilde{I}_{m}(y,x)$, we first note that $\tilde{I}_{m}(0,x)=0$ and
\be
\begin{aligned}
\frac{\partial \tilde{I}_{m}}{\partial y} 
&= -i  \int_0^{2\pi} {\rm d}\phi\> 
{\rm e}^{im\phi+iy(1+x \cos\zeta + \sqrt{1-x^2} \sin\zeta\cos\phi)} 
\\
&= -2 \pi i^{m+1} {\rm e}^{iy(1+x \cos\zeta)} J_m(y\sin\zeta\sqrt{1-x^2})\,.
\label{e:Itildedef}
\end{aligned}
\ee
This last equation can be integrated as follows.
For $1 + x \cos\zeta \neq \sin\zeta \sqrt{1-x^2}$
(which corresponds to $x + \cos\zeta \neq 0$) the integral to infinity
can be computed as
\be
\tilde{I}_{m} ( \infty,x) 
=  2 \pi (-1)^{m}\frac{1}{|\cos\zeta+x|} 
\left( \frac{\sin\zeta\sqrt{1-x^2}}{1+x\cos\zeta+|x+\cos\zeta|}\right)^{|m|}\,.
\label{e:Itildeinf}
\ee 
This is divergent at $x=-\cos\zeta$, but that is an
artefact of taking the limit $y\rightarrow \infty$. 
To evaluate $\tilde{I}_{m} (y,x)$ for finite $y$ we can write
\be
\tilde{I}_{m}(y,x) = 
\tilde{I}_{m} ( \infty,x) 
+ 2 \pi i^{m+1}\int_y^\infty {\rm d}\bar y\>
{\rm e}^{i\bar y(1+x \cos\zeta)} J_m(\bar y\sin\zeta\sqrt{1-x^2})\,.
\ee
For the range of $\bar y$ in the integral, we can approximate the Bessel
function using Eq.~(\ref{e:Jnasymp}). The corrections to this
approximation take the form of trigonometric functions times factors
of $1/{\bar y}^{3/2}$ and will contribute terms of order $1/\sqrt{y}$ and
smaller to the result. To obtain a result accurate to at least $O(y_1^0, y_2^0)$, we therefore just need to
evaluate
\begin{align}
&\sqrt{\frac{2}{\pi \sin\zeta\sqrt{1-x^2}}} \int_y^\infty {\rm d}\bar y\> 
\frac{1}{\sqrt{\bar y}} {\rm e}^{i\bar y(1+x \cos\zeta)} 
\cos\left(\bar y \sin\zeta\sqrt{1-x^2} - \frac{m \pi}{2} - \frac{\pi}{4}\right) 
\nonumber\\
&\hspace{1.5in}
=\sqrt{\frac{2}{\pi \sin\zeta\sqrt{1-x^2}}} 
\left[\frac{i^m{\rm e}^{i\pi/4}}{\sqrt{1+x_-}}F_c\left(\sqrt{y(1+x_-)}\right)
+\frac{(-i)^m{\rm e}^{-i\pi/4}}{\sqrt{1+x_+}}F_c\left(\sqrt{y(1+x_+)}\right)
\right]\,,
\end{align}
where $x_{\pm}$ is shorthand notation for 
\be
x_\pm \equiv x\cos\zeta\pm\sin\zeta\sqrt{1-x^2}\,,
\ee
and
\be
F_c(y) = \int_y^\infty {\rm d}u\>{\rm e}^{iu^2} 
= \frac{\sqrt{\pi}}{2} {\rm e}^{i\pi/4} 
- \sqrt{\frac{\pi}{2}}
\left[C\left( \sqrt{\frac{2}{\pi}} y\right) 
+iS\left( \sqrt{\frac{2}{\pi}} y\right)
\right]\,. 
\ee
Here $C(x)$ and $S(x)$ are the Fresnel cosine and sine integrals, defined by
\be
C(y) = \int_0^y {\rm d}u\>\cos\left(\frac{\pi}{2} u^2 \right)\,,
\qquad 
S(y) = \int_0^y {\rm d}u\>\sin\left(\frac{\pi}{2} u^2 \right)\,.
\ee
Thus,
\begin{multline}
\tilde I_{m}(y,x) 
= 2 \pi (-1)^{m}\Bigg\{
\frac{1}{|\cos\zeta+x|} 
\left( \frac{\sin\zeta\sqrt{1-x^2}}{1+x\cos\zeta+|x+\cos\zeta|}\right)^{|m|}
\\
+i \sqrt{\frac{2}{\pi \sin\zeta\sqrt{1-x^2}}} 
\left[\frac{{\rm e}^{i\pi/4}}{\sqrt{(1+x_-)}}F_c\left(\sqrt{y(1+x_-)}\right)
+\frac{(-1)^m {\rm e}^{-i\pi/4}}{\sqrt{(1+x_+)}}F_c\left(\sqrt{y(1+x_+)}\right)
\right]\Bigg\}\,.
\label{e:Itildeapprox}
\end{multline}
Although the first term above is singular at $x=-\cos\zeta$, it
becomes finite when combined with the term proportional to
$F_c\left(\sqrt{y(1+x_-)}\right)$.
To see this note that
\be
\label{e:ItildeAsymp}
\begin{aligned}
&\frac{1}{|\cos\zeta+x|} 
\left( \frac{\sin\zeta\sqrt{1-x^2}}{1+x\cos\zeta+|x+\cos\zeta|}\right)^{|m|} 
+i \sqrt{\frac{2}{\pi \sin\zeta\sqrt{1-x^2}}} 
\frac{{\rm e}^{i\pi/4}}{\sqrt{1+x_-}}F_c\left(\sqrt{y(1+x_-)}\right)
\\
&\hspace{.5in}
=\frac{1}{|\cos\zeta+x|} 
\left( \frac{\sin\zeta\sqrt{1-x^2}}{1+x\cos\zeta+|x+\cos\zeta|}\right)^{|m|} 
-\frac{1}{\sqrt{2\sin\zeta\sqrt{1-x^2}}\sqrt{1+x_-}}
+\cdots
\\
&\hspace{.5in}
=\frac{1}{|\cos\zeta+x|} 
\Bigg\{
\left( \frac{\sin\zeta\sqrt{1-x^2}}{1+x\cos\zeta+|x+\cos\zeta|}\right)^{|m|} 
-\sqrt{\frac{1+x_+}{2\sin\zeta\sqrt{1-x^2}}}\,
\Bigg\} + \cdots\,,
\end{aligned}
\ee
where we used 
\be
\sqrt{1+x_+}\sqrt{1+x_-} = |x+\cos\zeta|\,,
\ee
to get the last line,
and where the dots correspond to the Fresnel cosine and
sine integral terms from $F_c$.
Since, to leading order in $x+\cos\zeta$, the expression 
in curly brackets is $-|m||\cos\zeta+x|/\sin^2\zeta$, it 
follows that (\ref{e:Itildeapprox}) for 
$\tilde I_{m}(y,x)$ is actually finite at
$x=-\cos\zeta$ and therefore integrable. 
For small values of the
argument $C(y) \approx y$ and $S(y) \approx \pi y^3/6$,
so the terms in Eq.~(\ref{e:ItildeAsymp}) represented by the dots are also finite for all $x$, and
proportional to $\sqrt{y}$ near $x = -\cos\zeta$.

In deriving expression~(\ref{e:Itildeapprox}), we have neglected some terms of $O(1/\sqrt{y})$, but terms of that order and higher are present in Eq.~(\ref{e:Itildeapprox}) so these orders have been treated inconsistently. To obtain a consistent result at $O(y_1^0, y_2^0)$, we could expand this expression and drop terms of higher order. However, keeping the incomplete higher order terms in Eq.~(\ref{e:Itildeapprox}) was found empirically to give a better approximation to numerically computed overlap reduction functions.

\section{Analytic calculation of the overlap reduction function for 
co-directional pulsars for scalar-longitudinal backgrounds}
\label{s:appCoDirectional}

For two pulsars that lie along the same line of sight as 
seen from Earth (i.e., $\cos\zeta=1$), the calculation 
of $I_m(y,x)$ can be done analytically.
For this case
\be
I_{m}(y,x)\Big|_{\cos\zeta=1} 
=\int_0^{2\pi} {\rm d}\phi\>\frac{x^2}{1 + x} 
\left(1 - {\rm e}^{iy(1+x)}\right) {\rm e}^{im\phi} 
= 2\pi\delta_{m0}\frac{x^2}{1+ x}
\left(1-{\rm e}^{iy(1+ x)}\right)\,.
\ee
The integral for $\Gamma^L_{lm}(f)$ then takes the form
\be
\begin{aligned} 
\Gamma^L_{lm} (f)\Big|_{\cos\zeta=1}
&= \pi N_l^m \delta_{m0}  \int_{-1}^1 {\rm d}x\>
\left[ \frac{x^4}{(1+x)^2}  P_l(x)
\left(1 - {\rm e}^{-iy_1(1+x)}\right) \left(1 - {\rm e}^{iy_2(1+x)}\right) 
 \right]
\\
&= \pi N_l^m \delta_{m0}  
\left[G^L_l(y_1) + G^L_l(-y_2) - G^L_l(y_1-y_2)\right]\,,
\label{eq:coaligned}
\end{aligned}
\ee
where
\be
G^L_l(y)
=\int_{-1}^1 {\rm d}x\>
\left[ \frac{x^4}{(1+x)^2}  P_l(x)
\left(1 - {\rm e}^{-iy(1+x)}\right) 
\right]\,.
\ee
By making the expansion
\be
\begin{aligned}
\frac{x^4}{(1+x)^2} 
%&= (1+x)^2 - 4(1+x) + 6 - \frac{4}{(1+x)} + \frac{1}{(1+x)^2}\\
&= 3-2x+x^2- \frac{4}{(1+x)} + \frac{1}{(1+x)^2}\,,
\end{aligned}
\ee
we can write
\be 
\begin{aligned}
G^L_l(y)
&=\int_{-1}^1 {\rm d}x\>
\left[3-2x+x^2- \frac{4}{(1+x)} + \frac{1}{(1+x)^2}\right] P_l(x)
\left(1 - {\rm e}^{-iy(1+x)}\right) 
\\
&=\frac{20}{3} \delta_{l0} - \frac{4}{3} \delta_{l1}+\frac{4}{15} \delta_{l2} 
-2 (-i)^l {\rm e}^{-iy} \left[ 
-\left(\frac{(l-1)l}{y^2} +2i\frac{l}{y}-3\right)j_l(y)
+\left(\frac{2l+1}{y} + 2i\right)j_{l+1}(y) - j_{l+2}(y)\right] \label{eq:Gly}
\\
&\hspace{1in}-4 H_l(y) + K_l(y)\,,
\end{aligned}
\ee
where
\begin{align}
H_l(y)
&=\int_{-1}^1 {\rm d}x\>
\frac{1}{(1+x)} P_l(x)
\left(1 - {\rm e}^{-iy(1+x)}\right)\,,
\label{e:Hl}
\\
K_l(y)
&=\int_{-1}^1 {\rm d}x\>
\frac{1}{(1+x)^2} P_l(x)
\left(1 - {\rm e}^{-iy(1+x)}\right)\,.
\label{e:Kl}
\end{align}
These last two integrals 
are most easily evaluated using the recursion 
relation in Eq.~(\ref{e:xPl}) for Legendre polynomials,
which for this calculation is most conveniently 
rewritten as:
\be
P_l(x) = 
-\frac{(2l-1)}{l}P_{l-1}(x)
-\frac{(l-1)}{l}P_{l-2}(x)
+\frac{(2l-1)}{l}(1+x)P_{l-1}(x)\,,
\qquad{\rm for\ }l\ge 2.
\label{e:codirPlrec}
\ee
This leads to 
\be
\begin{aligned}
H_0(y) &= {\rm Cin}(2y)+i {\rm Si}(2y)\,,
\\
H_1(y) &= -H_0(y) +2 +\frac{i}{y}\left(1- {\rm e}^{-2iy}\right)\,,
\\
H_l(y) &= -\frac{(2l-1)}{l}H_{l-1}(y) - \frac{(l-1)}{l}H_{l-2}(y) 
- 2 (-i)^{l-1} \frac{(2l-1)}{l} {\rm e}^{-iy} j_{l-1}(y)\,,
\qquad{\rm for\ } l\ge 2\,,
\end{aligned}
\ee
and
\be
\begin{aligned}
K_0(y) &= 
\left(\frac{\cos(2y)-1}{2} + y\,{\rm Si}(2y)
\right)
+i\left(-\frac{1}{2}\sin(2y) - y\,{\rm Cin}(2y)+ y\left[1+\int_{-1}^1\frac{{\rm d}x}{1+x}\right]
\right)\,,
\\
K_1(y) &= H_0(y) - K_0(y)\,,
\\
K_l(y) &= -\frac{(2l-1)}{l}K_{l-1}(y) - \frac{(l-1)}{l}K_{l-2}(y) 
+\frac{(2l-1)}{l} H_{l-1}(y)\,,
\qquad {\rm for\ } l\ge 2\,,
\end{aligned}
\ee
where ${\rm Si}(x)$ and ${\rm Cin}(x)$ denote the sine and 
cosine integrals respectively, defined by
\be
{\rm Si}(x) = \int_{0}^x {\rm d}t\>\frac{\sin t}{t}\,.
\qquad
{\rm Cin}(x) = \int_{0}^x {\rm d}t\>\frac{1-\cos t}{t}\,.
\label{eq:cinsi}
\ee
Note that the last two terms (in square brackets) in the above
expression for $K_0(y)$ will cancel when forming the combination
$K_0(y_1)+K_0(-y_2)-K_0(y_1-y_2)$, which
enters the expression for $\Gamma^L_{lm}(f)$.
The above recursion relations are particularly useful when the 
values of $H_l(y)$ and $K_l(y)$ are required at fixed $y$ for 
all $l \leq l_{\rm max}$.

For isotropic backgrounds ($l=m=0$), 
an expression for the scalar-longitudinal overlap reduction 
function for equidistant ($y_1=y_2\equiv y$), co-directional 
($\cos\zeta=1$) pulsars valid in the limit $y \gg 1$ 
was given in~\citet{chamberlin2012}. 
Equation~(\ref{eq:coaligned}) reduces to that result in the 
appropriate limit, as we now show. 

For equidistant pulsars and $l=0, m=0$, 
the last term in Eq.~(\ref{eq:coaligned}) is $G^L_0(0)$, which is zero. 
This can be seen by direct evaluation or by noting that the last term 
in square brackets in Eq.~(\ref{eq:Gly}) reduces to 
$\left[3j_0(y)+(1/y + 2i)j_1(y) - j_2(y)\right]$ for $l=0$, 
which tends to $10/3$ as $y\rightarrow 0$. 
When multiplied by the pre-factor of $-2$, 
this cancels the first term in Eq.~(\ref{eq:Gly}). 
Likewise, $H_0(0)=0$ and $K_0(0)=0$, so $G^L_0(0)=0$. 
The equidistant, co-aligned, isotropic overlap reduction 
function is therefore
\be
\Gamma^L_{00}(f)|_{\cos\zeta=1} 
= \frac{\sqrt\pi}{2}\left[G^L_0(y) + G^L_0(-y)\right]
= \frac{\sqrt\pi}{2}\left[G^L_0(y) + G^L_0(y)^*\right]
= \sqrt\pi\,{\rm Re}\{G^L_0(y)\}\,.
\ee
We now evaluate this expression in the limit $y\gg 1$. 
All spherical Bessel functions decay to zero as $y\rightarrow\infty$, 
so the term in square brackets in Eq.~(\ref{eq:Gly}) 
makes no contribution in this limit. 
Hence, we focus on the behaviour of $H_0(y)$ and $K_0(y)$ for large $y$. 
We make use of the following asymptotic expressions:
\be
\begin{aligned}
\mathrm{Si}(y) \approx \frac{\pi}{2}\,,
&\qquad y\gg 1\,,
\\
\mathrm{Cin}(y) \approx \gamma + \ln(y)\,,
&\qquad y\gg 1\,,
\end{aligned}
\ee
where $\gamma$ is the Euler-Masheroni constant, $\gamma=0.57722\cdots$.
We deduce that, for large $y$,
\begin{equation}
G^L_0(y) 
\approx 
\frac{20}{3} 
- 4(\gamma  +\ln(2y)) 
- i2\pi -\frac{1}{2} + \frac{\pi y}{2} 
- iy(\gamma+\ln(2y)) 
+ \frac{1}{2}{\rm e}^{-2iy}\,,
\end{equation}
so
\be
\begin{aligned}
{\rm Re}\{G^L_0(y)\} 
&\approx
\frac{37}{6} - 4\gamma  - 4\ln(2y) + \frac{\pi y}{2} + \frac{1}{2}\cos(2y)\,,
\\
&\approx
\frac{37}{6} - 4\gamma  - 4\ln(2y) + \frac{\pi y}{2}\,,
\end{aligned}
\ee
and
\be
\begin{aligned}
\Gamma^L_{00}(f)|_{\cos\zeta=1} 
&\approx \sqrt\pi\left[\frac{37}{6} 
- 4\gamma  - 4\ln(2y) + \frac{\pi y}{2}\right]\,,
\\
&\approx \sqrt\pi\left[\frac{37}{6} 
- 4\gamma  - 4\ln\left(\frac{4\pi fL}{c}\right) + 
\frac{\pi^2 fL}{c}\right]\,,
\end{aligned}
\ee
where $f$ is the gravitational-wave frequency and 
$L$ is the distance of the two pulsars from the Earth. 
This agrees with Eq.~(40) of \citet{chamberlin2012}, 
apart from a factor of $\sqrt\pi$, which comes from a 
difference in our normalization convention.

\section{Analytic calculation of the overlap reduction function 
for anti-directional pulsars for scalar-longitudinal backgrounds}
\label{s:appAntiDirectional}

For two pulsars that lie in antipodal positions along the same line of sight as 
seen from Earth (i.e., $\cos\zeta=-1$), the calculation 
of $I_m(y,x)$ can also be done analytically.
For this case
\be
I_{m}(y,x)\Big|_{\cos\zeta=-1} 
=\int_0^{2\pi} {\rm d}\phi\>\frac{x^2}{1 - x} 
\left(1 - {\rm e}^{iy(1-x)}\right) {\rm e}^{im\phi} 
= 2\pi\delta_{m0}\frac{x^2}{1- x}
\left(1-{\rm e}^{iy(1- x)}\right)\,.
\ee
The integral for $\Gamma^L_{lm}(f)$ then takes the form
\begin{align}
\Gamma^L_{lm} (f)\Big|_{\cos\zeta=-1}
&= \pi N_l^m \delta_{m0}  \int_{-1}^1 {\rm d}x\>
\left[ \frac{x^4}{1-x^2}  P_l(x)
\left(1 - {\rm e}^{-iy_1(1+x)}\right) \left(1 - {\rm e}^{iy_2(1-x)}\right) 
 \right]\,.
 \end{align}
By making the expansion
\be
\begin{aligned}
\frac{x^4}{(1-x^2)} 
&= -1-x^2 + \frac{1}{2(1+x)} + \frac{1}{2(1-x)}\,,
\end{aligned}
\ee
we can write
\be
\begin{aligned}
\Gamma^L_{lm} (f)\Big|_{\cos\zeta=-1}
&= \pi N_l^m \delta_{m0} \int_{-1}^1 {\rm d}x\>
\left[-1-x^2 +\frac{1}{2(1+x)} + \frac{1}{2(1-x)}\right] P_l(x)
\left(1 - {\rm e}^{-iy_1(1+x)}\right) \left(1 - {\rm e}^{iy_2(1-x)}\right) 
\\
&=\pi N_l^m \delta_{m0} \left[-\frac{8}{3} \delta_{l0} - \frac{4}{15} \delta_{l2} + 2(-i)^l {\rm e}^{-i y_1} \left( \left[1 - \frac{l(l-1)}{y_1^2} \right] j_l(y_1) +\frac{2l+1}{y_1} j_{l+1}(y_1) - j_{l+2}(y_1) \right) \right.
\\
&\qquad+2(-i)^l  {\rm e}^{i y_2} \left(  \left[1 - \frac{l(l-1)}{y_2^2} \right] j_l(y_2) +\frac{2l+1}{y_2} j_{l+1}(y_2) - j_{l+2}(y_2) \right) \\
&\qquad- 2(-i)^l  {\rm e}^{i (y_2-y_1)} \left(  \left[1 - \frac{l(l-1)}{(y_1+y_2)^2} \right] j_l(y_1+y_2) +\frac{2l+1}{y_1+y_2} j_{l+1}(y_1+y_2) - j_{l+2}(y_1+y_2)\right)
\\
&\left.\hspace{1in}+\frac{1}{2} \tilde{H}_l(y_1,y_2) + \frac{1}{2}\tilde{H}^*_l(y_2,y_1)\right]\,,
\end{aligned}
\ee
where
\begin{align}
\tilde{H}_l(y_1,y_2)
&=\int_{-1}^1 {\rm d}x\>
\frac{1}{(1+x)} P_l(x)
\left(1 - {\rm e}^{-iy_1(1+x)}\right)\,\left(1 - {\rm e}^{iy_2(1-x)}\right)\,.
\label{e:Hltilde}
\end{align}
This final integral can be obtained via a recurrence relation using Eq.~(\ref{e:codirPlrec}) from App.~\ref{s:appCoDirectional}. We find
\be
\begin{aligned}
\tilde{H}_0(y_1,y_2) &= {\rm Cin}(2y_1)+i {\rm Si}(2y_1) + {\rm e}^{2iy_2} 
\left({\rm Cin}(2y_1)+i {\rm Si}(2y_1)  - {\rm Cin}\left[2(y_1+y_2)\right]
-i {\rm Si}\left[2(y_1+y_2)\right] \right)\,,  
\\
\tilde{H}_1(y_1,y_2) &= 
- \tilde{H}_0(y_1,y_2)
+2 \left( 1 - \frac{\sin y_1}{y_1} {\rm e}^{-i y_1}- \frac{\sin y_2}{y_2} {\rm e}^{i y_2} + \frac{\sin(y_1+y_2)}{y_1+y_2} {\rm e}^{i (y_2-y_1)} \right)\,, 
\\
\tilde{H}_l(y_1,y_2) &= -\frac{(2l-1)}{l}\tilde{H}_{l-1}(y_1,y_2) - \frac{(l-1)}{l}\tilde{H}_{l-2}(y_1,y_2) 
\\&\qquad- 2 (-i)^{l-1} \frac{(2l-1)}{l} \left[{\rm e}^{-iy_1} j_{l-1}(y_1) + {\rm e}^{iy_2} j_{l-1}(y_2) - {\rm e}^{i(y_2-y_1)} j_{l-1}(y_1+y_2)\right]\,,
\qquad{\rm for\ } l\ge 2\,,
\end{aligned}
\ee
where, as before, ${\rm Si}(x)$ and ${\rm Cin}(x)$ denote the sine and 
cosine integrals, which were defined in Eq.~(\ref{eq:cinsi}). 
The result for $\tilde{H}_0(y_1,y_2)$ can be obtained by rewriting 
Eq.~(\ref{e:Hltilde}) as a combination of integrals of the following four forms:
\be
\begin{aligned}
\int_0^{2y} {\rm d}u\>
\left( \frac{1-\cos u}{u} \right) \cos (a u) 
&= \frac{1}{2} {\rm Cin}[2(a+1)y] + \frac{1}{2} {\rm Cin}[2(a-1)y]  
- {\rm Cin}(2ay)\,,
\\
\int_0^{2y} {\rm d}u\>
\left( \frac{1-\cos u}{u} \right) \sin (a u) 
&= {\rm Si}(2ay) - \frac{1}{2} {\rm Si}[2(a+1)y]  
- \frac{1}{2} {\rm Si}[2(a-1)y]\,, 
\\
\int_0^{2y} {\rm d}u\>
\frac{\sin u}{u} \cos(a u) 
&=  \frac{1}{2} {\rm Si}[2(a+1)y]  - \frac{1}{2} {\rm Si}[2(a-1)y]\,,
\\
\int_0^{2y} {\rm d}u\>
\frac{\sin u}{u} \sin(a u) 
&=  \frac{1}{2} {\rm Cin}[2(a+1)y]  - \frac{1}{2} {\rm Cin}[2(a-1)y]\,.
\end{aligned}
\ee

\section{Analytic calculation of the overlap reduction functions
for vector-longitudinal backgrounds}
\label{s:vector-ORF}

Ignoring the pulsar terms, the overlap reduction functions for 
an uncorrelated, unpolarised, anisotropic vector-longitudinal 
background are given by $\Gamma^Y_{lm} (f) =0$ and
\be
\begin{aligned}
&\Gamma^X_{lm} (f) 
\\
&\qquad = -N_l^m \int_{-1}^1{\rm d}x \int_{0}^{2\pi}{\rm d}\phi 
\left[ \frac{x\sqrt{1-x^2}}{1+x}\frac{(\sin\zeta\cos\phi\sqrt{1-x^2}+x\cos\zeta)
(x\sin\zeta\cos\phi-\sqrt{1-x^2}\cos\zeta)}{1+x \cos\zeta + \sqrt{1-x^2} \sin\zeta\cos\phi} \right] 
P_l^m(x) {\rm e}^{im\phi}
\\
&\qquad =  N_l^m \int_{-1}^1{\rm d}x \int_{0}^{2\pi} {\rm d}\phi\>
\frac{x}{1+x}
\left[\left( x + \cos\zeta(1-x^2) - x \sqrt{1-x^2} \sin\zeta \cos\phi \right) \right.
\\
&\hspace{2in} \left. -\frac{(x+\cos\zeta)}{(1+x \cos\zeta + \sqrt{1-x^2} \sin\zeta\cos\phi)}\right] 
P_l^m(x) {\rm e}^{im\phi}
\\
&\qquad= 2 \pi N_l^m\left( I_{lm} + J_{lm} \right),
\end{aligned}
\ee
where
\be
\begin{aligned}
I_{lm} &= -\frac{\sin\zeta}{2} \left(\delta_{m,1}+\delta_{m,-1}\right) 
\int_{-1}^1 {\rm d}x\>x^2 \sqrt{\frac{1-x}{1+x}} P_l^m(x)\,,
\\
J_{lm} &= \int_{-1}^1 {\rm d}x\>\left[ (x+(1-x^2)\cos\zeta) \delta_{m,0} 
- (x+\cos\zeta) K_{lm}(x) \right] \frac{x}{1+x} P_l^m(x)\,,
\\
K_{lm}(x) &= \frac{1}{2\pi} \int_0^{2\pi} {\rm d}\phi\>
\frac{{\rm e}^{im\phi}}{1+x \cos\zeta + \sqrt{1-x^2} \sin\zeta\cos\phi}\,.
\end{aligned}
\ee
The integral $K_{lm}(x)$ can be evaluated using contour integration, as
described in \cite{gair-2014} for the response of a PTA to anisotropic
backgrounds with GR polarisations. The result is 
\be
\begin{aligned}
K_{lm}(x) &= \frac{1}{|x+\cos\zeta|} 
\left( \frac{|x+\cos\zeta|-1-x\cos\zeta}{\sqrt{1-x^2}\sin\zeta}\right)^{|m|} 
\\
&= \left\{
\begin{array}{ll} 
\frac{(-1)^{|m|}}{x+\cos\zeta} 
\left( \frac{(1-x)(1-\cos\zeta)}{(1+x)(1+\cos\zeta)}\right)^{\frac{|m|}{2}}\,, 
\quad -\cos\zeta < \cos\theta < 1 
\\
\frac{(-1)^{|m|+1}}{x+\cos\zeta} \left( \frac{(1+x)(1+\cos\zeta)}{(1-x)(1-\cos\zeta)}\right)^{\frac{|m|}{2}}\,,
\quad -1 < \cos\theta < -\cos\zeta\,. 
\end{array} \right.
\end{aligned}
\ee
The non-zero $I_{lm}$'s can be straightforwardly evaluated:
\be
N_l^1 I_{l1}
=-N_l^{-1} I_{l,-1}
= \frac{\sin\zeta N_l^1}{2} \left( 2(-1)^{l+1} +2\delta_{l0} 
- \frac{4}{3}\delta_{l1} + \frac{4}{5} \delta_{l2} \right)\,.
\ee
The $J_{lm}$'s can be written in terms of the 
${\mathcal F}^{\pm}_{q,r,L,m}(\cos\zeta)$ functions defined in \cite{gair-2014}:
\be
\begin{aligned}
{\mathcal F}^-_{q,r,L,m}(\cos\zeta)
&\equiv\int_{-1}^{-\cos\zeta} dx\> \frac{(1+x)^q}{(1-x)^r}\frac{d^m}{dx^m}P_L(x)\,,
\\
{\mathcal F}^+_{q,r,L,m}(\cos\zeta)
&\equiv\int_{-\cos\zeta}^1 dx\> \frac{(1-x)^q}{(1+x)^r}\frac{d^m}{dx^m}P_L(x)\, .
\end{aligned}
\ee
For $m=0$ we have
\be
J_{l0} = \frac{2}{3} \cos\zeta \left( -\delta_{l0}+\delta_{l1}-\frac{2}{5} \delta_{l2}\right) 
-2\delta_{l0} + {\mathcal F}^-_{1,0,l,0}(\cos\zeta) 
+ 2{\mathcal F}^+_{0,1,l,0}(\cos\zeta) - {\mathcal F}^+_{1,0,l,0}(\cos\zeta)\,,
\ee
while for $m > 0$ we have
\be
J_{lm} 
= \left( \frac{1+\cos\zeta}{1-\cos\zeta}\right)^{\frac{m}{2}} 
\left({\mathcal F}^-_{m,0,l,m}(\cos\zeta) -{\mathcal
          F}^-_{m-1,0,l,m}(\cos\zeta) \right) 
- \left( \frac{1-\cos\zeta}{1+\cos\zeta}\right)^{\frac{|m|}{2}} 
\left({\mathcal F}^+_{m,0,l,m}(\cos\zeta) - {\mathcal F}^+_{m,1,l,m}(\cos\zeta) 
\right)\,,
\ee
and $N_l^{-m} J_{l,-m} = (-1)^m N_l^m J_{lm}$.
Explicit expressions for the ${\mathcal F}^{\pm}_{q,r,L,m}(\cos\zeta)$ 
functions are given in App.~\ref{s:Fpmfncs}.

%%%%%%%%%%%%%%%%%%%%%%%%%%%%%%%%%%%%%%%%%%
\subsection{Limiting case: $\cos\zeta=1$}
\label{s:vector-ORF-coszeta1}

As noted in the main text, in the limit $\cos\zeta \rightarrow 1$, 
the $m=0$ overlap reduction functions calculated above diverge.
This singularity is eliminated if the pulsar terms are included 
in the integrand, and the pulsars are assumed to be at finite distance. 
Proceeding in a fashion identical to the case of 
co-directional pulsars in scalar-longitudinal backgrounds, we find
\be
\begin{aligned}
\Gamma^X_{lm}(f)\Big|_{\cos\zeta=1} 
&= 2 \pi N_l^0 \delta_{m0}\int_{-1}^1 {\rm d}x\> \frac{x^2 (1-x)}{1+x} P_l(x)  
\left(1- {\rm e}^{-i y_1 (1+x)} \right)  \left(1- {\rm e}^{i y_2 (1+x)} \right) 
\\
&= 2\pi N_l^0 \delta_{m0} \left[ G^X_l(y_1) + G^X_l(-y_2) - G^X_l(y_1-y_2)\right]\,,
\label{e:gammaxlm}
\end{aligned}
\ee
where
\be
\begin{aligned}
G^X_l(y) 
&= \int_{-1}^{1}{\rm d}x\>
\frac{x^2 (1-x)}{1+x} P_l(x) \left(1- {\rm e}^{-i y (1+x)} \right).
\\
&= \int_{-1}^{1}{\rm d}x\>
\left[-2 + 2x -x^2 +\frac{2}{1+x}\right]
P_l(x) \left(1- {\rm e}^{-i y (1+x)} \right) 
\\
&=-\frac{14}{3} \delta_{l0} + \frac{4}{3} \delta_{l1}-\frac{4}{15} \delta_{l2} 
-2 (-i)^l {\rm e}^{-iy} \left[ 
\left(\frac{(l-1)l}{y^2} +2i\frac{l}{y}-2\right)j_l(y)
\right.
\\
&\hspace{2.5in}
\left.
-\left(\frac{2l+1}{y} + 2i\right)j_{l+1}(y) + j_{l+2}(y)\right]+ 2 H_l(y)\,,
\label{eq:GlX}
\end{aligned}
\ee 
with $H_l(y)$ defined as in Eq.~(\ref{e:Hl}). 
This is a finite expression provided $y_1$ and $y_2$ are finite.

\section{Evaluating the ${\mathcal F}^\pm$ integrals for
transverse tensor and vector-longitudinal backgrounds}
\label{s:Fpmfncs}

The overlap reduction functions for both the standard transverse tensor 
and vector-longitudinal backgrounds can be written in terms of the functions
\begin{align}
{\mathcal F}^-_{q,r,L,m}(\cos\zeta)
&\equiv\int_{-1}^{-\cos\zeta} dx\> \frac{(1+x)^q}{(1-x)^r}\frac{d^m}{dx^m}P_L(x)\,,
\\
{\mathcal F}^+_{q,r,L,m}(\cos\zeta)
&\equiv\int_{-\cos\zeta}^1 dx\> \frac{(1-x)^q}{(1+x)^r}\frac{d^m}{dx^m}P_L(x)\, .
\end{align}
These integrals can be 
evaluated using the series representation of the Legendre polynomials
\begin{equation}
P_l(x) = \sum_{k=0}^{l} (-1)^k \frac{ (l+k)! }{ (k!)^2 (l-k)! } 
\left(\frac{1-x}{2}\right)^k 
=\sum_{k=0}^l (-1)^{l+k} \frac{ (l+k)! }{ (k!)^2 (l-k)! } 
\left(\frac{1+x}{2}\right)^k .
\end{equation}
Explicitly, we find
\be
\begin{aligned}
{\mathcal F}^-_{q,r,L,m}(\cos\zeta)
&\equiv\int_{-1}^{-\cos\zeta} {\rm d}x\> 
\frac{(1+x)^q}{(1-x)^r}\frac{{\rm d}^m}{{\rm d}x^m}P_L(x) 
\\ 
&= \sum_{i=0}^q \sum_{j=m}^L 2^{i-j} (-1)^{q-i+j+m} 
\frac{ q! (L+j)!} {i! (q-i)! j! (L-j)! (j-m)! } 
\int_{-1}^{-\cos\zeta} {\rm d}x\>(1-x)^{q-i-r+j-m}\,,
\end{aligned}
\ee
for which
\be
\begin{aligned}
{\mathcal F}^-_{q,0,L,m}(\cos\zeta) 
&=\sum_{i=0}^q \sum_{j=m}^L 2^{i-j} (-1)^{q-i+j+m} 
\frac{ q! (L+j)! \left(2^{q-i+j-m+1} - (1+\cos\zeta)^{q-i+j-m+1}\right)} {i! (q-i)! j! (L-j)! (j-m)! (q-i+j-m+1)}\,,
\\
{\mathcal F}^-_{q,1,L,m}(\cos\zeta) 
&=\sum_{i=0}^{q-1} \sum_{j=m}^L 2^{i-j} (-1)^{q-i+j+m} 
\frac{ q! (L+j)! \left(2^{q-i+j-m} - (1+\cos\zeta)^{q-i+j-m}\right)} {i! (q-i)! j! (L-j)! (j-m)! (q-i+j-m)} 
\\
& \hspace{0.4cm} + \sum_{j=m+1}^L 2^{q-j} (-1)^{j+m} 
\frac{ (L+j)! \left(2^{j-m} - (1+\cos\zeta)^{j-m}\right)} {j! (L-j)! (j-m)! (j-m)} 
\\
& \hspace{0.8cm} + \frac{2^{q-m} (L+m)!}{m! (L-m)!} \ln\left(\frac{2}{1+\cos\zeta}\right)\,.
\end{aligned}
\ee
Similarly,
\be
\begin{aligned}
{\mathcal F}^+_{q,r,L,m}(\cos\zeta)
&\equiv\int_{-\cos\zeta}^1 {\rm d}x\> 
\frac{(1-x)^q}{(1+x)^r}\frac{{\rm d}^m}{{\rm d}x^m}P_L(x)
\\ 
&= \sum_{i=0}^q \sum_{j=m}^L 2^{i-j} (-1)^{L+q-i+j} \frac{ q! (L+j)!} {i! (q-i)! j! (L-j)! (j-m)! } 
\int_{-\cos\zeta}^1 {\rm d}x\>(1+x)^{q-i-r+j-m}\,,
\end{aligned}
\ee
for which
\be
\begin{aligned}
{\mathcal F}^+_{q,0,L,m}(\cos\zeta) 
&=\sum_{i=0}^q \sum_{j=m}^L 2^{i-j} (-1)^{L+q-i+j} 
\frac{ q! (L+j)! \left(2^{q-i+j-m+1} - (1-\cos\zeta)^{q-i+j-m+1}\right)} {i! (q-i)! j! (L-j)! (j-m)! (q-i+j-m+1)}\,,
\\
{\mathcal F}^+_{q,1,L,m}(\cos\zeta) 
&=\sum_{i=0}^{q-1} \sum_{j=m}^L 2^{i-j} (-1)^{L+q-i+j} 
\frac{ q! (L+j)! \left(2^{q-i+j-m} - (1-\cos\zeta)^{q-i+j-m}\right)} {i! (q-i)! j! (L-j)! (j-m)! (q-i+j-m)} 
\\
& \hspace{0.4cm} + \sum_{j=m+1}^L 2^{q-j} (-1)^{L+j} 
\frac{ (L+j)! \left(2^{j-m} - (1-\cos\zeta)^{j-m}\right)} {j! (L-j)! (j-m)! (j-m)} 
\\
& \hspace{0.8cm} + \frac{(-1)^{L+m} 2^{q-m} (L+m)!}{m! (L-m)!} \ln\left(\frac{2}{1-\cos\zeta}\right)\,.
\end{aligned}
\ee
For the standard transverse tensor backgrounds, 
we also need to evaluate ${\mathcal F}^-_{q,r,l,m}(\cos\zeta)$ for $r=-1$.
This can be reduced to combinations
of ${\mathcal F}^-_{q,0,l,m}(\cos\zeta)$ and 
${\mathcal F}^-_{q+1,0,l,m}(\cos\zeta)$ by writing $(1-x)=2-(1+x)$:
\be
{\mathcal F}^-_{q,-1,l,m}(\cos\zeta) =
2{\mathcal F}^-_{q,0,l,m}(\cos\zeta) -{\mathcal F}^-_{q+1,0,l,m}(\cos\zeta)\,.
\ee
Alternatively, we can just evaluate this integral directly, finding
\be
{\mathcal F}^-_{q,-1,l,m}(\cos\zeta) 
=\sum_{i=0}^q \sum_{j=m}^l 2^{i-j} (-1)^{q-i+j+m} 
\frac{ q! (l+j)! \left(2^{q-i+j-m+2} - (1+\cos\zeta)^{q-i+j-m+2}\right)} 
{i! (q-i)! j! (l-j)! (j-m)! (q-i+j-m+2)}\,.
\ee

\section{Recovering the overlap reduction function for
an uncorrelated, anisotropic scalar-transverse background}
\label{s:appRecoverBreathingORF}

Ignoring the pulsar term, we can show that the response of a pulsar to
the indiviudal modes of a scalar-transverse 
gravitational-wave background can be used to 
recover the overlap reduction function for an arbitrary uncorrelated,
anisotropic background.  
Inverting Eq.~(\ref{e:hB}) to find $a_{(lm)}^{B}(f)$ gives
\be
a_{(lm)}^{B}(f) = \sqrt{2}\int_{S^2}{\rm d}^2\Omega_{\hat k}\> 
h_{B}(f,\hat{k})Y^*_{lm}(\hat{k})\,,
\ee
from which we deduce the following quadratic expectation values:
\begin{align} 
C^B_{lml'm'}(f,f') 
\equiv \langle a_{(lm)}^{B}(f)a_{(l'm')}^{B*}(f') \rangle
= 2\int_{S^2}{\rm d}^2\Omega_{\hat k}\int_{S'^2}{\rm d}^2\Omega_{\hat{k}'}\>
\langle h_{B}(f,\hat{k})h^*_{B}(f',\hat{k}') \rangle Y^*_{lm}(\hat{k})Y_{l'm'}(\hat{k}')\,,
\label{eq:<aa>}
\end{align}
where 
$C^{B}_{lml'm'}(f,f') = C^{B}_{lml'm'}H_B(f)\delta(f-f')$ assuming
stationarity. For a Gaussian-stationary, uncorrelated, anisotropic
background, the quadratic expectation value of
breathing mode amplitudes is given by Eq.~(\ref{eq:<hh>}):
\be
\langle h_{B}(f,\hat{k})h^*_{B}(f',\hat{k}') \rangle 
= H_B(f)P_B(\hat{k})\delta^2(\hat{k},\hat{k}')\delta(f-f')\,.
\ee
The angular distribution of gravitational-wave power can be expanded in terms of
scalar spherical harmonics (see Eq.~(\ref{eq:Pexpand})). Hence the
integrals over the sphere in Eq.~(\ref{eq:<aa>}) can be explicitly evaluated:
\be
\begin{aligned}
%\langle a_{(lm)}^{B}(f)a_{(l'm')}^{B*}(f') \rangle 
C^B_{lml'm'}(f,f') 
&= H_B(f)\delta(f-f')\sum_{L=0}^{\infty}\sum_{M=-L}^{L}2 P^B_{LM}
\int_{S^2}{\rm d}^2\Omega_{\hat k}\;Y_{LM}(\hat{k})Y^*_{lm}(\hat{k})Y_{l'm'}(\hat{k})
\\
&= H_B(f)\delta(f-f')\sum_{L=0}^{\infty}\sum_{M=-L}^{L}2 P^B_{LM}(-1)^m
\int_{S^2}{\rm d}^2\Omega_{\hat k}\;Y_{LM}(\hat{k})Y_{l,-m}(\hat{k})Y_{l'm'}(\hat{k})
\\
&= H_B(f)\delta(f-f')\sum_{L=0}^{\infty}\sum_{M=-L}^{L}2 P^B_{LM}(-1)^m
\sqrt{\frac{(2L+1)(2l+1)(2l'+1)}{4\pi}}\left( 
\begin{array}{ccc}L&l&l'\\M&-m&m'\end{array} \right)
\left( \begin{array}{ccc}L&l&l'\\0&0&0\end{array} \right)\,.
\end{aligned}
\ee
Now the overlap reduction function between pulsars $1$ and $2$ is given by
\be
\begin{aligned} 
\Gamma^B &= \sum_{(lm)}\sum_{(l'm')}C^{B}_{lml'm'}R^{B}_{1(lm)}R^{B*}_{2(l'm')}
\\
&= \sum_{(LM)}\sum_{(lm)}\sum_{(l'm')}2 P^B_{LM}(-1)^m\sqrt{\frac{(2L+1)(2l+1)(2l'+1)}{4\pi}}\left( 
\begin{array}{ccc}L&l&l'\\M&-m&m'\end{array} \right)\left( 
\begin{array}{ccc}L&l&l'\\0&0&0\end{array} \right)R^{B}_{1(lm)}R^{B*}_{2(l'm')}
\\
&= \sum_{(LM)}P^B_{LM}\Gamma^B_{LM}.
\label{eq:breatheoverlap reduction function}
\end{aligned}
\ee
Note that the breathing response is limited to $l=0,1$. Hence, Wigner-$3j$
selection rules restrict the sensitivity of the breathing mode 
overlap reduction function to $L\leq2$. By
substituting the breathing response function from Eq.~(\ref{e:RB-mapping-approx})
into Eq.~(\ref{eq:breatheoverlap reduction function}), we fully recover the form of
$\Gamma^B_{LM}$ obtained by direct calculation in Eq.~(\ref{eq:gammaST}). 
For example, with $L=0, M=0$, Eq.~(\ref{eq:breatheoverlap reduction function}) gives
${\Gamma^B_{00}=(\sqrt{\pi}/2)(1+\frac{1}{3}\cos\zeta)}$, 
where $\zeta$ is the
angular separation between the two pulsars. 
(Recall that $P^B_{00}=\sqrt{4\pi}/2$ for an isotropic uncorrelated background,
as described at the end of Sec.~\ref{s:ORF-mapping}.)
This exactly matches the
expression given by Eq.~(\ref{eq:gammaST}), as do the remaining expressions for $L=1,2$.
\medskip
\\

\end{widetext}
\end{appendix}
\bibliography{altpolpaper}

\end{document}